\documentclass{article}
\PassOptionsToPackage{numbers,compress}{natbib}

\usepackage[final]{neurips_2021}

\usepackage{soul}
\usepackage[utf8]{inputenc} 
\usepackage[T1]{fontenc}    
\usepackage{hyperref}       
\usepackage{url}            
\usepackage{booktabs}       
\usepackage{amsfonts}       
\usepackage{nicefrac}       
\usepackage{microtype}      
\usepackage{xcolor}         
\usepackage{graphicx}
\usepackage[subrefformat=parens]{subcaption}
\usepackage{wrapfig}
\usepackage{pgfplots}
\newcommand{\websiteref}[1][]{\href{https://galgreshler.github.io/Catch-A-Waveform/#1}{\textcolor{blue}{website}}}

\usepackage{amsmath,amsfonts,bm,relsize,dsfont}

\usepackage{color}
\usepackage{xcolor}
\usepackage{wrapfig}

\newcommand{\myparagraph}[1]{\vspace{0pt} \noindent \textbf{#1}}

\newcommand{\eg}{\textit{e.g.\ }}

\def\real{x} 
\def\fake{\tilde{x}} 

\title{Catch-A-Waveform: Learning to Generate Audio from a Single Short Example}

\author{
Gal Greshler\\ 
Technion -- Israel Institute of Technology\\
\texttt{galgreshler@gmail.com}\\
\And
Tamar Rott Shaham\\ 
Technion -- Israel Institute of Technology\\
\texttt{stamarot@campus.technion.ac.il}
\And
Tomer Michaeli\\ 
Technion -- Israel Institute of Technology\\
\texttt{tomer.m@ee.technion.ac.il}
}

\begin{document}

\maketitle

\begin{abstract}
Models for audio generation are typically trained on hours of recordings. Here, we illustrate that capturing the essence of an audio source is typically possible from as little as a few tens of seconds from a single training signal. Specifically, we present a GAN-based generative model that can be trained on one short audio signal from any domain (\eg speech, music, etc.) and does not require pre-training or any other form of external supervision. Once trained, our model can generate random  samples of arbitrary duration that maintain semantic similarity to the training waveform, yet exhibit new compositions of its audio primitives. This enables a long line of interesting applications, including generating new jazz improvisations or new a-cappella rap variants based on a single short example, producing coherent modifications to famous songs (\eg adding a new verse to a Beatles song based solely on the original recording), filling-in of missing parts (inpainting), extending the bandwidth of a speech signal (super-resolution), and enhancing old recordings without access to any clean training example. We show that in most cases, no more than 20 seconds of training audio suffice for our model to achieve state-of-the-art  
results. This is despite its complete lack of prior knowledge about the nature of audio signals in general.
\end{abstract}

\section{Introduction}
In recent years, deep models for audio generation have had an immense impact on a wide range of applications, including text-to-speech synthesis~\cite{donahue2018adversarial,marafioti2019adversarial,engel2019gansynth, binkowski2019high}, voice-to-voice translation~\cite{chandna2019wgansing, sisman2020generative}, music generation~\cite{liu2020unconditional, dhariwal2020jukebox}, singing voice conversion~\cite{chandna2019wgansing, sisman2020generative}, timbre transfer~\cite{engel2020ddsp, michelashvili2020hierarchical}, bandwidth-extension~\cite{kuleshov2017audio, birnbaum2019temporal}, and audio inpainting~\cite{marafioti2020gacela}. Existing generative models require large datasets of training signals from the domain of interest. However, there are practical scenarios in which such datasets are extremely hard to collect, or are even nonexistent. Examples include a speaker that has only recorded a few sentences, an artist that had the chance to record only a few songs, or a unique jazz improvisation appearing in one particular recording. A natural question to ask, then, is whether large amounts of training data are a necessity for training a generative model.

\begin{figure*}[t]
	\centering
	\captionsetup[subfigure]{labelformat=empty,justification=centering,aboveskip=1pt,belowskip=1pt}
	\begin{subfigure}[t]{0.06\textheight}
	    \vspace{9mm}
		\vspace{.3mm}
		\includegraphics[height=55pt]{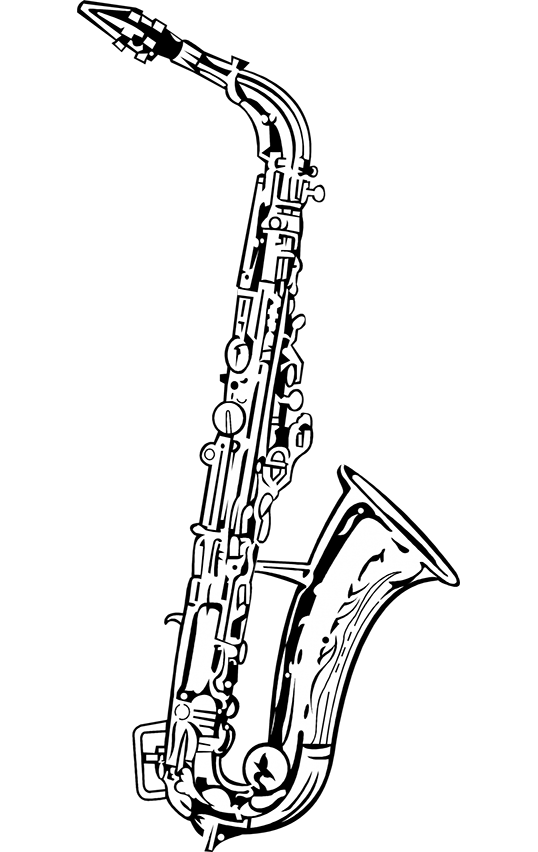}
	\end{subfigure}
	\begin{subfigure}[t]{0.11\textheight}
	    \vspace{-.3mm}
		\centering
		\caption{Single training example (20 sec.)}
		\includegraphics[height=75pt]{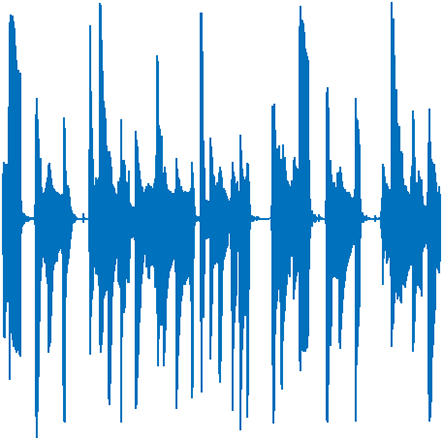}
	\end{subfigure}
	\begin{subfigure}[t]{0.03\textheight}
		\centering
	    \vspace{56pt}
	    \Large$\, \, \rightarrow$
	\end{subfigure}
	\begin{subfigure}[t]{0.55\textwidth}
	    \vspace{3.1mm}
		\centering
		\caption{Random sample (60 sec.)}
		\includegraphics[height=75pt]{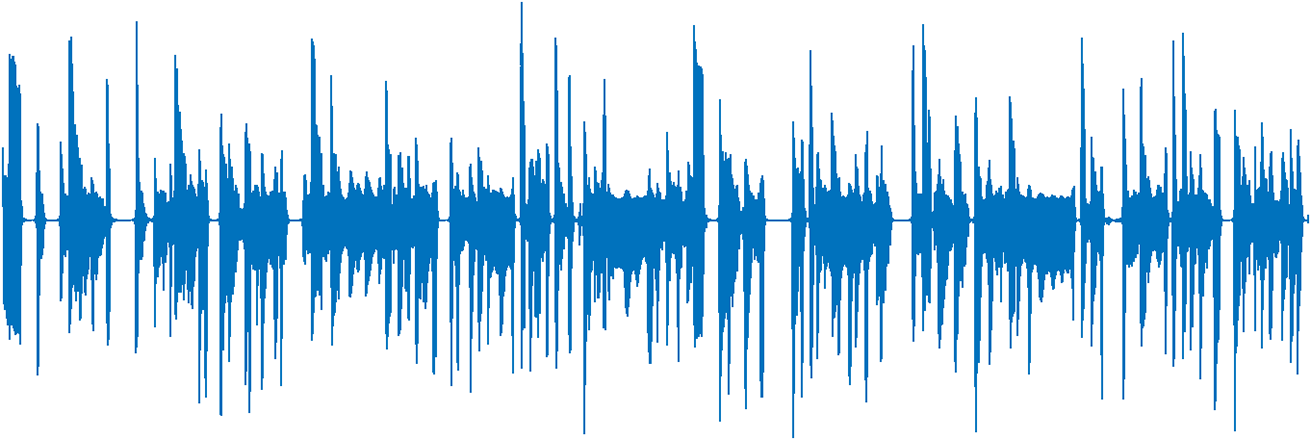}
	\end{subfigure}
	\par\smallskip
	\begin{subfigure}[t]{0.06\textheight}
	    \vspace{3mm}
		\vspace{.3mm}
		\includegraphics[height=55pt]{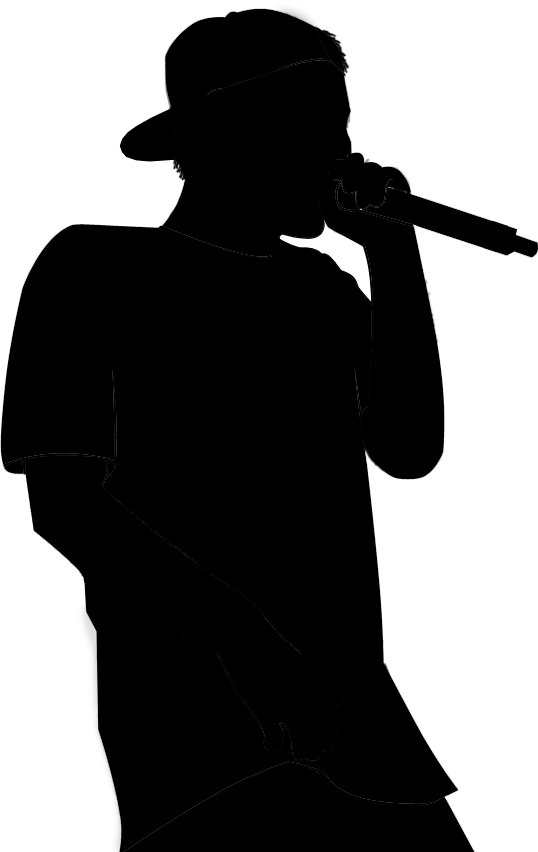}
	\end{subfigure}
	\begin{subfigure}[t]{0.11\textheight}
	    \vspace{-.3mm}
		\centering
		\includegraphics[height=75pt]{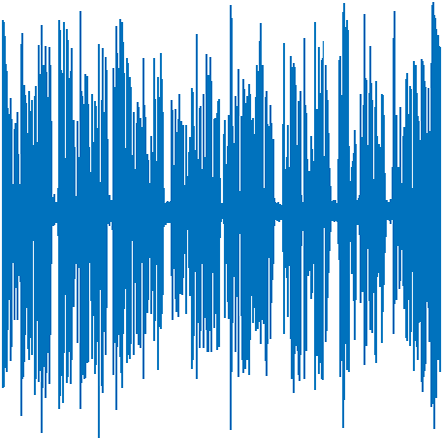}
	\end{subfigure}
	\begin{subfigure}[t]{0.03\textheight}
		\centering
	    \vspace{33pt}
	    \Large$\, \, \rightarrow$
	\end{subfigure}
	\begin{subfigure}[t]{0.55\textwidth}
	    \vspace{-.3mm}
		\centering
		\includegraphics[height=75pt]{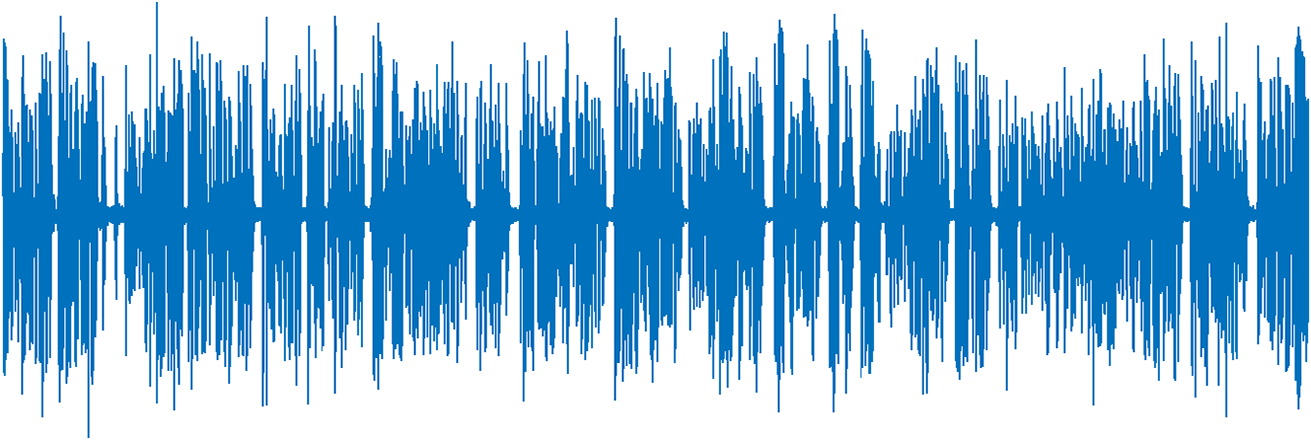}
	\end{subfigure}
	\caption{\label{fig:teaser} \textbf{Catch-A-Waveform.} We present a generative model that is able to capture the statistics of a single short audio recording (20 seconds in these examples). At inference, it can generate new diverse samples of arbitrary length, that exhibit new interesting compositions. The figure illustrates generation of new jazz improvisations and new freestyle rap variants. All examples can be listened to in our \websiteref.}
\end{figure*}

Here, we take this question to the extreme. We illustrate that capturing the essence of an audio source is possible from as little as a few tens of seconds from a single training recording. Specifically, we present a generative adversarial network (GAN) based model that can be trained on one short raw waveform and does not require pre-training or any other type of external supervision \footnote{code is available at \href{https://github.com/galgreshler/Catch-A-Waveform}{\textcolor{blue}{https://github.com/galgreshler/Catch-A-Waveform}}}. Once the model is trained, it is capable of generating diverse new signals that are semantically similar to the training recording, but contain new compositions and structures. Our model can handle different types of audio signals, from instrumental music to speech. For example, after training on 20 seconds of a saxophone solo, our model is able to generate new similar improvisations. The same can be done with a-capella rap, or old famous speeches, as exemplified in Fig.\ref{fig:teaser}. Our model can also generate samples conditioned on the low frequencies of some signal (be it the training signal or a similar one). This constraints the global structure of the generated signals, allowing to generate \eg new versions of a Beatles song (all audio samples mentioned in the paper can be found in our \websiteref).

It is important to note that a short snippet of an audio signal is insufficient for learning language (for speech) or rules of harmony (for music). Therefore, our generated signals lack the linguistic semantics or long-range harmonic structure that can be potentially achieved with externally-trained models. However, surprisingly, the coherence of our generated signals over short time scales, typically suffices for confusing listeners to believe they are real, as we confirm through extensive user studies.

\begin{figure*}
	\centering
	\includegraphics[width=\textwidth]{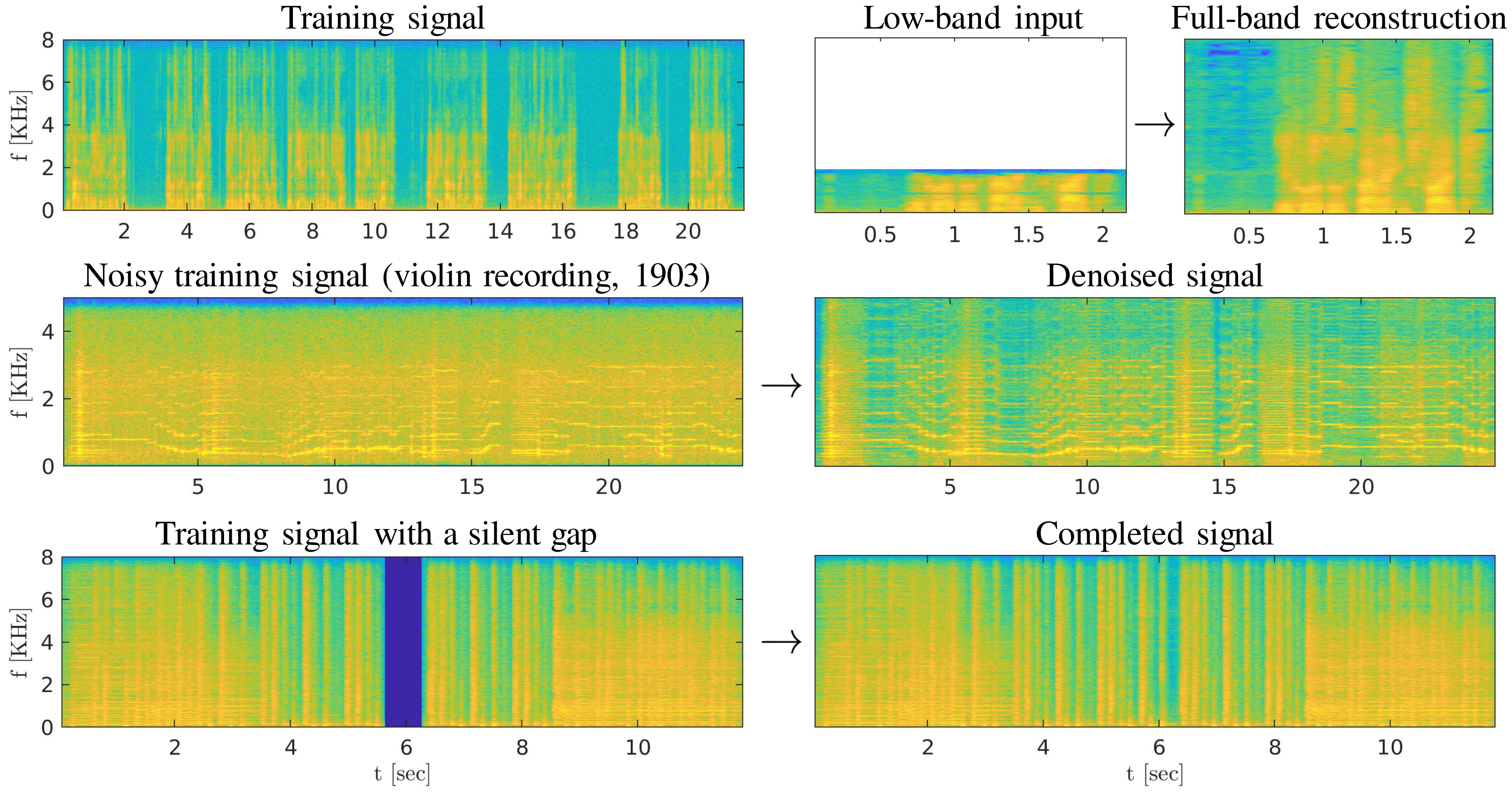}
	\caption{\textbf{Applications.}
	Our method can be used for a variety of tasks, including extending the bandwidth of a low-resolution signal, enhancing a noisy signal (without any prior knowledge on the signal or the noise), and completing missing parts.
	} 
	\label{fig:applications}
\end{figure*}

Besides generating random samples, we illustrate the utility of our approach in the common tasks of \emph{bandwidth extension}, \emph{inpainting} and \emph{denoising} (see Fig.~\ref{fig:applications}). We show that in the latter two tasks, no training signal whatsoever is required beyond the input itself. This allows handling sources for which no training data exist, like old recordings of famous musicians. In fact, our evaluation suggests that for the tasks of bandwidth extension and inpainting (sections \ref{sec:app:bandwidth} and \ref{sec:app:inpainting}), limiting the training to a single short signal is actually beneficial, and can lead to results that outperform models trained on hours of recordings.

Our work is inspired by generative models for visual data, which have been recently explored in the context of learning from a single image~\cite{shaham2019singan,shocher2019ingan} or a single short video~\cite{gur2020hierarchical}. Similarly to those works, we present a multi-scale GAN architecture that generates signals in their raw (time domain) representation. Audio signals, however, are very different from visual data; they are of high temporal resolution (usually at least 16,000 samples per second), they exhibit correlations at very long timescales, and they have diverse frequency contents. As we discuss, this necessitates dedicated architectures, losses, and adaptive selection of the multi-scale pyramid levels.

{}

\section{Related Work}
\label{sec:related}

\myparagraph{Generative models for audio.} 
Audio generation models have been extensively studied in the past few years. Some utilize autoregressive architectures~\cite{oord2016wavenet,mehri2016samplernn}, including the computationally efficient inverse autoregressive flow (IAF) scheme~\cite{oord2018parallel, ping2018clarinet} and other flow based models~\cite{kim2018flowavenet, ping2020waveflow, kim2020glow, prenger2019waveglow}. Others use GANs and variational autoencoders (VAEs), which have been found effective for many applications, including text-to-speech~\cite{donahue2018adversarial,marafioti2019adversarial,engel2019gansynth, binkowski2019high}, unconditional generation~\cite{liu2020unconditional, dhariwal2020jukebox}, singing voice conversion~\cite{chandna2019wgansing, sisman2020generative}, timbre transfer~\cite{michelashvili2020hierarchical}, inpainting~\cite{ebner2020audio, marafioti2020gacela}, bandwith-extension~\cite{kim2019bandwidth}, and denoising~\cite{pascual2017segan, li2020learning}. Several pipelines also integrate classical signal processing blocks to obtain improved results \cite{engel2020ddsp}. All these models rely on large training sets with hours of recordings. In contrast, here we focus on settings where only a single short signal is available for training.

\myparagraph{Few shot audio learning.} Audio generation models have also been taken to the few-shot regime, mainly in the context of voice cloning for speech~\cite{arik2018neural, chen2020again} and singing~\cite{nercessian2020zero}. In this setting, only a few examples are provided at \emph{test-time}. However, a \emph{large training set} is still used for learning to perform the task. Here, on the other hand, we study the use of a single short waveform for training.

\myparagraph{Internal generative learning.} Exploiting the internal statistics of a single audio example by training a deep neural network (DNN) was recently explored for the tasks of audio restoration, source separation, audio editing, and ambient sound synthesis~\cite{dap, tian2019deep}. These methods, however, cannot generate fake signals of complex structure (like music or speech). In the visual domain, recent \emph{generative} models, like SinGAN~\cite{shaham2019singan} and InGAN~\cite{shocher2019ingan}, were developed for learning from a single natural image. These approaches were later extended to other domains, including videos~\cite{gur2020hierarchical}, medical imaging~\cite{zhang2020learning}, and 3D graphics~\cite{hertz2020deep}. Here we adapt some of these ideas to the audio domain.
\section{Method}
\label{sec:method}

Consider a short sample $\real$ from a stationary audio source. 
Our goal is to learn a generative model that can draw new random samples $\fake$ from the source's distribution. Our approach is inspired by the single image GAN (SinGAN) model~\cite{shaham2019singan}. Specifically, we aim at matching the distribution of length-$T$ segments of $\fake$ to that of length-$T$ segments of $\real$, at multiple resolutions. 

\begin{figure*}[t]
\centering
\includegraphics[width=1.0\columnwidth]{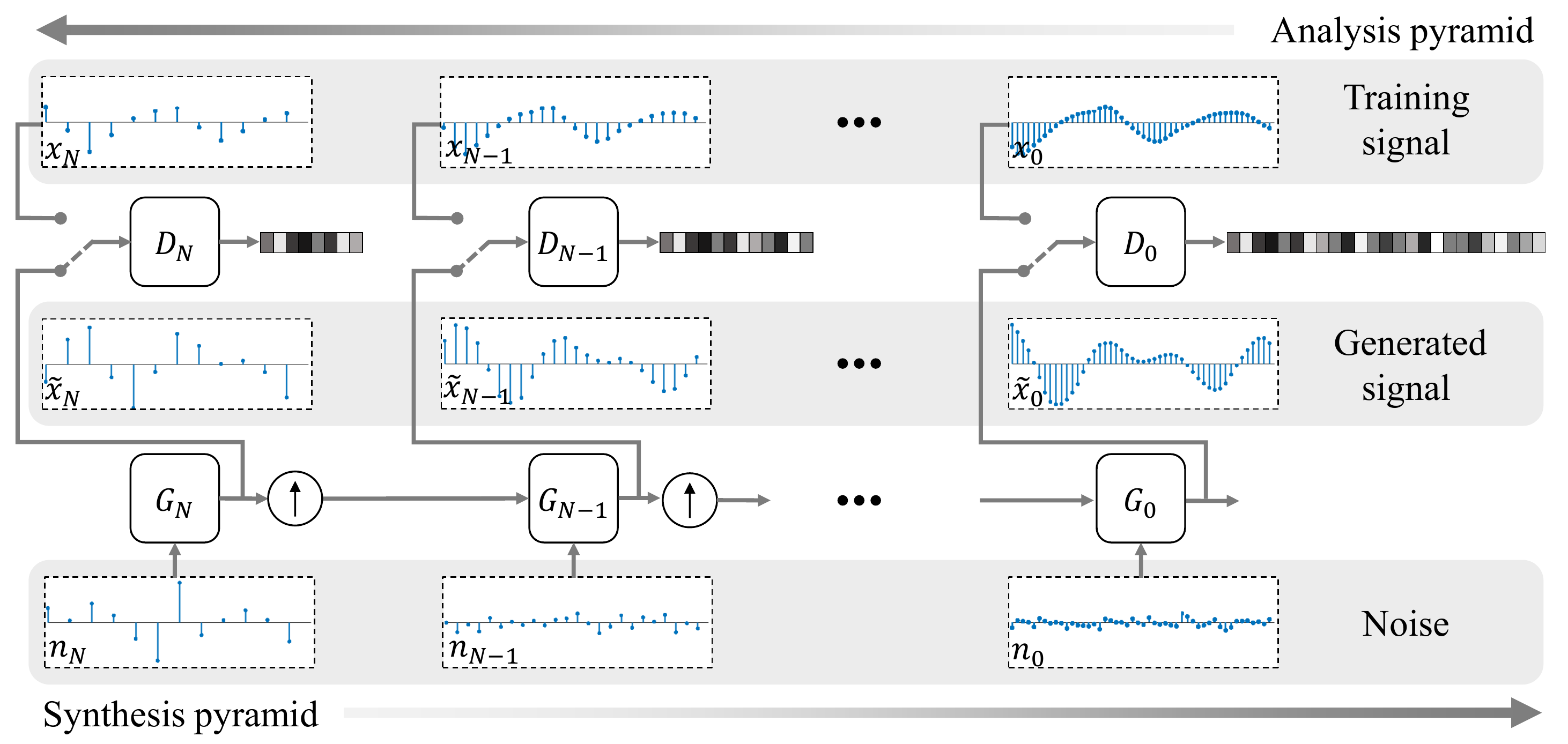}
\caption{\textbf{Model illustration.} Our model is built from a pyramid of generators that operate at gradually increasing sampling rates, each fed by the preceding one. Adversarial training is performed sequentially in a coarse-to-fine manner, using a corresponding pyramid of discriminators.}
\label{fig:model_illustration}
\end{figure*}

\myparagraph{Analysis pyramid.}
We start by constructing an analysis pyramid of the training signal,
\begin{align}
\real_0 &= \real,\nonumber\\
\real_n&=(\real*h_n)\downarrow_{d_n},\quad n=1,\ldots,N,
\end{align}
where $d_1<d_2<\dots<d_N$ are down-sampling factors and $h_1,\ldots,h_n$ are the corresponding anti-aliasing filters. This is illustrated at the top of Fig.~\ref{fig:model_illustration}. Denoting the sampling rate of $\real$ by $f^{\text{s}}$ (usually $16$Khz in our experiments), we have that the sampling rate at the $n$th pyramid level is $f^{\text{s}}_n=f^{\text{s}}/d_n$. Similarly, we denote by $\fake_n$ the $n$th level of the multi-scale representation of the fake signal $\fake$.

\begin{figure*}[t]
\centering
\includegraphics[width=1.0\columnwidth]{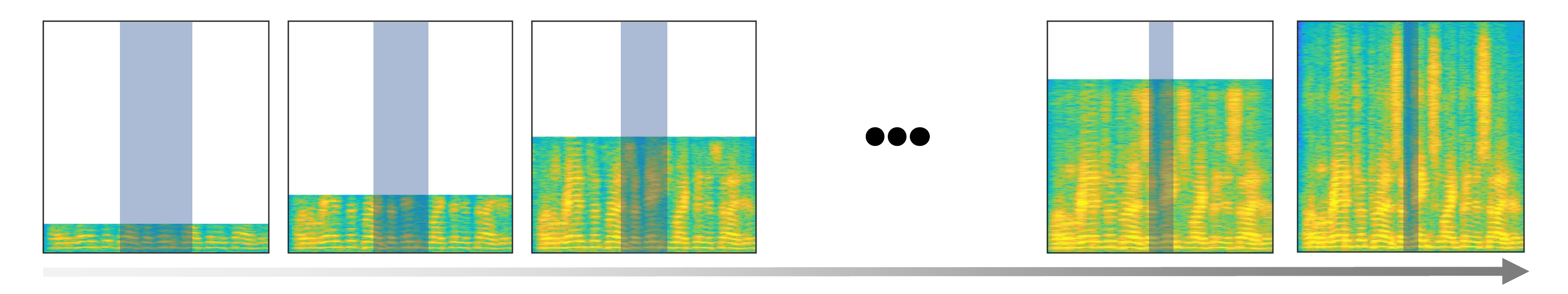}
\caption{\textbf{Generation process.} Our generation process gradually increases the frequency range of the signal. The receptive field of all generators is the same. This translates to larger \emph{effective} receptive fields (shaded rectangles) at the lower sampling rates, which shape the global structure of the signal.}
\label{fig:generation_process}
\end{figure*}

\myparagraph{Synthesis pyramid.}
The generation of a fake sample $\fake$ is performed sequentially by generating each of its pyramid levels conditioned on the previous one, from coarse to fine. Specifically, 
\begin{align}
\fake_N&=G_N(z_N),\nonumber\\
\fake_n&=G_n\left(z_n,\,(\fake_{n+1})\uparrow^{\alpha_n}\right),\quad n=N-1,\ldots,0,
\end{align}
where $z_n$ is white Gaussian noise, $G_n$ is a convolutional neural network generator, $\alpha_n=d_{n+1}/d_n$ is the resolution ratio between scales $n+1$ and $n$, and $(\cdot)\uparrow^\alpha$ stands for up-sampling by a factor of~$\alpha$ using cubic interpolation~\cite{keys1981cubic}. The signal $\tilde{x}_0$ at the end of this process is the generated fake sample $\tilde{x}$. This synthesis pyramid is shown at the middle and bottom rows of Fig.~\ref{fig:model_illustration}. All generators have the same receptive field, as measured in samples. This translates to larger effective receptive fields (in seconds) for the coarser levels than for the finer ones. As a result, the coarsest scale can capture the long-range dependencies that are typical of low frequencies of audio signals. Each subsequent generator, then, only needs to add a narrow band of frequencies to the signal generated at the previous scale (see Fig.~\ref{fig:generation_process}). The higher the frequency band, the smaller the receptive field that suffices to achieve this goal. Following this understanding, we take the variance of $z_n$ to be proportional to the energy of $\real$ in the frequency band $[\frac{1}{2}f^\text{s}_n,\frac{1}{2}f^\text{s}_{n-1}]$, which is at the responsibility of the generator $G_n$ to synthesize. It is important to note that as opposed to images, audio signals tend to exhibit long-range dependencies even at the highest frequency bands. Therefore, we take the receptive field in samples to be three orders of magnitude larger than the resolution factor $\alpha$ between scales (see below). This is as opposed to the SinGAN image model \cite{shaham2019singan}, which uses only one order of magnitude.

\myparagraph{Training.}
We train our model in a coarse to fine manner as well. At each stage, a single generator in the pyramid is trained while the generators of all coarser levels are kept fixed. When training the $n$th level, the goal is to drive the distribution of length-$T$ segments within $\fake_n$ to become as close as possible to the distribution of length-$T$ segments within $x_n$. To this end, we use a patch-GAN framework~\cite{li2016precomputed,isola2017image}, which employs a convolutional discriminator network $D_n$ with receptive field $T$. The discriminator is tasked with classifying each of the overlapping length-$T$ windows in its input as real or fake, so that its output is a classification sequence of the same length as the input (minus $T-1$ samples). The final score of the discriminator is the mean of this classification sequence. We specifically use the Wasserstein GAN loss~\cite{arjovsky2017wasserstein}, 
\begin{equation}\label{eq:adv}
\mathcal{L}_{\text{adv}}(D_n, G_n) = \underset{x\sim\mathbb{P}_{\real_n}}{\mathbb{E}}[D_n(\real_n)]-\underset{\fake_n\sim\mathbb{P}_{\fake_n}}{\mathbb{E}}[D_n(\fake_n)],
\end{equation} together with a gradient penalty~\cite{gulrajani2017improved}. Additionally, we pick a particular input at each scale, $z_n^\text{r}$, and enforce that its corresponding reconstructed signal, $\fake_n^\text{r} = G_n(z_n^\text{r})$, be close to the real signal $\real_n$ at that scale. This ensures that there is at least one point in the latent space of our model that maps to the real signal. We do this via a reconstruction loss, \begin{equation}\label{eq:rec}
    \mathcal{L}_{\text{rec}}(G_n)=\alpha_1\Vert \real_n-\fake_n^r\Vert_2^2+\alpha_2 \,\text{MSS}(\real_n, \fake_n^\text{r}),
\end{equation}
where the second term is the multi-scale spectrogram (MSS) loss~\cite{arik2018fast, oord2018parallel}, which penalizes for differences between spectograms (thus disregarding phase). We use the particular MSS formulation of~\cite{dhariwal2020jukebox} (see SM). 
For the reconstruction sequences, we choose $\{z_N^\text{r},z_{N-1}^\text{r},...,z_0^\text{r}\} = \{z^\star,0,...,0\}$, where $z^\star$ is a fixed white Gaussian noise realization. Therefore, overall, we solve \begin{equation}
    \min_{G_n}\max_{D_n} \mathcal{L}_{\text{adv}}(D_n,G_n)+\mathcal{L}_{\text{rec}}(G_n),
\end{equation}
where we alternate between performing one update step for $D_n$, which also involves minimizing the gradient penalty term, and one update step for $G_n$. In practice, we typically use only one of the terms in \eqref{eq:rec} (setting the other coefficient to $0$), depending on the application (see Sec.~\ref{sec:exp}).

\begin{figure*}[t]
\centering
\includegraphics[width=1\columnwidth]{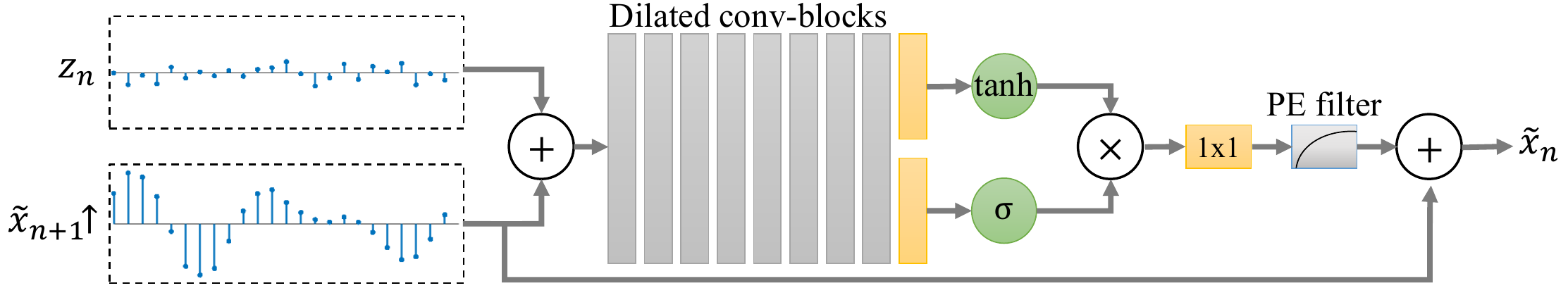}
\caption{\textbf{Single synthesis scale.} The generator at the $n$th scale gets an up-sampled version of the signal generated at the previous scale, $(\fake_{n+1})\uparrow^\alpha$, which has frequency contents in the range $[0,\frac{1}{2}f^\text{s}_{n+1}]$. Together with a noise realization $z_n$, it generates a signal $\fake_n$ with frequency contents in $[0,\frac{1}{2}f^\text{s}_n]$. This is done with a residual architecture involving 8 dilated convolution blocks; 7 of the form conv-BN-leakyReLU, 1 convolutional only. Dilation grows by a factor of $2$ in each block. We add a gated activation at the end of the generator, followed by a fixed pre-emphasis filter. 
}
\label{fig:G_illustration}
\end{figure*}

\myparagraph{Architecture.}
The generators and discriminators at all scales have the same fully-convolutional architecture. We use stacked blocks of 8 dilated convolutions, followed by batch normalization and leaky ReLU with slope $0.2$. The dilation factor grows by a factor of 2 in each layer, which is known to be an effective way for increasing of receptive field \cite{yu2015multi, oord2016wavenet}. At the end of the generator we use the gated activation unit~\cite{van2016conditional}, which is an element-wise product of tanh and sigmoid, each fed by an extra non-dilated convolution. All of our convolution layers have a kernel size of 9, which leads to a total receptive field of 2040 samples at each scale. We use weight normalization~\cite{salimans2016weight}, which we found to improve results and training stability. At the end of the trainable blocks, we add a fixed pre-emphasis (PE) filter with impulse response $[1,-0.97]$, which amplifies the high frequencies, as common in similar tasks \cite{wright2019real, wright2020perceptual}. An illustration of the generator's architecture is shown in Fig.~\ref{fig:G_illustration}.

\begin{figure*}[t]
\centering
\includegraphics[width=1\columnwidth]{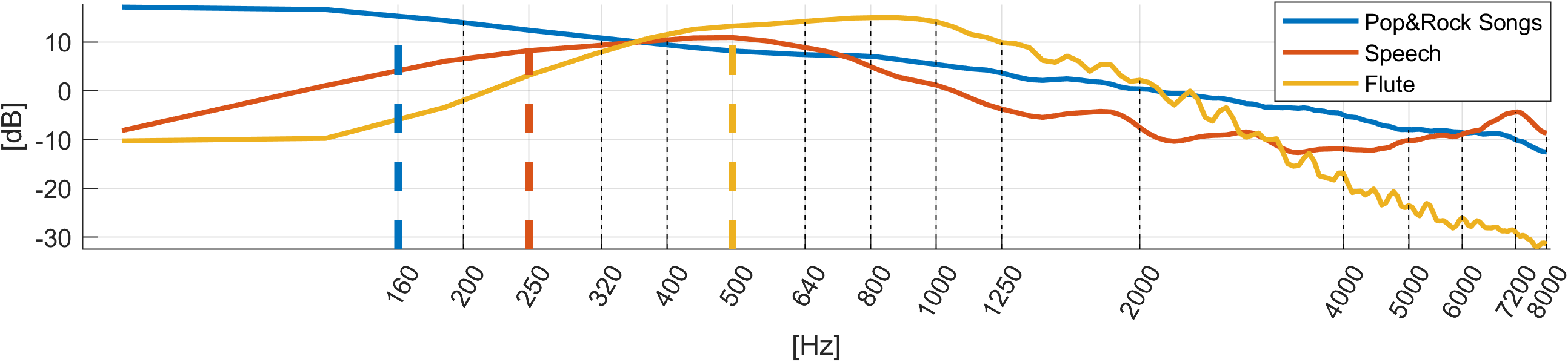}
\caption{\textbf{Scales selection.} The plot depicts the power spectral densities of three different audio datasets (rock and pop songs \cite{tzanetakis2002musical}, speech \cite{ljspeech17} and monophonic flute \cite{solo_audio19}). The dashed black lines show the predefined band partitions (note the logarithmic axis). The first band is adaptively chosen to contain enough energy. The bold colored lines show the typical first scale chosen for each dataset.} 
\label{fig:scales_illustration}
\end{figure*}

\myparagraph{Automatic scales selection.}
Different types of audio signals can have very different power spectra, as we illustrate in Fig.~\ref{fig:scales_illustration}. This suggests that the frequency bands of the pyramid should be adaptively chosen based on the spectrum of the training signal. However, to allow for efficient implementations of resampling techniques, we also want the sampling rates of all scales to be rational factors (with small denominators) of $f^\text{s}$ \cite[ch.~9]{eldar2014sampling}. We therefore use a predefined discrete set of potential sampling rates, and choose our bands adapitively only from this set. As can be seen in Fig.~\ref{fig:scales_illustration}, up to $2$Khz, where most of the audible energy resides, the predefined scales grow at a factor of around $1.25$. The mid-range, $2$-$4$Khz, typically contains less energy and so the scales are sparser there. Finally, to be able to capture the energy of non-vocal syllables in speech signals, the scales become denser again from $4$Khz. In practice, the most significant effect is due to the automatic selection of the first band, which shapes the global structure of the signal. We therefore choose automatically only this band, such that it contains enough energy (see SM). Fig.~\ref{fig:scales_illustration} shows typical selections of the first band for different types of audio signals. Additional spectra are presented in the SM. 

\section{Experiments}
\label{sec:exp}
We test our \emph{catch-a-waveform (CAW)} method in several applications and evaluate it both qualitatively and quantitatively. Our training examples contain a variety of audio types, including polyphonic rock and pop music, monophonic instrumental music, speech, and ambient sounds. Unless noted otherwise, all training signals have a sampling rate of $16$Khz. For training, we use the Adam optimizer~\cite{kingma2014adam} with $(\beta_1, \beta_2)=(0.5, 0.999)$ and learning rate $0.0015$, which we reduce by a factor of $10$ after two thirds of the epochs (we run a total of $3000$ epochs). Training on a $25$ second long signal takes about $10$ hours on Nvidia GeForce RTX 2080. Inference is $60$ times faster than real-time.

\subsection{Unconditional generation}

\myparagraph{Monophonic music.}
We trained CAW models on monophonic music played by various instruments, including cello, violin, saxophone, trumpet, and electric guitar. Here, we used $\alpha_1=0$ and $\alpha_2=10^{-4}$ in \eqref{eq:rec}. We trained on signals of length $25$ to $100$ seconds, and at test time generated signals of various lengths by simply injecting input noise signals of appropriate length (see SM for additional details). The generated signals sound like variations or naive improvisations on the original piece (see \websiteref[\#unconditional_music].

\myparagraph{Speech signals.}
We further trained CAW models on various human voice recordings, with lengths varying from $20$ to 
$40$ seconds. These include short segments from speeches of American presidents Trump and Obama and a-capella rap. Here we used $\alpha_1=10, \alpha_2=0$ in \eqref{eq:rec}. At inference, 
we generated random samples of lengths between $20$ and $60$ seconds. As exemplified in our \websiteref[\#unconditional_speech], the generated signals preserve the voice of the speaker, but exhibit new compositions of syllables, words, intonations and silent gaps. Note that since our model has no notion of language, the generated signals are not necessarily interpretable. The temporal coherence of the generated signals can be controlled by changing the receptive field of the model. As we illustrate in the \websiteref[\#rf], reducing the receptive field from $4$ to $2$ seconds (by removing one convolutional block), causes the structure to become less coherent and makes the generated speech sound like mumbling. Increasing the receptive field to $8$ seconds, on the other hand, preserves short sequences of words from the training signal.

\myparagraph{Human perception tests.}
In order to evaluate the perceptual quality of our generated signals, 
we conducted auditory studies through Amazon Mechanical Turk (AMT). The studies were performed on solo signals of $8$ different 
instruments (saxophone, trumpet, violin, flute, clarinet, cello, accordion and distorted electric guitar) randomly chosen from the Medly-solos-Db dataset~\cite{lostanlen2016deep} and the solo-audio dataset~\cite{solo_audio19}. For each instrument, we randomly cropped $25$ second long segments from $7$ different parts of the recording to serve as our real signals ($56$ segments in total) and used our method to generate a $10$ second long fake version for each of them. We performed two types of user studies: (i)~\emph{a paired test}, where the real signal and its fake version were played sequentially, and the user was asked to choose which is the fake, and (ii)~\emph{an unpaired test}, where the user listened to a single signal and had to determine whether it is real or fake. 
\begin{wrapfigure}{r}{0.42\textwidth}
    \centering
    \scalebox{0.33}
    {\input{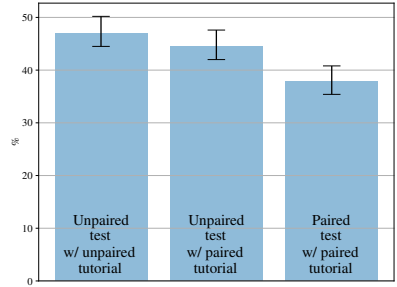}}
    \caption{\textbf{Real-vs.-Fake AMT studies.}
    Users were asked to discriminate between our generated signals and real ones, both in paired and in unpaired tests. We present confusion rates and $[5\%, 95\%]$ confidence intervals. As can be seen, the confusion rates in all cases are close to the ultimate rate, which is $50\%$.}
    \label{fig:amt_study}
\end{wrapfigure}
Each test opened with a tutorial of 5-8 questions identical to the structure of the main test, but with a feedback to the user. We also had an additional version, (iii)~\emph{an unpaired test with a paired tutorial}. In this case users were exposed to examples of paired real and fake signals during the training phase, but the test itself was unpaired. In each of the tests, we had $50$ different users answer $25$ questions each. 
The results are summarized in Fig.~\ref{fig:amt_study}. As can be seen, in all the studies the confusion rates are relatively high (the ultimate rate being $50\%$). As expected, the confusion rate of the paired test is lower than the unpaired test, as this setting is less challenging. But there is no significant difference between the results of the two unpaired tests, suggesting that a paired tutorial does not help the listener perform better discrimination.

\begin{figure*}[h]
	\centering
	\captionsetup[subfigure]{labelformat=empty,justification=centering,aboveskip=1pt,belowskip=1pt}
	\begin{subfigure}{0.48\textwidth}
		\centering
		\includegraphics[width=0.9\linewidth]{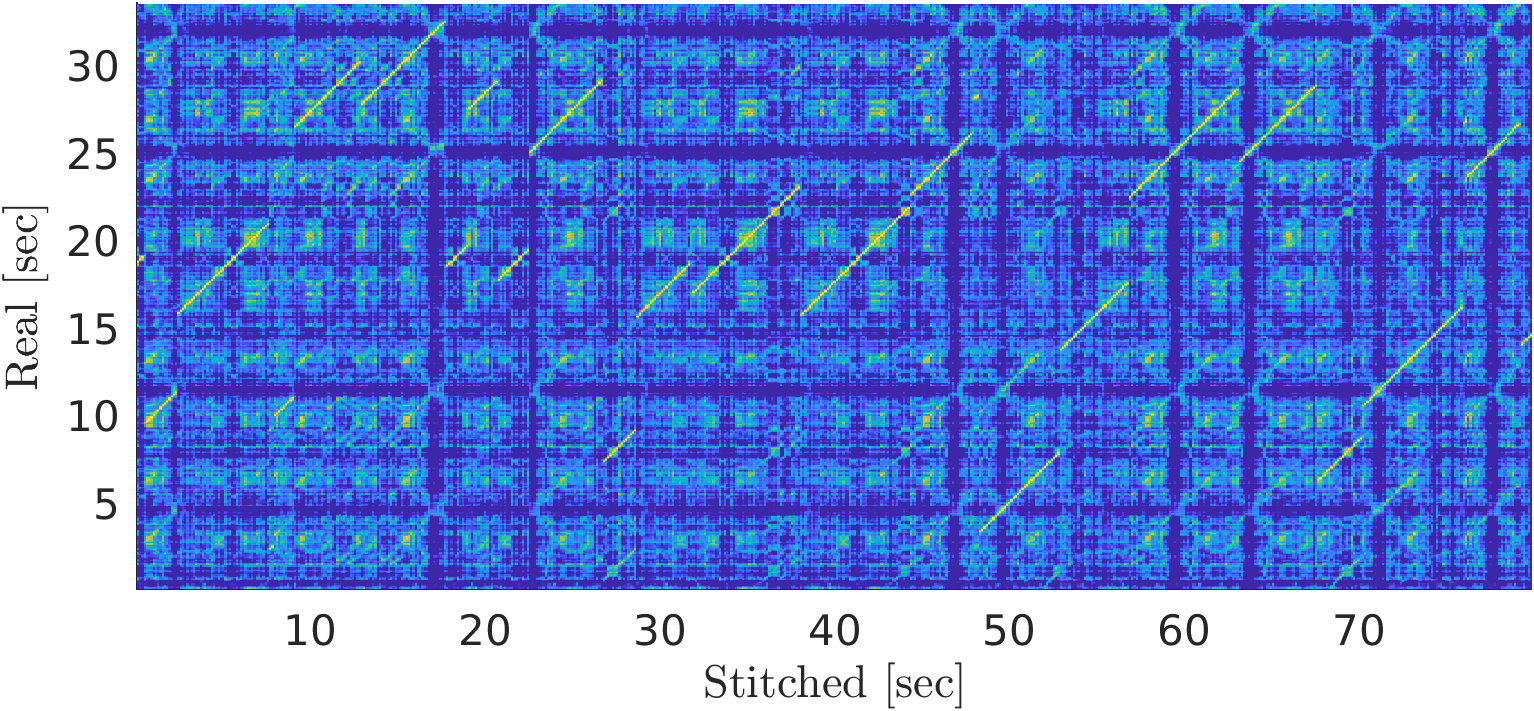}	
	\end{subfigure}
	\begin{subfigure}{0.48\textwidth}
		\centering
		\includegraphics[width=0.9\linewidth]{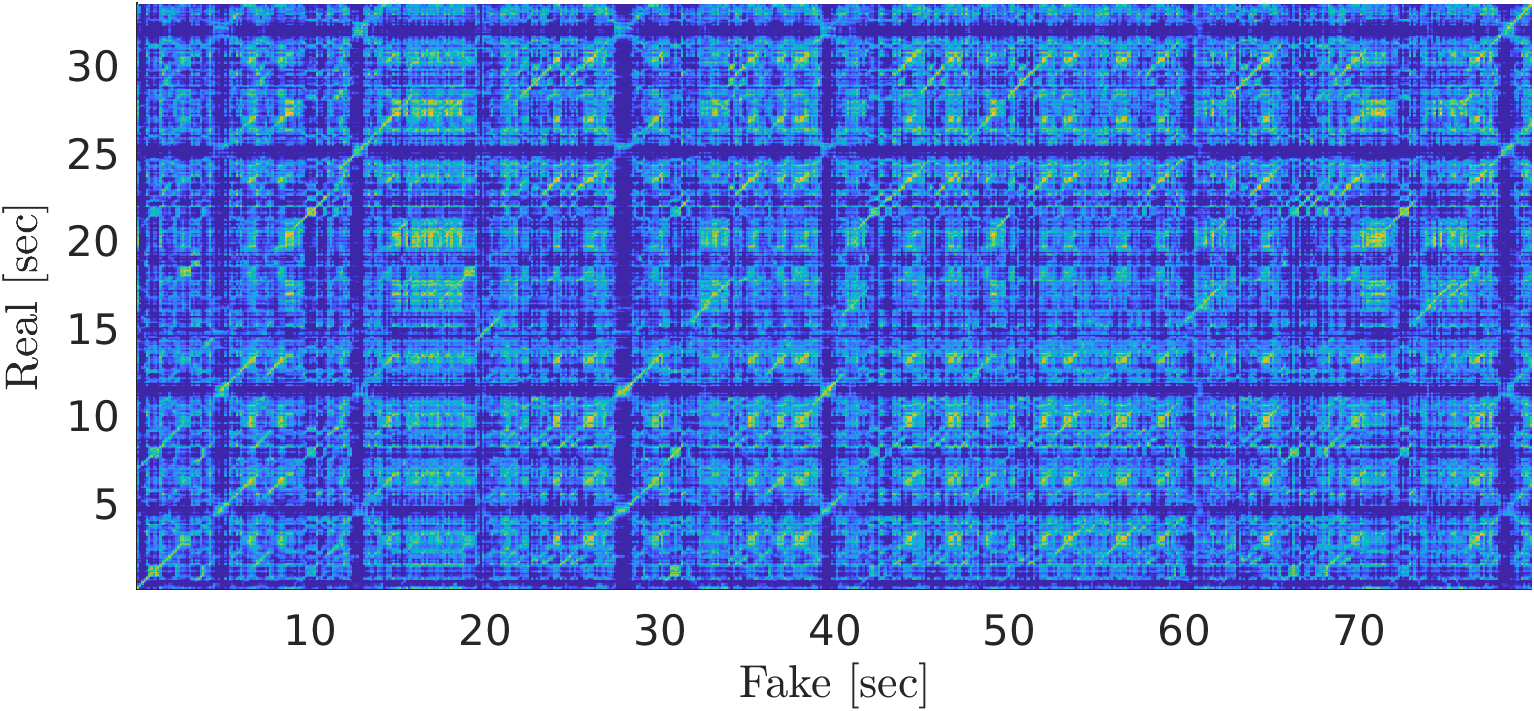}
	\end{subfigure}
	\caption{\textbf{Similarity matrix.}\label{fig:sim_matrix} Distinct diagonal lines correspond to segments in the generated signal that were `copied' from the real signal.	Naively stitched signals show clear lines in the copied parts (left). In our generated signal (right) more lines can be seen as the model can maintain global temporal structure while combining higher frequencies from different temporal locations.} 
\end{figure*}

\myparagraph{Comparison to naive copy-and-paste.} As our model is trained on a very short signal, its ability to generate new semantic contents is limited. For example, it will most likely not generate syllables or notes that did not exist in the training signal. However, does that mean it naively copy and pastes segments from the training signal? To examine this, we depict in Fig.~\ref{fig:sim_matrix} (right pane) a similarity matrix between small overlapping patches of a generated signal and the training signal. We specifically use the cosine similarity between the vectors of absolute values of the discrete Fourier transforms of the patches (see more details about the similarity matrix calculation in the SM). In this visualization, matching segments appear as diagonal lines. For comparison, we also show the same visualization for a naively generated signal constructed by stitching together cropped segments from the training signal, with crop sizes randomly chosen between the receptive fields of our finest and coarsest scales. Stitching is done using cross fading. We ensured that the naively stitched signals sound realistic by performing another unpaired user study, as in the left bar of Fig.~\ref{fig:amt_study}, and got a confusion rate of $47.88\%$. While the similarity matrix of the naively stitched signal shows clear solid lines, in our generated signal we can observe sometimes two or more parallel, weaker, lines one above the other. This suggests that our model often mixes several parts from the training signal. This happens thanks to our multi-scale architecture that allows each generated frequency band to contain contents from a different temporal location in the training signal.
{}

\subsection{Conditional generation of music variations}
\label{sec:app:music}
\begin{figure*}
	\centering
	\captionsetup[subfigure]{labelformat=empty,justification=centering,aboveskip=1pt,belowskip=1pt}
	\begin{subfigure}[t]{0.48\textwidth}
	    \vspace{-.3mm}
		\centering
		\caption{Original verse (used for training)}
		\includegraphics[width=1\linewidth, ,clip]{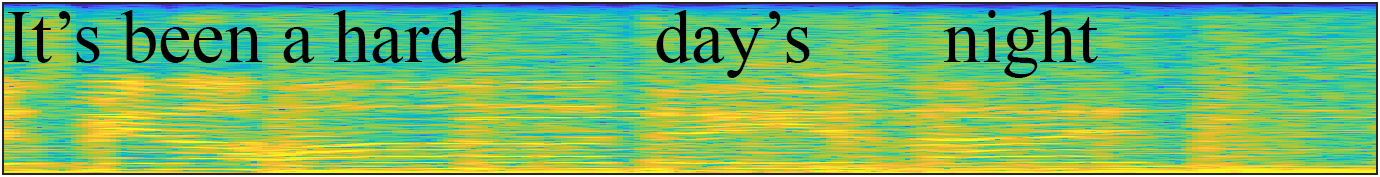}	
	\end{subfigure}
	\begin{subfigure}[t]{0.48\textwidth}
	    \vspace{-.3mm}
		\centering
		\caption{Generated verse}
		\includegraphics[width=1\linewidth, ,clip]{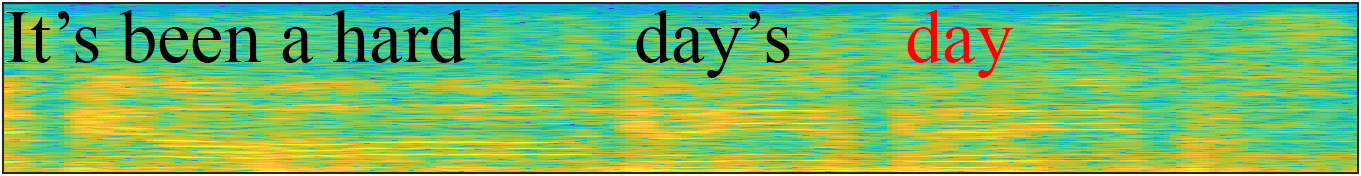}
	\end{subfigure}
	\begin{subfigure}[t]{0.48\textwidth}
	    \vspace{-.3mm}
		\centering
		\includegraphics[width=1\linewidth, ,clip]{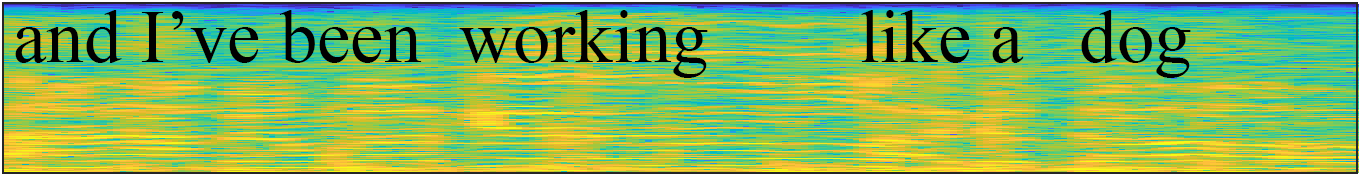}
	\end{subfigure}
	\begin{subfigure}[t]{0.48\textwidth}
	    \vspace{-.3mm}
		\centering
		\includegraphics[width=1\linewidth, ,clip]{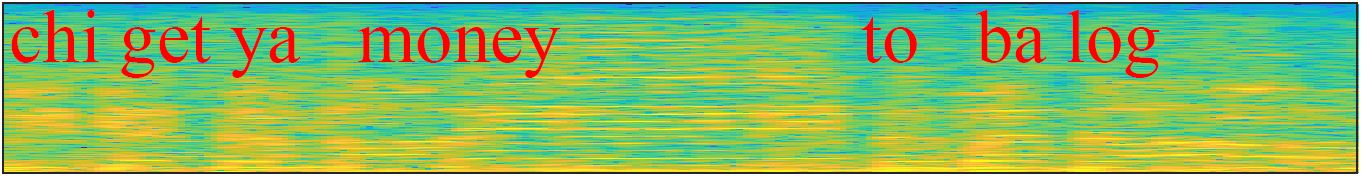}
	\end{subfigure}
	\begin{subfigure}[t]{0.48\textwidth}
	    \vspace{-.3mm}
		\centering
		\includegraphics[width=1\linewidth, ,clip]{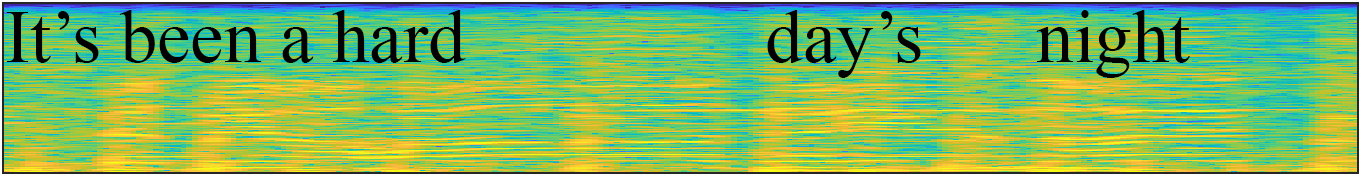}
	\end{subfigure}
	\begin{subfigure}[t]{0.48\textwidth}
	    \vspace{-.3mm}
		\centering
		\includegraphics[width=1\linewidth, ,clip]{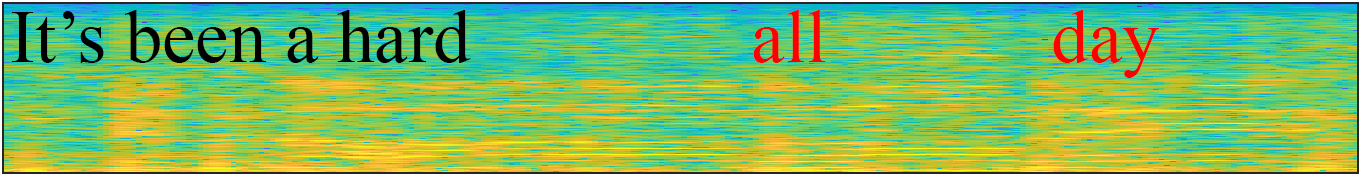}
	\end{subfigure}
	\begin{subfigure}[t]{0.48\textwidth}
	    \vspace{-.3mm}
		\centering
		\includegraphics[width=1\linewidth, ,clip]{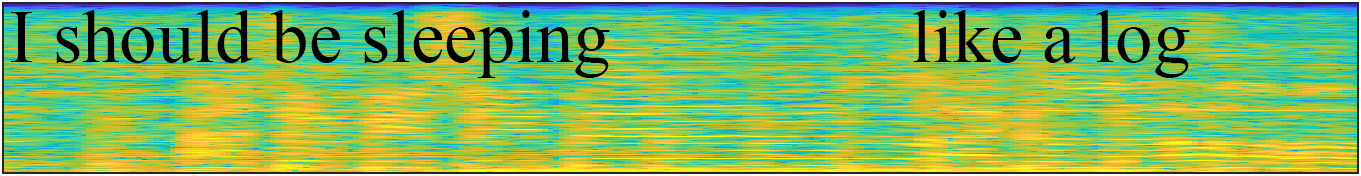}
	\end{subfigure}
	\begin{subfigure}[t]{0.48\textwidth}
	    \vspace{-.3mm}
		\centering
		\includegraphics[width=1\linewidth, ,clip]{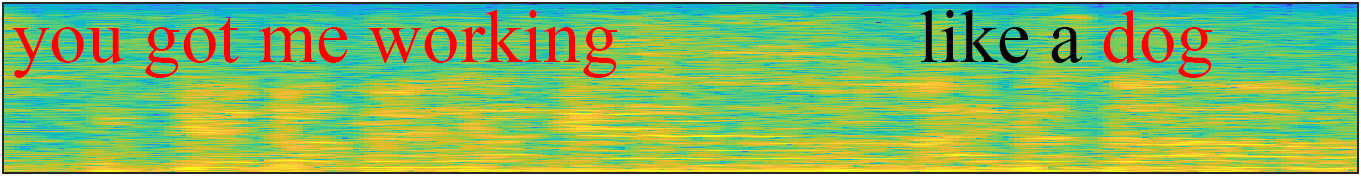}
	\end{subfigure}
	\caption{\textbf{Music variations.} After training on a specific song, we can inject a down-sampled version of the song to the second coarsest scale of the model. This way, our model generates a signal having the same structure as the original song, but with randomly generated finer details, like lyrics. In this example we generate a new verse to ``A Hard Day's Night'' by The Beatles, after training on its two first verses. Modified lyrics are shown in red.}
	\label{fig:music}
\end{figure*}

Another interesting application is generating variations or extensions of existing songs (\eg adding a new verse). To do so, we first train our model on a popular rock or pop song. Then, at inference time, we start the generation from the second coarsest scale by injecting the real (training) signal as input to that scale. This ensures that the generated signal maintains the global structure of the real signal, as its low frequencies are constrained to be the same. But finer details, like the lyrics, are randomly generated. Here we take $\alpha_1=0$ and $\alpha_2=10^{-4}$ in \eqref{eq:rec}. Also, to encourage large variability between different random samples, we set the input noise in the second coarsest scale to have the same energy as that of the real signal at that scale. Examples can be found in the \websiteref[\#music] and in Fig.~\ref{fig:music}, which shows a new verse generated by our model to the famous Beatles song ``A Hard Day’s Night''. 

\subsection{Bandwidth extension}
\label{sec:app:bandwidth}
\begin{figure*}
	\centering
	\captionsetup[subfigure]{labelformat=empty,justification=centering,aboveskip=1pt,belowskip=1pt}
	\begin{subfigure}[t]{0.48\textwidth}
	    \vspace{-.3mm}
		\centering
		\caption{Low-resolution audio}
		\includegraphics[width=1\linewidth]{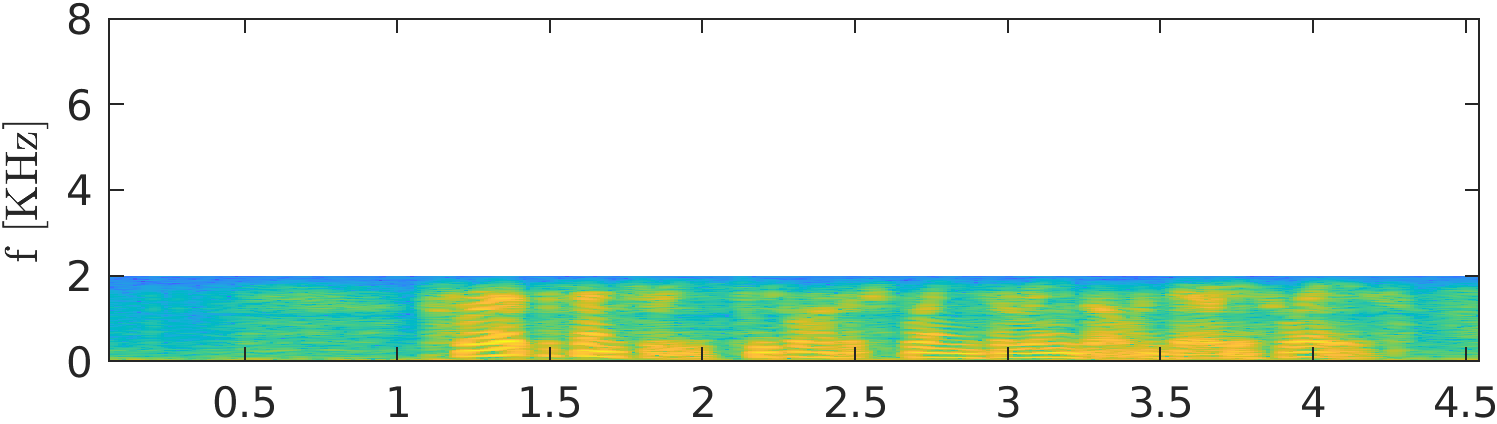}	
	\end{subfigure}
	\begin{subfigure}[t]{0.48\textwidth}
	    \vspace{-.3mm}
		\centering
		\caption{High-resolution audio (GT)}
		\includegraphics[width=1\linewidth]{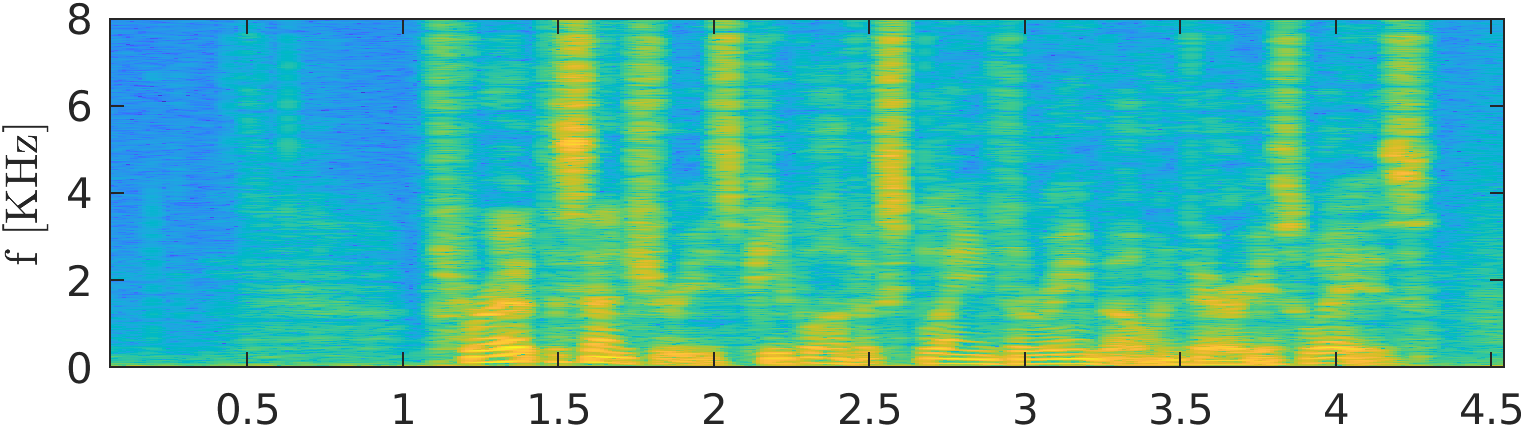}
	\end{subfigure}
	\par\smallskip
	\begin{subfigure}[t]{0.48\textwidth}
	    \vspace{-.3mm}
		\centering
		\caption{Extended by TFiLM~\cite{birnbaum2019temporal}}
		\includegraphics[width=1\linewidth]{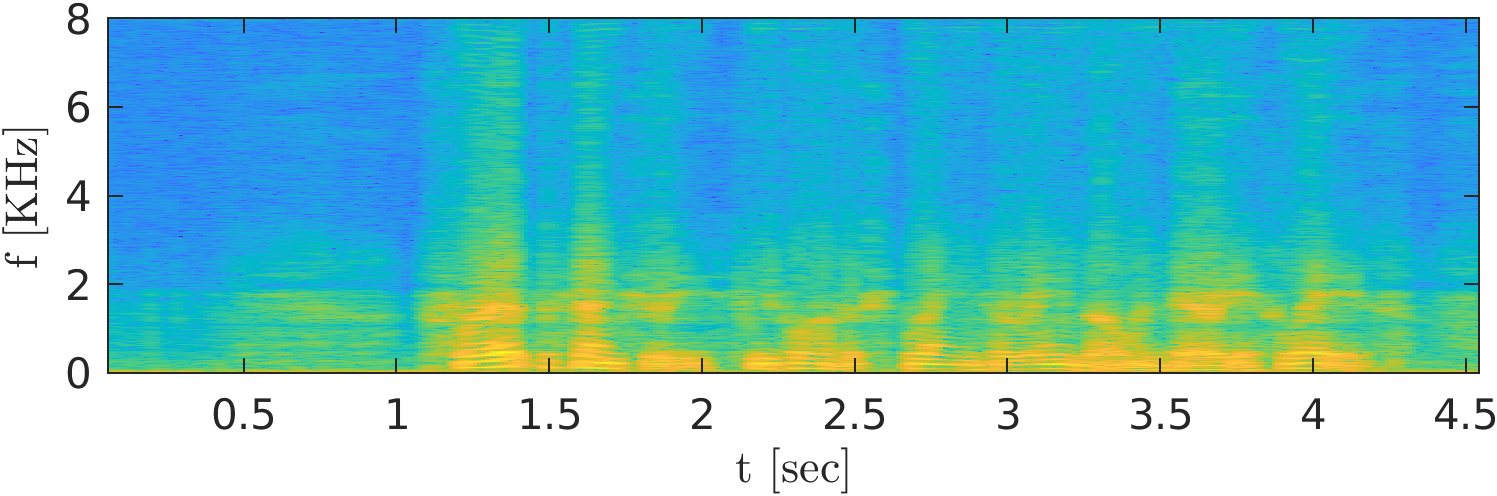}
	\end{subfigure}
	\begin{subfigure}[t]{0.48\textwidth}
	    \vspace{-.3mm}
		\centering
		\caption{Extended by CAW (ours)}
		\includegraphics[width=1\linewidth]{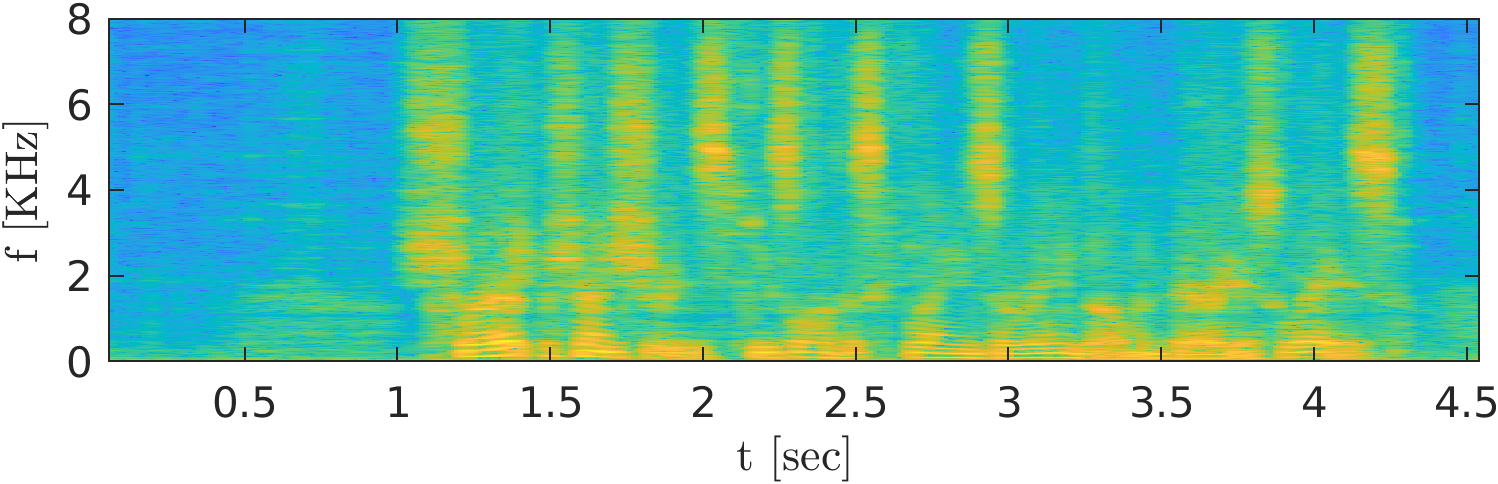}
	\end{subfigure}
	\caption{\label{fig:sr_example} 
	\textbf{Bandwidth extension.} We use our model for speech bandwidth-extension. A model trained on one short high-bandwidth signal ($25$ seconds in this case) can be used at test time to increase the bandwidth of any low-bandwidth signal of the same speaker (by injecting it to a coarse scale of the model). This results in sharper reconstructions than those obtained with TFiLM, which was trained on hours of examples.}
\end{figure*}

Bandwidth Extension (BE) is the task of reconstructing a high-bandwidth signal from its 
low-bandwidth version, and is usually demonstrated on speech~\cite{kuleshov2017audio,birnbaum2019temporal,kim2019bandwidth,gupta2019speech, wang2018speech} and music~\cite{lagrange2020bandwidth,sulun2020filter}. To perform BE using CAW, we first train it on a high-bandwidth short audio example of a specific speaker. At inference time, we can then inject any other low-bandwidth signal of the same speaker to a coarse scale of the model (we choose the scale whose sampling rate matches that of the input signal). Here we use $\alpha_1=10$ and $\alpha_2=0$ in \eqref{eq:rec}. We then stitch the reconstructed higher frequencies generated by our model with the low frequency range of the input signal to obtain our final full-bandwidth reconstruction. Figure~\ref{fig:sr_example} shows a BE example, where the sampling rate of a speech signal is increased from $4$KHz to $16$KHz. Our bandwidth-extended signals contain realistic high frequency contents, which makes them sound sharp (see examples and comparisons in our \websiteref[\#be], including for the easier task of extension from $8$KHz to $16$KHz).

{}

We compare our BE results to the the state-of-the-art temporal FiLM (TFiLM) method ~\cite{birnbaum2019temporal}, which requires a large training set to perform this task. We use the VCTK dataset, and report both the signal to noise ratio (SNRs) and the log spectral distance (LSD)~\cite{gray1976distance} between the recovered signal and the ground-truth one, averaged over a test set. LSD is known to better correlate with human perception. 
We perform comparisons to several TFiLM variants, following the protocols of~\cite{birnbaum2019temporal}.

\myparagraph{Single-speaker baseline.}
In this setting, we train a separate CAW model for each of $9$ speakers, and then test each of the models on a set of held-out sentences of the same speaker. For TFiLM, we use $30$ minutes of training data for each speaker, and for our model we use only $25$ seconds. As can be seen in Table~\ref{tab:BE}, our model outperforms TFiLM in LSD, but achieves a slightly lower SNR (we report mean and standard deviation over $50$ different trained models).

\myparagraph{Multi-speaker baselines.}
Here, we train TFiLM on $99$ speakers from the VTCK dataset and test it on the remaining $9$ speakers. We have three variants, corresponding to training sets of $25$ minutes, $4$ hours, and $10$ hours. Our model is trained as in the single-speaker case. We use the same test set for evaluating both methods. As can be seen in Table~\ref{tab:BE}, our model is again superior in terms of LSD compared to all TFiLM variants, and is slightly worse in terms SNR.
\begin{table}
\center
\caption{\textbf{Bandwidth extension quantitative evaluation.} We compare our method to TFiLM~\cite{birnbaum2019temporal} using SNR (higher is better) and LSD (lower is better) both for multi-speaker test and single-speaker test. In all cases our model achieves better LSD scores, indicating of higher perceptual quality.}
\begin{tabular}{c|cccc|cc|} 
\toprule
\hline
 & \multicolumn{4}{c|}{Multi speaker test} &
 \multicolumn{2}{c|}{Single speaker test} \\
 [0.5ex] 
 \cline{2-7}
  & \multicolumn{3}{c}{TFiLM
 ~\cite{birnbaum2019temporal}} & CAW (ours) & TFiLM
 ~\cite{birnbaum2019temporal} & CAW (ours) \\ 
 Training set size [min] & 25 & 240 & 600 & 0.4 & 30 & 0.4 \\
 \hline 
 SNR [dB] $\uparrow$ & $14.66$ & $14.83$ & $15.45$ & $13.8 \pm0.94$ & $14.77$ & $13.03\pm0.83$ \\
 LSD $\downarrow$ & $4.96$ & $3.89$ & $3.79$ & $2.97\pm0.26$ & $3.92$ & $3.03\pm0.26$ \\
\toprule
\end{tabular}
\label{tab:BE}
\end{table}

\begin{wraptable}{r}{6.6cm}
\centering
\caption{\textbf{Inpainting AMT study.} Users chose between our model (trained on $12$ seconds), GACELA (trained on $8$ hours), and the ground-truth signals. The preference rates indicate that our results are at least comparable to GACELA, and are often confused to be real signals.}
\label{tab:inpainting}
\begin{tabular}{cc} 
\toprule
Study & Preference rate \\
\hline
Ours vs. GACELA & $55.3\% \pm 2.4 \% $ \\
Ours vs. Real & $44.3\% \pm 2.3 \% $ \\
\toprule
\end{tabular}
\end{wraptable}

\subsection{Audio inpainting}
\label{sec:app:inpainting}
\begin{figure*}
	\centering
	\captionsetup[subfigure]{labelformat=empty,justification=centering,aboveskip=1pt,belowskip=1pt}
	\begin{subfigure}[t]{0.49\textwidth}
	    \vspace{-.3mm}
		\centering
		\caption{Spectogram of a signal with a silent gap}
		\includegraphics[width=1\linewidth]{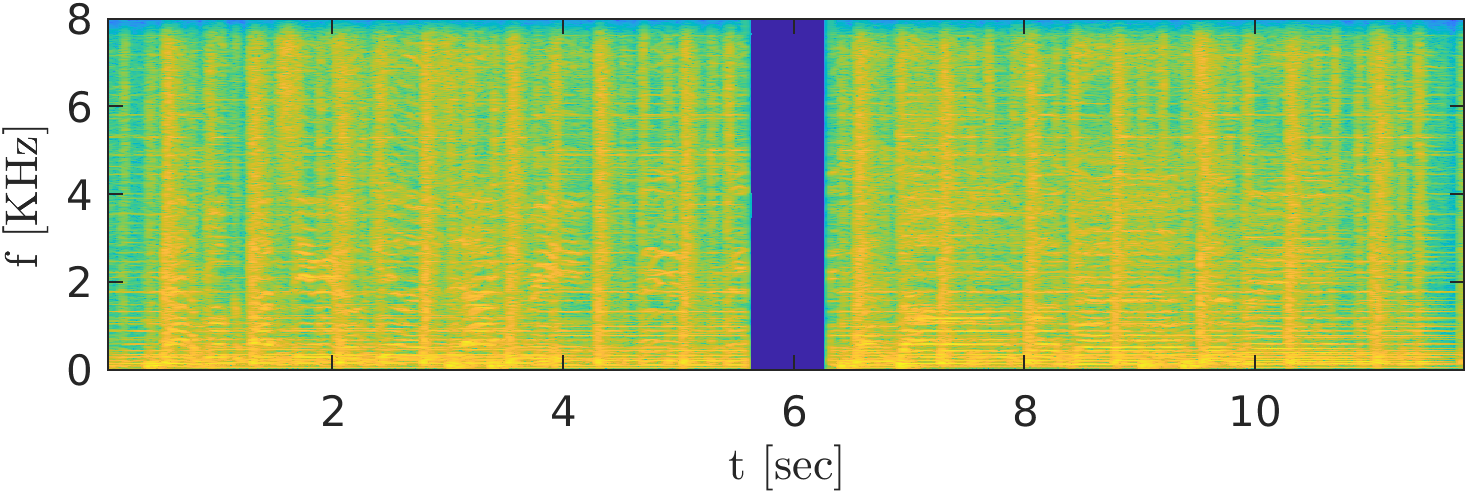}	
	\end{subfigure}
	\begin{subfigure}[t]{0.49\textwidth}
	    \vspace{.3mm}
		\centering
		\caption{Spectogram of a signal inpainted by our model}
		\includegraphics[width=0.93\linewidth]{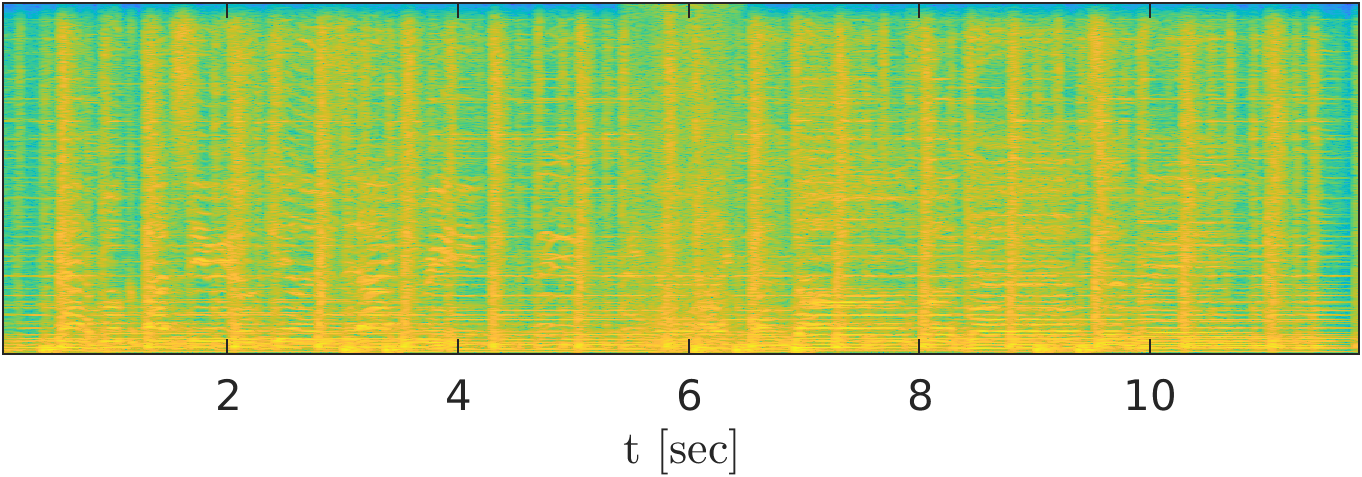}
	\end{subfigure}
	\caption{\textbf{Audio inpainting.}\label{fig:inpainting_example} Our model is able to complete a mising silent gap in a given signal without any additional information other than the signal with the gap itself. We train our model on the valid region of the signal excluding the missing part, to learn its internal statistics, and then at test time we generate the missing gap, which results in a coherent realistic completion.} 
\end{figure*}

Audio inpainting refers to the task of completing a missing part of a given audio signal. It has been previously addressed using classical signal processing methods~\cite{adler2011audio,bahat2015self,manilow2017leveraging}, graph-based approaches~\cite{perraudin2018inpainting} and neural networks~
\cite{ebner2020audio, marafioti2019audio, marafioti2020gacela}. Here, we address the long-inpainting task, where several hundred milliseconds are missing. We do this by training CAW with slight adaptations: (i) we calculate the loss with respect to only the valid parts of the signal (excluding the gap), and (ii)~we sample a new reconstruction noise realization for the missing part at each iteration. Here we use $\alpha_1=10$ and $\alpha_2=0$ in \eqref{eq:rec}. After training, we take the completed part from the reconstruction
, and stitch it with the input. As can be seen in Fig. \ref{fig:inpainting_example}, our model coherently completes the missing part, and thanks to its relatively large receptive field, the completion smoothly fuses with the valid parts. Examples of completed rock songs can be found in our \websiteref[\#inpainting]. 

{}

\myparagraph{Human perception tests.} We evaluated our results using an AMT user preference test. We took $64$ rock songs from the FMA-small dataset~\cite{fma_dataset}, extracted a $12$ second long segment from each, and masked a $750$ms window. We compared our results with those of GACELA ~\cite{marafioti2020gacela}, a GAN based context encoder trained on roughly $8$ hours of rock songs from the same dataset. In each query, raters listened to a $5$-$9$ second long segment containing the missing gap, as well as to our and to GACELA's completions. They could re-listen to all signals as many times as they wanted, and eventually had to pick the completion that sounded better. In total, $50$ raters answered $20$ queries each. As can be seen in Table~\ref{tab:inpainting}, raters preferred our completions over $50\%$ of the times, suggesting that the performance of our method is at least comparable to GACELA's. We also performed a user study that compared our completions to the real signals. Interestingly, the preference rate for our completion was still relatively high (see Table~\ref{tab:inpainting}), indicating that our results are often indistinguishable from real signals. 

\subsection{Audio denoising}
\label{sec:app:enhancement}
\begin{figure*}
	\centering
	\captionsetup[subfigure]{labelformat=empty,justification=centering,aboveskip=1pt,belowskip=1pt}
	\begin{subfigure}[t]{0.49\textwidth}
	    \vspace{-.3mm}
		\centering
		\caption{Noisy}
		\includegraphics[width=1\linewidth]{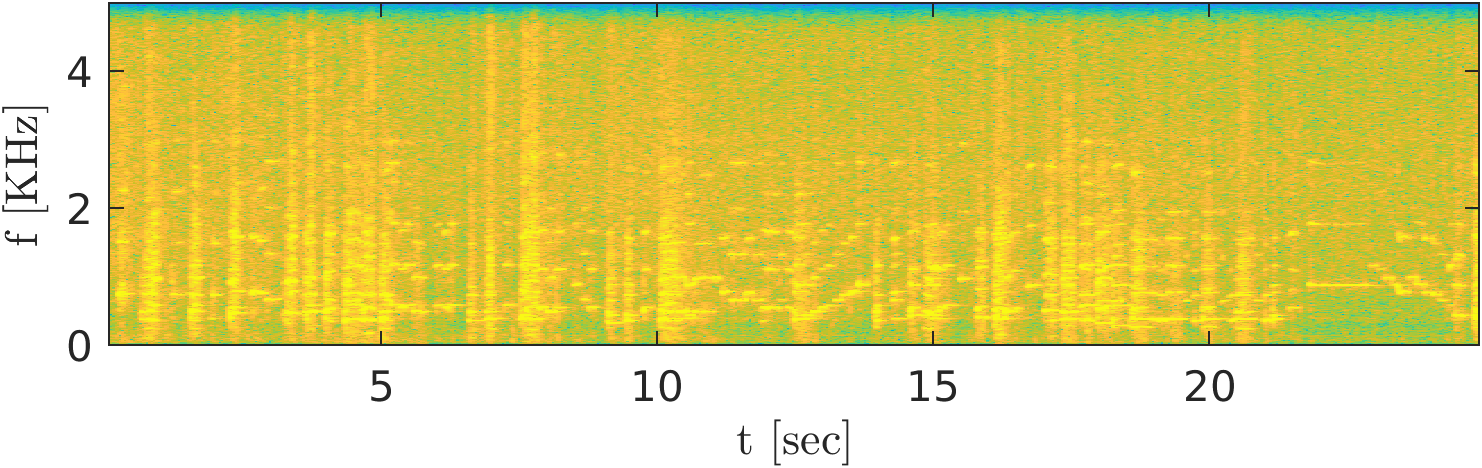}	
	\end{subfigure}
	\begin{subfigure}[t]{0.49\textwidth}
	    \vspace{-.3mm}
		\centering
		\caption{Denoised}
		\includegraphics[width=0.93\linewidth]{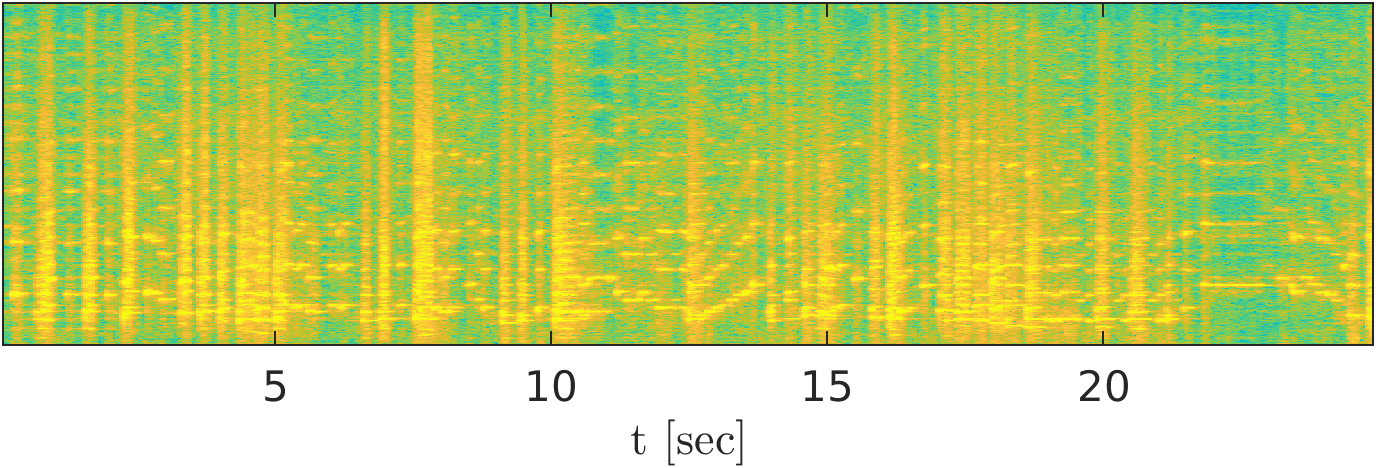}
	\end{subfigure}
	\caption{\label{fig:denoising_example} 
	\textbf{Denoising.} When trained on a single noisy example, our model produces a clean reconstruction. This enables to denoise \eg old recordings, illustrated here on one of the first violin recordings, from 1903. Here, no ground-truth data is available for computing SNR. However, in controlled experiments with modern recordings of the same musical piece, we measure improvement of $1.5\text{dB}-4\text{dB}$ in SNR (see text for details).}
\end{figure*}

An interesting side effect of CAW's training process, is that it can be used for audio denoising. As explained in Sec.~\ref{sec:method}, during training we enforce a certain noise hypothesis to generate a reconstruction of the training signal. We found that when training our model on a noisy signal, this reconstruction often preserves harmonic parts while suppressing ambient noise. This effect is more distinct when using $\alpha_1 = 10$ and $\alpha_2 = 0$ in~\eqref{eq:rec}. This enables to perform denoising without access to a any clean example for training, and without any prior knowledge about the noise distribution. This is in contrast to externally supervised approaches, which require many pairs of noisy-clean examples, \eg~\cite{li2020learning}. As an example, we demonstrate denoising 
of old recordings of the violinist Joseph Joachim from 1903 (for which obviously no clean training examples can be collected). As seen in Fig.~\ref{fig:denoising_example} and on our \websiteref[\#denoising], the reconstructed signals are notably cleaner than the original ones. To obtain some quantitative measure of the denoising performance in this experiment, we took a (clean) modern recording of the same musical piece (Bach's Adagio), preformed by famous violinist Hilary Hahn\footnote{taken from \href{https://www.youtube.com/watch?v=c3mwVaQIZ1c}{\textcolor{blue}{https://www.youtube.com/watch?v=c3mwVaQIZ1c}}}. We synthetically generated noisy versions of this recording using both white noise and old gramophone noise recordings\footnote{taken from \href{https://freesound.org/people/lollosound/sounds/387005/}{\textcolor{blue}{https://freesound.org/people/lollosound/sounds/387005/}}} (see the SM and our \websiteref[\#denoising]). For each noise type, we examined noise levels of $5$dB and $10$dB. 
For the white noise, our denoising increased the SNR from $5\text{dB}\rightarrow9.78\text{dB}$ and from $10\text{dB}\rightarrow11.53\text{dB}$. For the gramophone noise, it increased SNR from $5\text{dB}\rightarrow6.89\text{dB}$ and from $10$dB$\rightarrow11.56$dB. We believe these results can be further improved in the future by optimizing the model for the specific task of denoising. 
\section{Conclusion and limitations}
\label{sec:conclusion}
We presented a new GAN-based model for audio generation that can be trained on a single short example. Our model works on raw waveforms, and is useful for a variety of tasks. As we illustrated, learning from a single waveform often has advantages over learning from large datasets. We believe this new learning scheme can be further developed to a general framework for training `personalized' deep learning models, where personal small data would be sufficient for a variety of tasks. This approach, however, is not free of limitations. Models trained on small data cannot learn high level semantics. Moreover, they can sometimes struggle even with low-level attributes. For example, our model is occasionally less successful in 
handling high-pitched speech signals, like that of a female or a
child. Another challenging scenario is speech recorded in reverbrant environments (\eg in a large conference room), where our model tends to transform the reverberations into high-pitched noise. Examples for a variety of such failure cases can be found in our \websiteref[\#limitations].
\paragraph{Acknowledgements}
This research was supported by the Israel Science Foundation (grant 852/17) and by the Technion Ollendorff Minerva Center.
\small
\bibliographystyle{splncs04}
\bibliography{bibliography}
\newpage
\appendix{\Large{\textbf{Supplementary Material}}}

Code is available \href{https://github.com/galgreshler/Catch-A-Waveform}{\textcolor{blue}{here}}. Audio samples and additional figures can be found on the project's \href{https://galgreshler.github.io/Catch-A-Waveform/}{\textcolor{blue}{website}}.

\section{Model and training details}
\subsection{Training details}
\myparagraph{Gradient penalty.}
In each update step of the discriminator, we minimize the generator's loss with an additional gradient penalty regularization term~\cite{gulrajani2017improved}, defined as
\begin{equation}
    \lambda\underset{\hat{x}\sim\mathbb{P}_{\hat{x}}}{\mathbb{E}}[(\Vert\nabla_{\hat{x}}D_n(\hat{x})\Vert_2-1)^2],
\end{equation}
where $\lambda=0.01$ and $\hat{x}$ is a convex combination of the real signal $\real_n$ and a generated one $\fake_n$, with random weights.

\myparagraph{MSS loss.} The MSS reconstruction loss we use, is given by
\begin{equation}
    \text{MSS}(\real_n,\fake_n^\text{r})=\frac{1}{M}\sum_{m=0}^{M-1}{\Vert|\text{STFT}_m(\real_n)|-|\text{STFT}_m(\fake_n^\text{r})|\Vert_2}
\end{equation}
where STFT is the short-time Fourier transform and $M$ is the number of different STFT parameter sets. The set of parameters we use are as follows:
\begin{center}
\begin{tabular}{ c c c } 
\toprule
 window size & hop length & fft size \\
 \hline
 240 & 50 & 512 \\ 
 600 & 120 & 1024 \\ 
 1200 & 240 & 2048 \\ 
 \hline
\end{tabular}
\end{center}

\subsection{Scales selection}
As explained in the main text, we have a set of predefined sampling rates. The first scale of the model is chosen automatically among them, according to the signal energy at that scale. Specifically, we normalize the input signal $\real$ such that $\max_{n}{|\real[n]|} = 1$. Then, we choose the coarsest scale (namely scale $N$) to be the first that satisfies
\begin{equation}
    \frac{1}{K}\sum_{n=0}^{K-1}{{x_N^2[n]}} \geq 0.0025,
\end{equation}
where $x_N$ is the real signal at that scale and $K$ is the number of samples in $x_N$. 

\begin{figure*}[t]
\centering
\includegraphics[width=1\columnwidth]{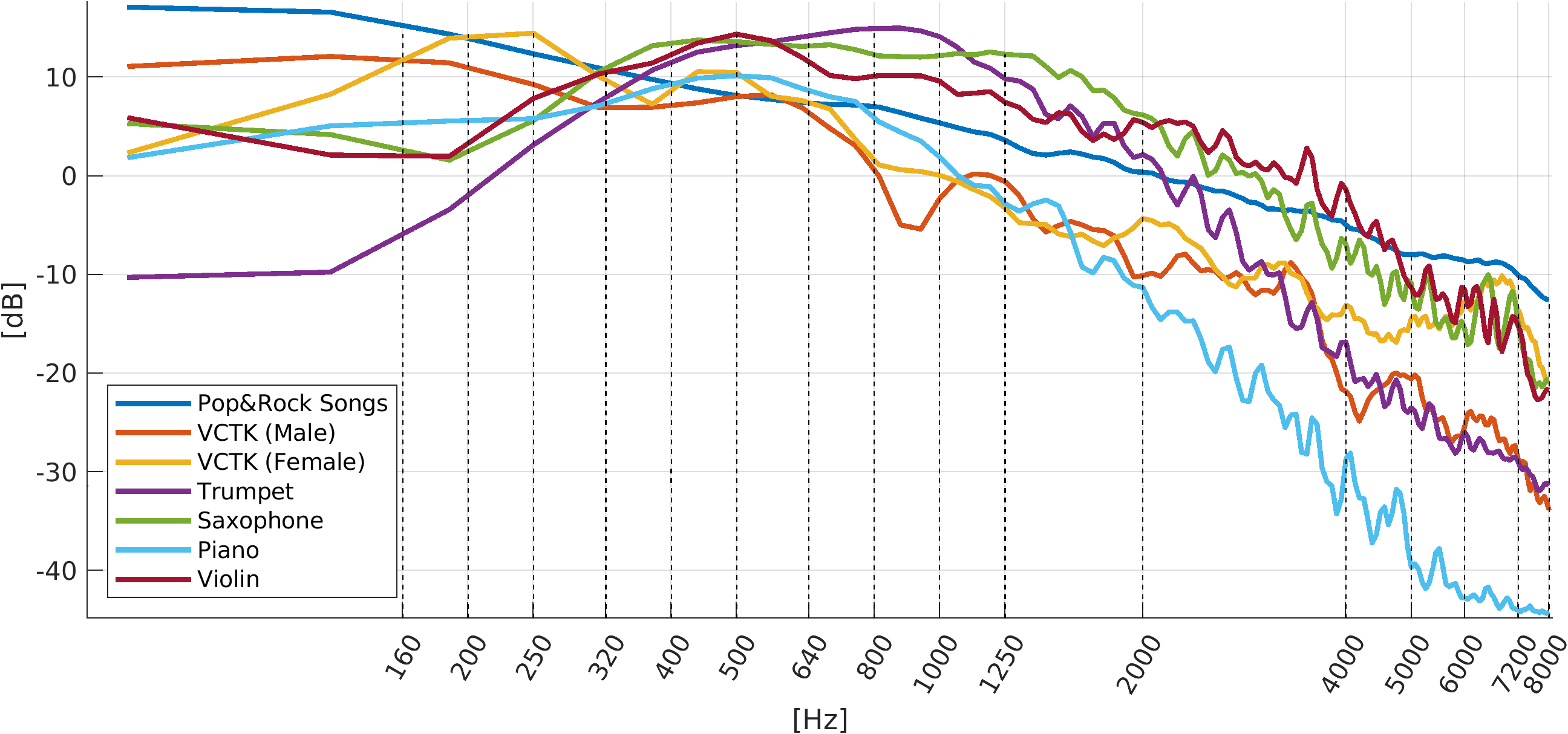}
\caption{\textbf{Frequency contents of different datasets.} Note that complex music (here rock and pop songs) have a more energy in the low frequencies than speech and monophonic music.}
\label{fig:scales_selection_SM}
\end{figure*}

Figure~\ref{fig:scales_selection_SM} shows the average frequency contents of several datasets, and the predefined scales. Note that the graph shows averages on entire datasets, while the actual first scale is chosen for each signal individually. The coarsest scale defines the receptive field (in seconds) for the entire model. Therefore, for inpainting tasks, we also make sure that the signal at the coarsest scale has more samples than the missing gap plus the receptive field.

\subsection{Architecture}
\begin{figure*}
	\centering
	\captionsetup[subfigure]{labelformat=empty,justification=centering,aboveskip=1pt,belowskip=1pt}
	\begin{subfigure}[t]{0.98\textwidth}
	    \vspace{-.3mm}
		\centering
		\caption{\textbf{Generator}}
		\includegraphics[width=1\linewidth]{figs/G_illustration_3.pdf}	
	\end{subfigure}
	\par\smallskip
	\begin{subfigure}[t]{0.98\textwidth}
	    \vspace{-.3mm}
		\centering
		\caption{\textbf{Discriminator}}
		\includegraphics[width=1\linewidth]{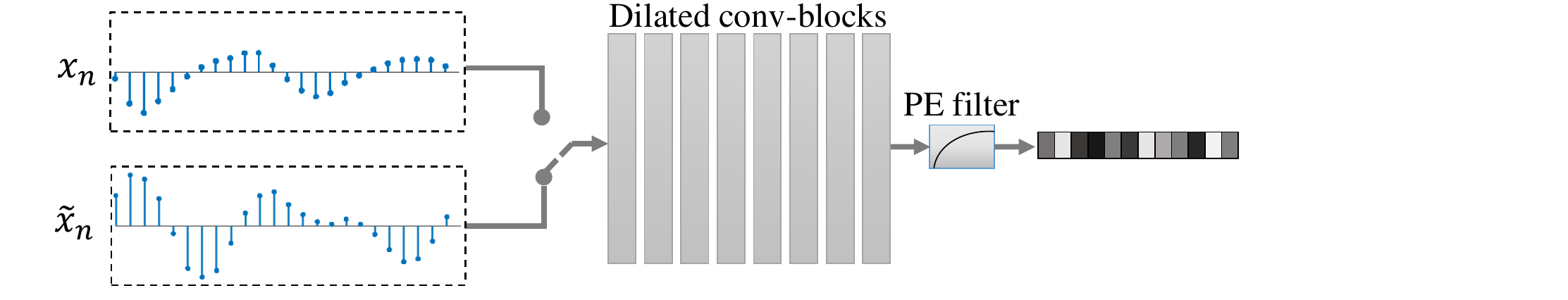}
	\end{subfigure}
	\caption{\label{fig:architectures} \textbf{Architectures.}
	}
\end{figure*}
The generator at each scale is built from 8 dilated convolutional blocks. The first 7 blocks contain dilated convolution, Batch-Norm and leakyReLU with slope 0.2. The last block is convolutional only. The dilation factor grows exponentially from 1 in the first block to 128 in the last one. The convolutional blocks feed a gated activation unit, which is followed by an extra $1\times1$ convolution and PE filter. The discriminator at each scale is similar to the generator, except it does not have the gated activation unit. The number of channels in each layer, for both the generator and the discriminator, is 16 at the coarsest scale and 96 at the rest of the scales. Figure \ref{fig:architectures} shows an illustration of the generator and discriminator architectures.

\subsection{Positional encoding}
In order to ensure that the generator’s output is of the same length as the real signal during training, we use zero-padding at its input. This zero padding functions as a positional encoding \cite{xu2021positional}, which allows the generator to know the absolute location with respect to the beginning and ending of the signal. This encoding extends up to one receptive field from the beginning and one receptive field from the ending of the signal. Therefore, the generator manages to ``remember'' these parts and to ``paste'' them at the borders of the generated signals. This makes the beginning and ending of the generated signals sound like those of the input. If desired, this phenomenon can be avoided by simply trimming the generated signal by one receptive field from each side.

\section{Additional experimental details}
\subsection{Unconditional generation}
\begin{figure*}
	\centering
	\captionsetup[subfigure]{labelformat=empty,justification=centering,aboveskip=1pt,belowskip=1pt}
	\begin{subfigure}[t]{0.48\textwidth}
	    \vspace{-.3mm}
		\centering
		\caption{Real waveform}
		\includegraphics[width=1\linewidth]{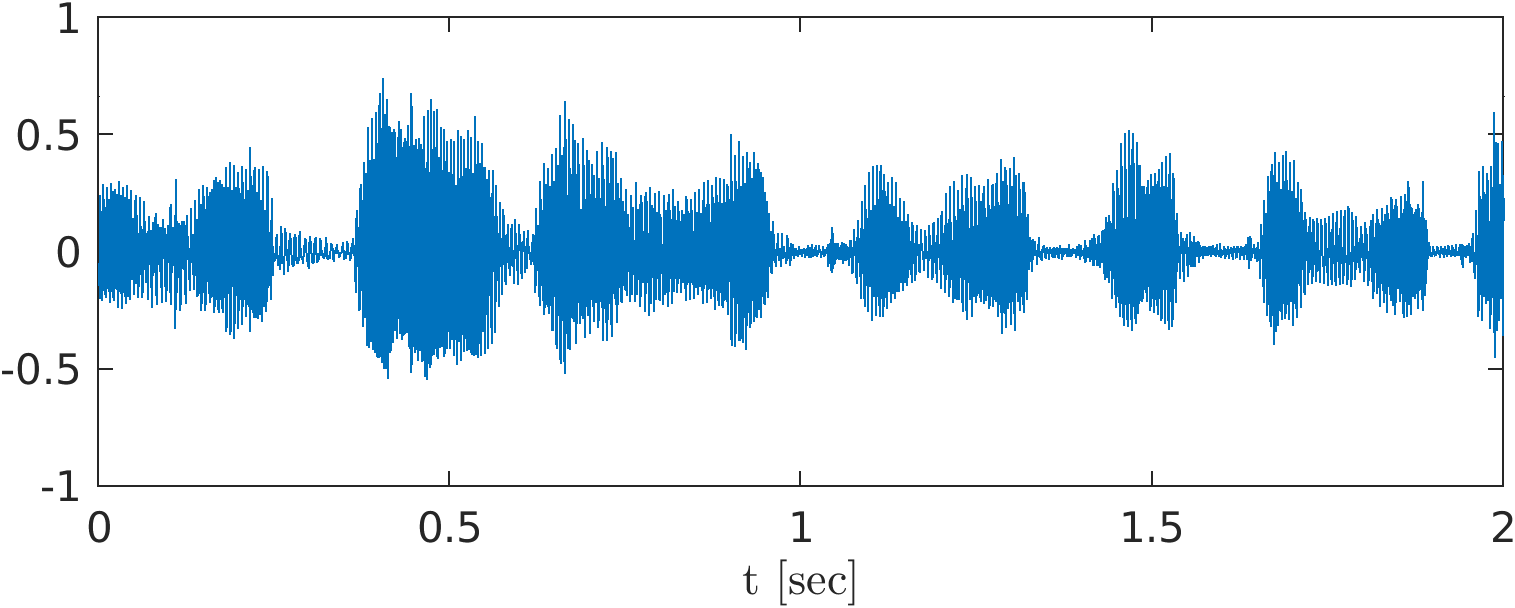}	
	\end{subfigure}
	\begin{subfigure}[t]{0.48\textwidth}
	    \vspace{-.3mm}
		\centering
		\caption{Spectrogram of real signal}
		\includegraphics[width=1\linewidth]{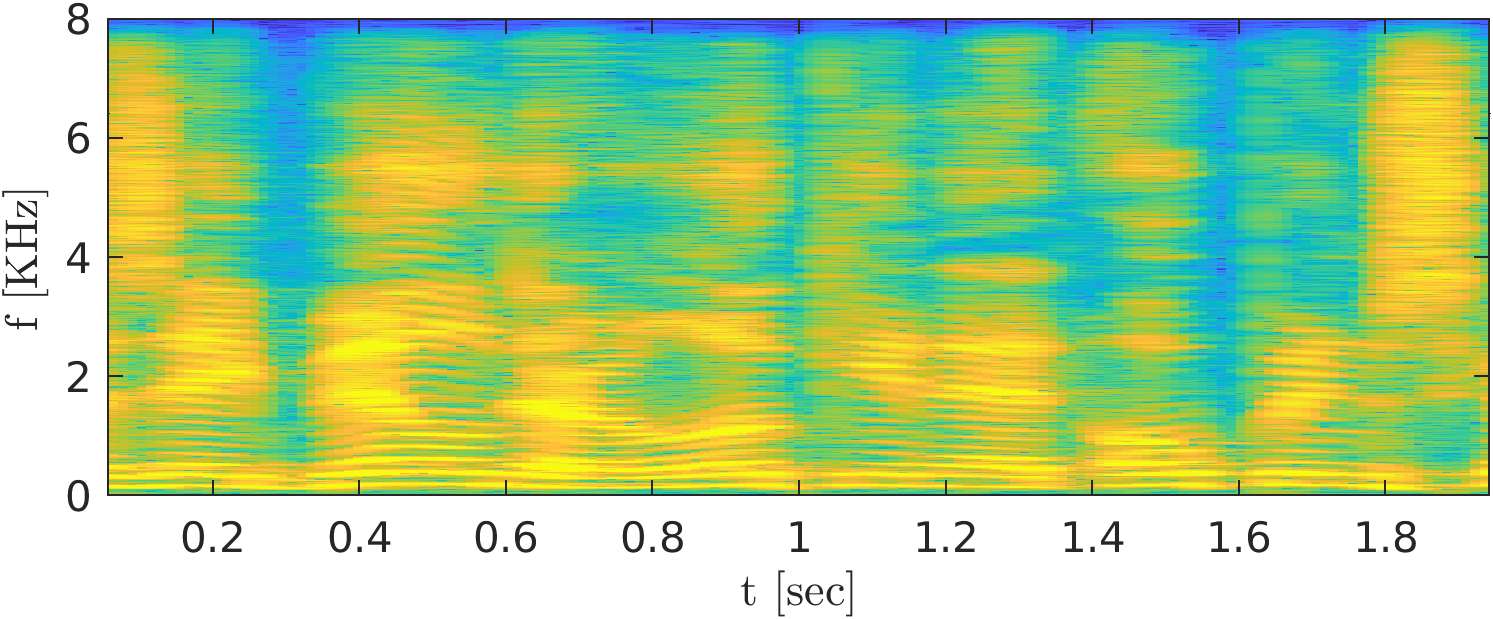}
	\end{subfigure}
	\par\smallskip
	\begin{subfigure}[t]{0.48\textwidth}
	    \vspace{-.3mm}
		\centering
		\caption{Generated waveform}
		\includegraphics[width=1\linewidth]{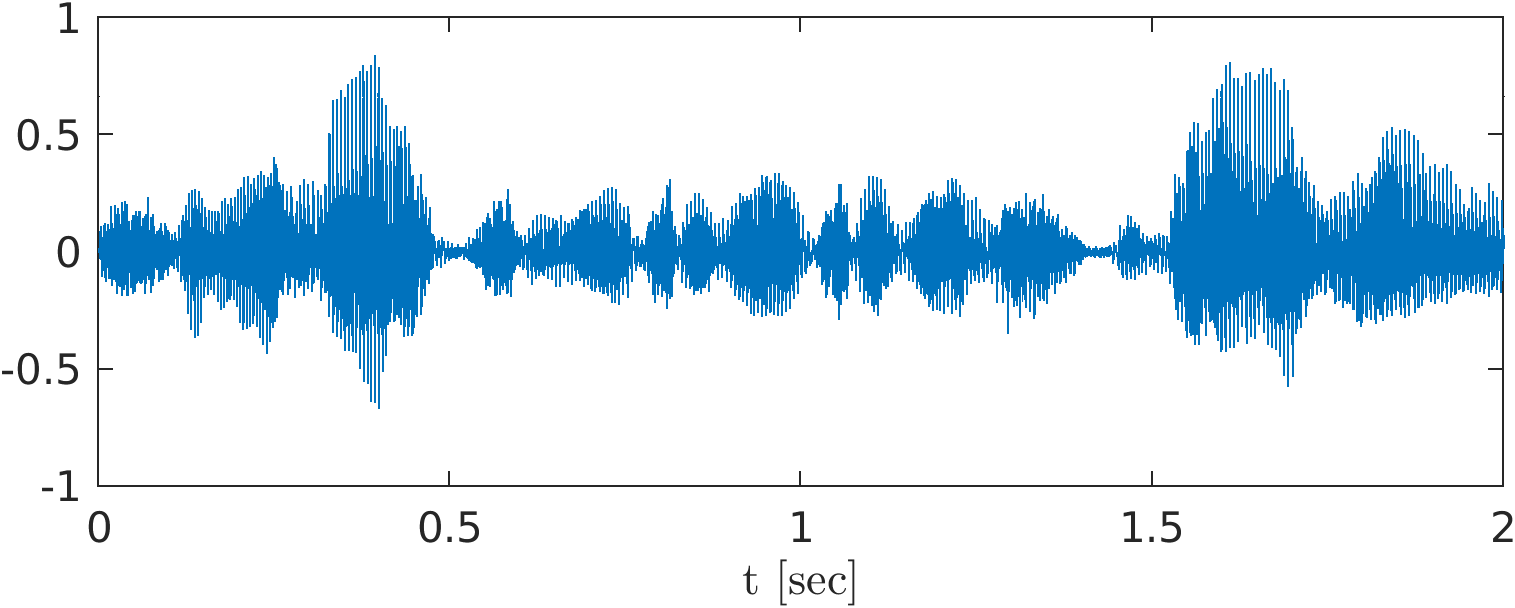}
	\end{subfigure}
	\begin{subfigure}[t]{0.48\textwidth}
	    \vspace{-.3mm}
		\centering
		\caption{Spectrogram of generated signal}
		\includegraphics[width=1\linewidth]{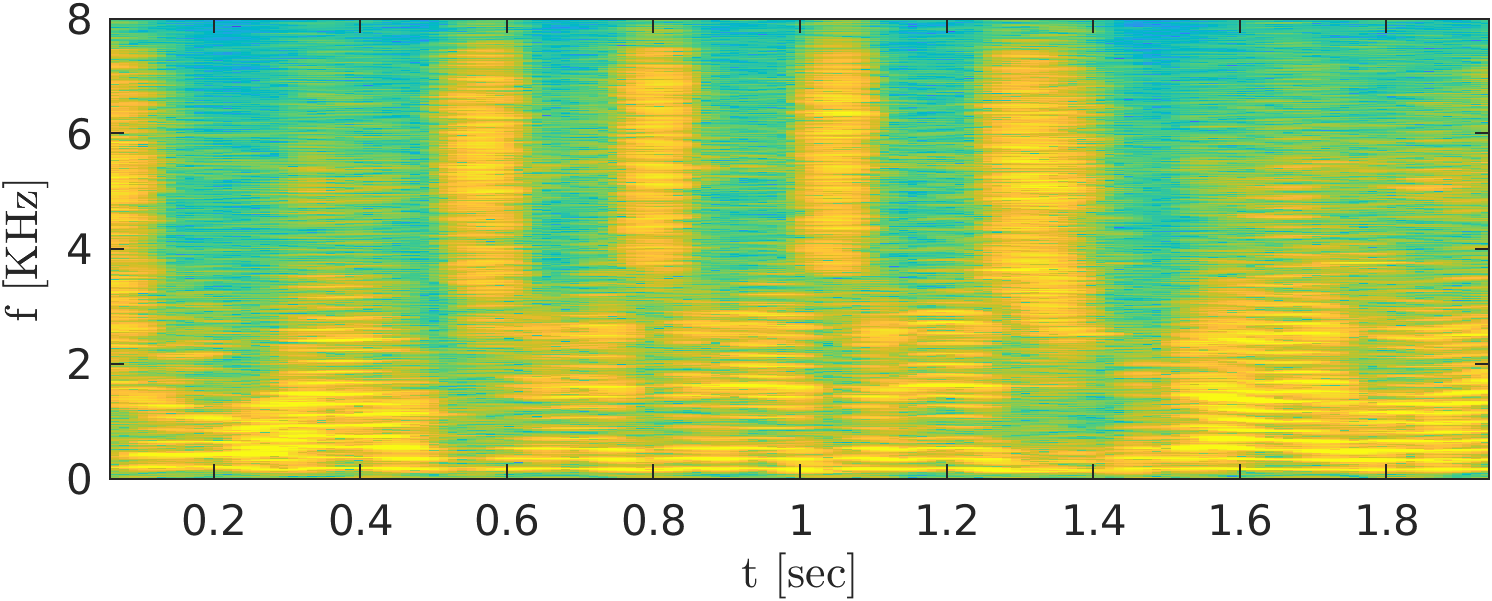}
	\end{subfigure}
	\caption{\label{fig:unconditional_example_speech} \textbf{Unconditional generation of speech signal}
	}
\end{figure*}

\begin{figure*}
	\centering
	\captionsetup[subfigure]{labelformat=empty,justification=centering,aboveskip=1pt,belowskip=1pt}
	\begin{subfigure}[t]{0.48\textwidth}
	    \vspace{-.3mm}
		\centering
		\caption{Real waveform}
		\includegraphics[width=1\linewidth]{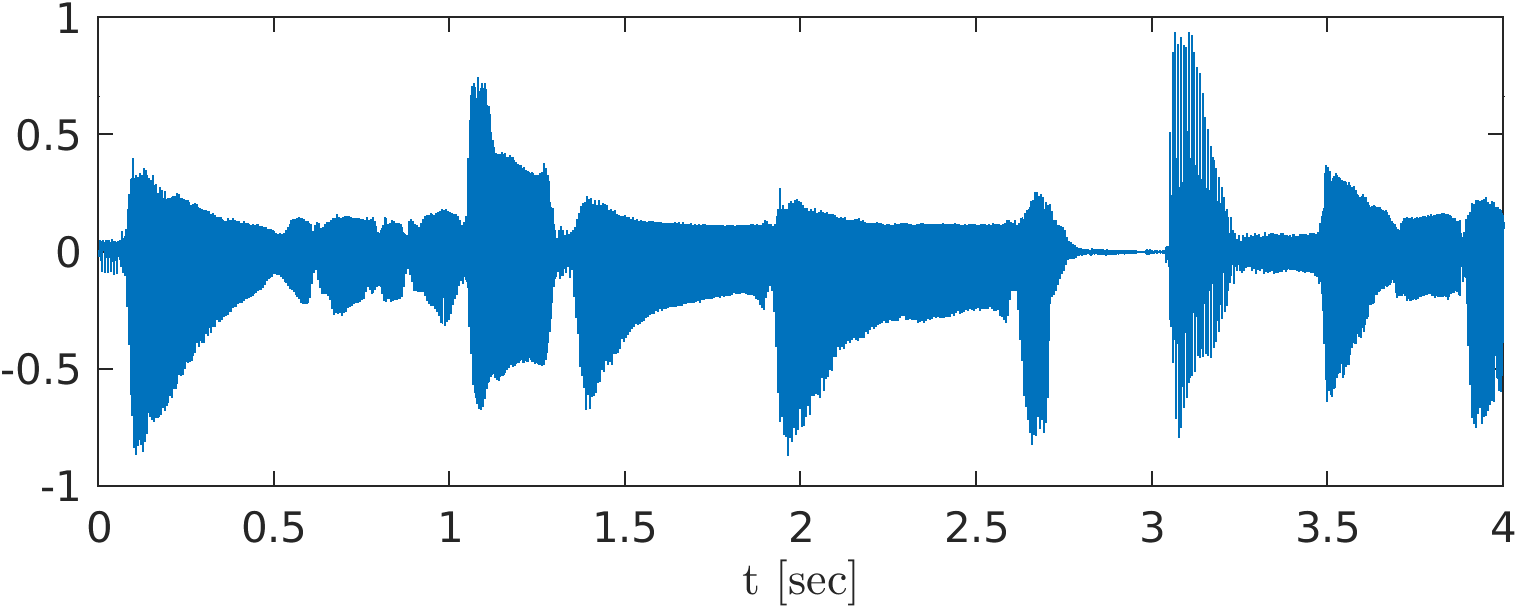}	
	\end{subfigure}
	\begin{subfigure}[t]{0.48\textwidth}
	    \vspace{-.3mm}
		\centering
		\caption{Spectrogram of real signal}
		\includegraphics[width=1\linewidth]{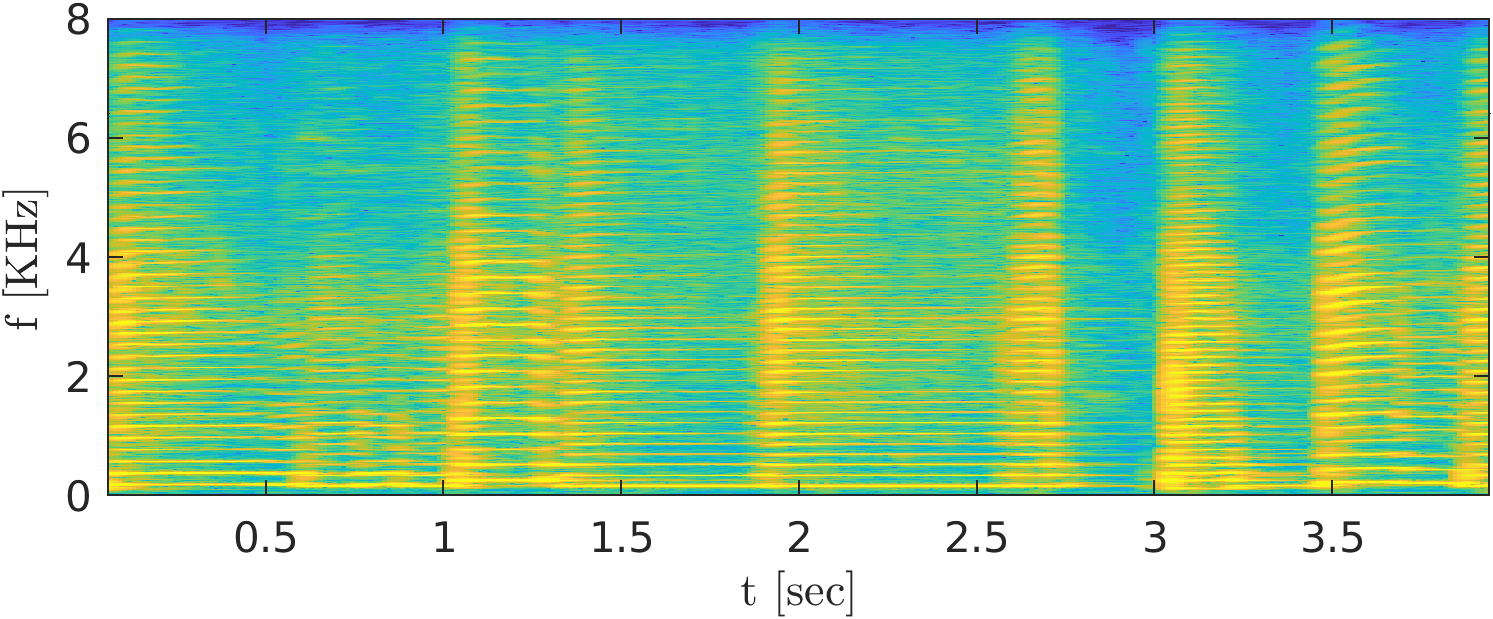}
	\end{subfigure}
	\par\smallskip
	\begin{subfigure}[t]{0.48\textwidth}
	    \vspace{-.3mm}
		\centering
		\caption{Generated waveform}
		\includegraphics[width=1\linewidth]{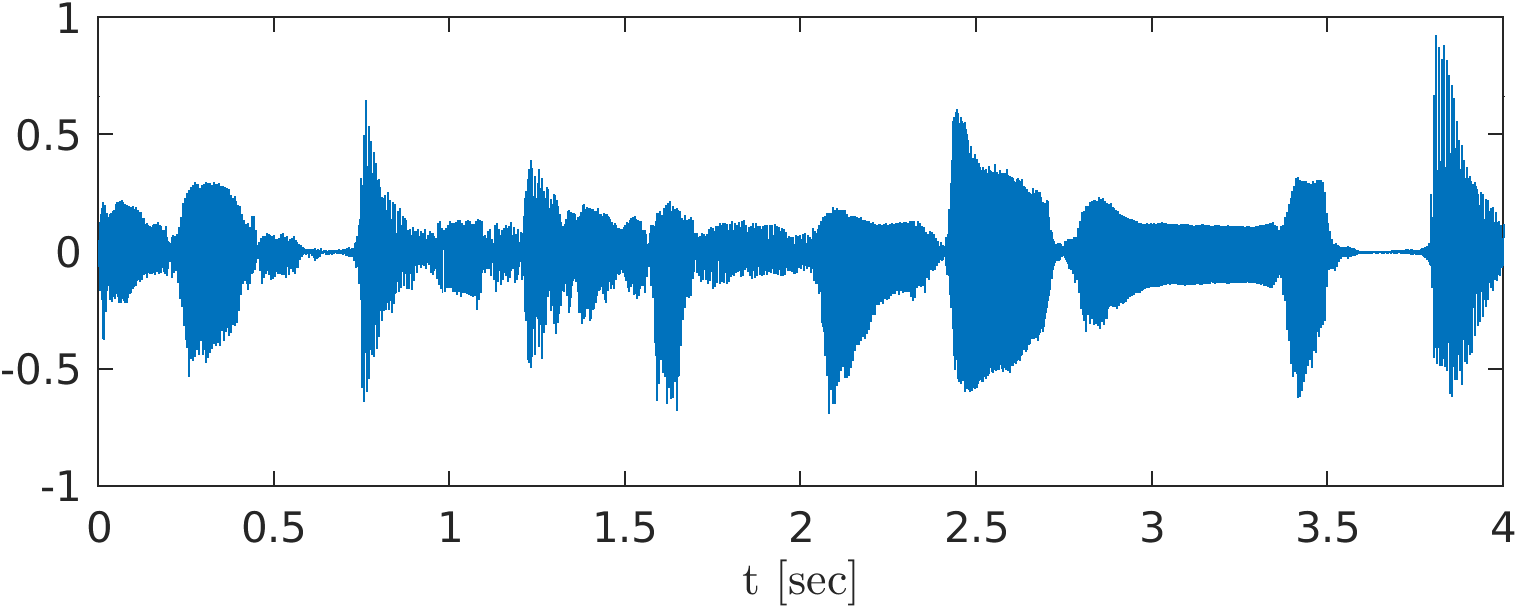}
	\end{subfigure}
	\begin{subfigure}[t]{0.48\textwidth}
	    \vspace{-.3mm}
		\centering
		\caption{Spectrogram of generated signal}
		\includegraphics[width=1\linewidth]{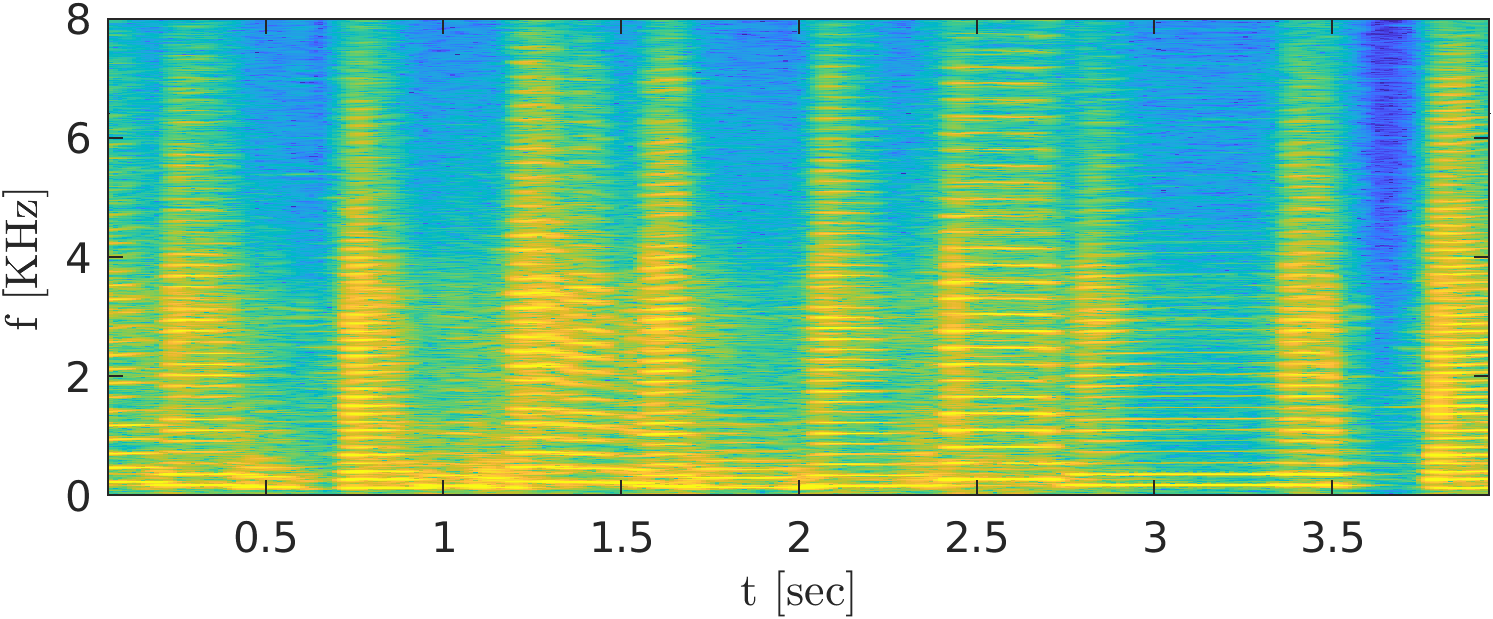}
	\end{subfigure}
	\caption{\label{fig:unconditional_example_sax} \textbf{Unconditional generation of saxophone signal}
	}
\end{figure*}

\begin{figure*}
	\centering
	\captionsetup[subfigure]{labelformat=empty,justification=centering,aboveskip=1pt,belowskip=1pt}
	\begin{subfigure}[t]{0.48\textwidth}
	    \vspace{-.3mm}
		\centering
		\caption{Real waveform}
		\includegraphics[width=1\linewidth]{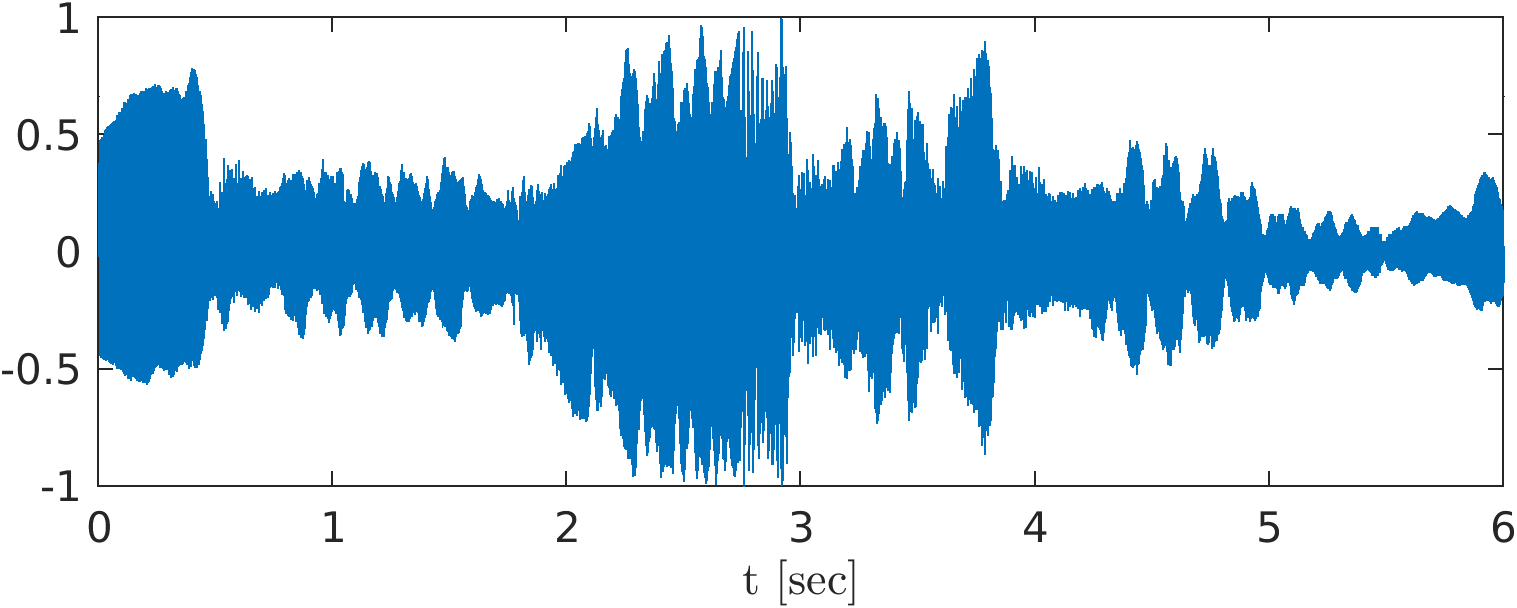}	
	\end{subfigure}
	\begin{subfigure}[t]{0.48\textwidth}
	    \vspace{-.3mm}
		\centering
		\caption{Spectrogram of real signal}
		\includegraphics[width=1\linewidth]{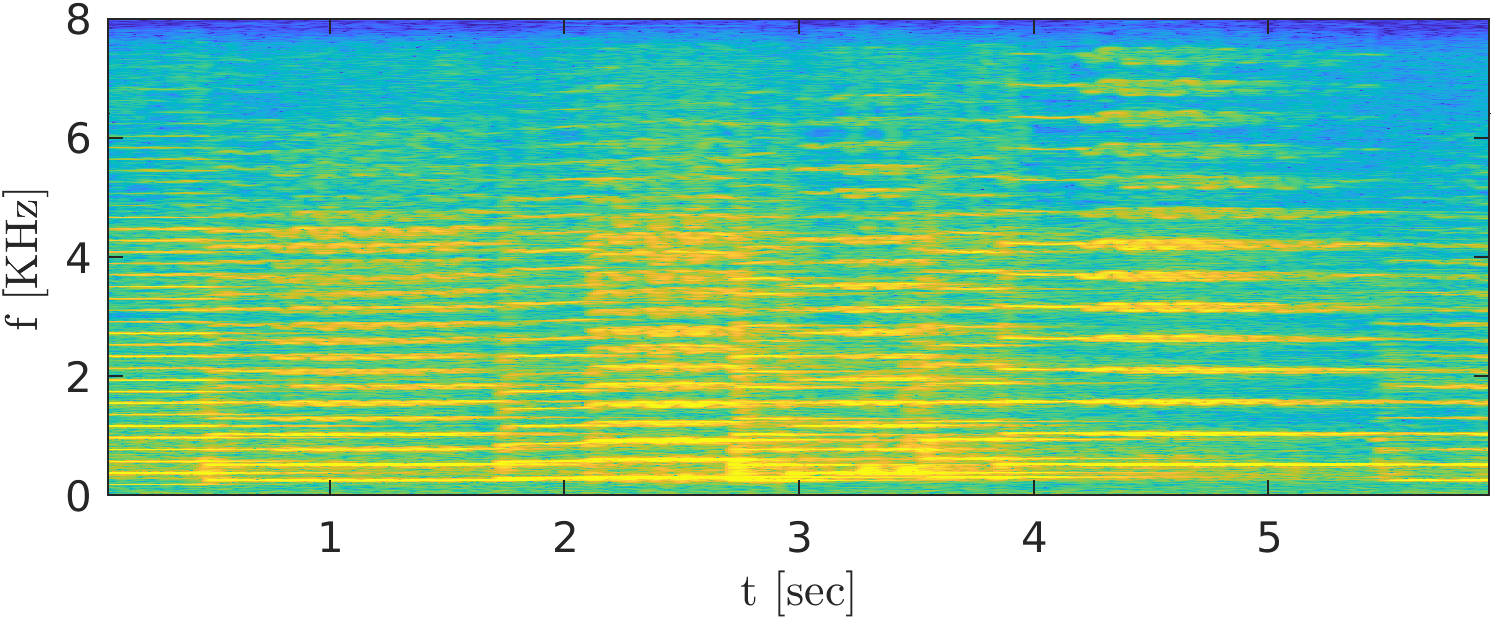}
	\end{subfigure}
	\par\smallskip
	\begin{subfigure}[t]{0.48\textwidth}
	    \vspace{-.3mm}
		\centering
		\caption{Generated waveform}
		\includegraphics[width=1\linewidth]{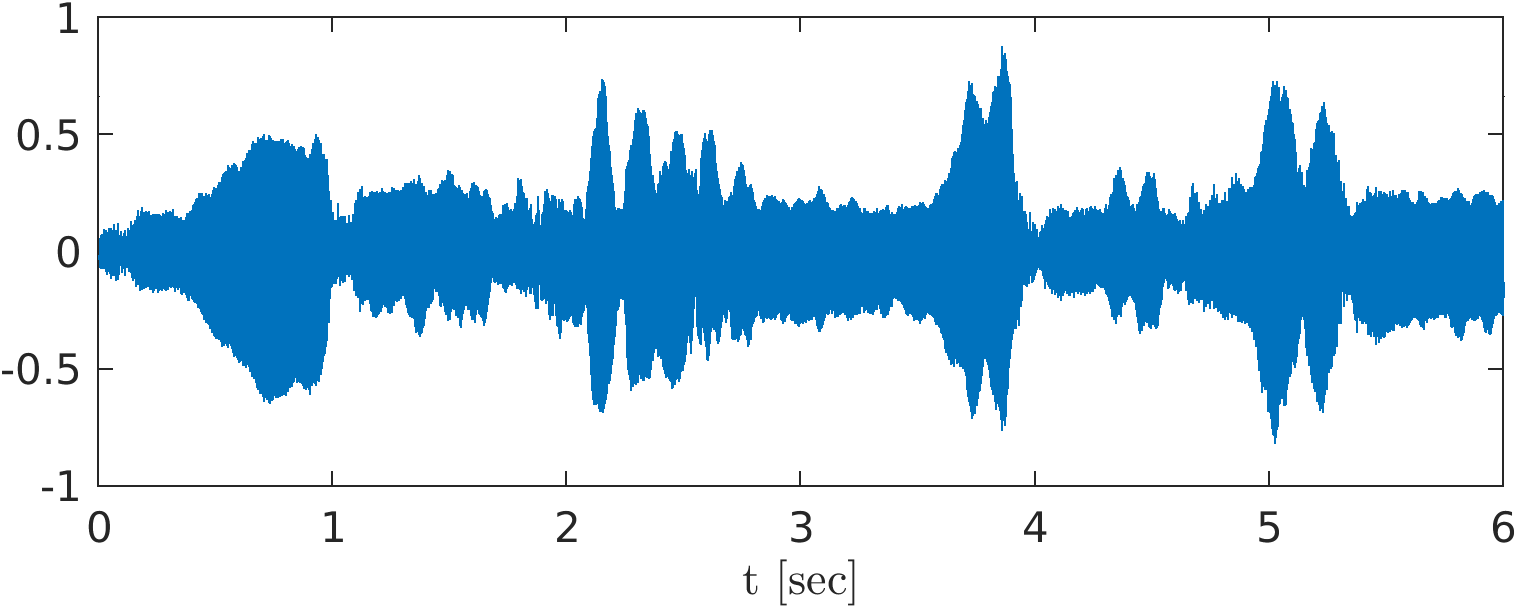}
	\end{subfigure}
	\begin{subfigure}[t]{0.48\textwidth}
	    \vspace{-.3mm}
		\centering
		\caption{Spectrogram of generated signal}
		\includegraphics[width=1\linewidth]{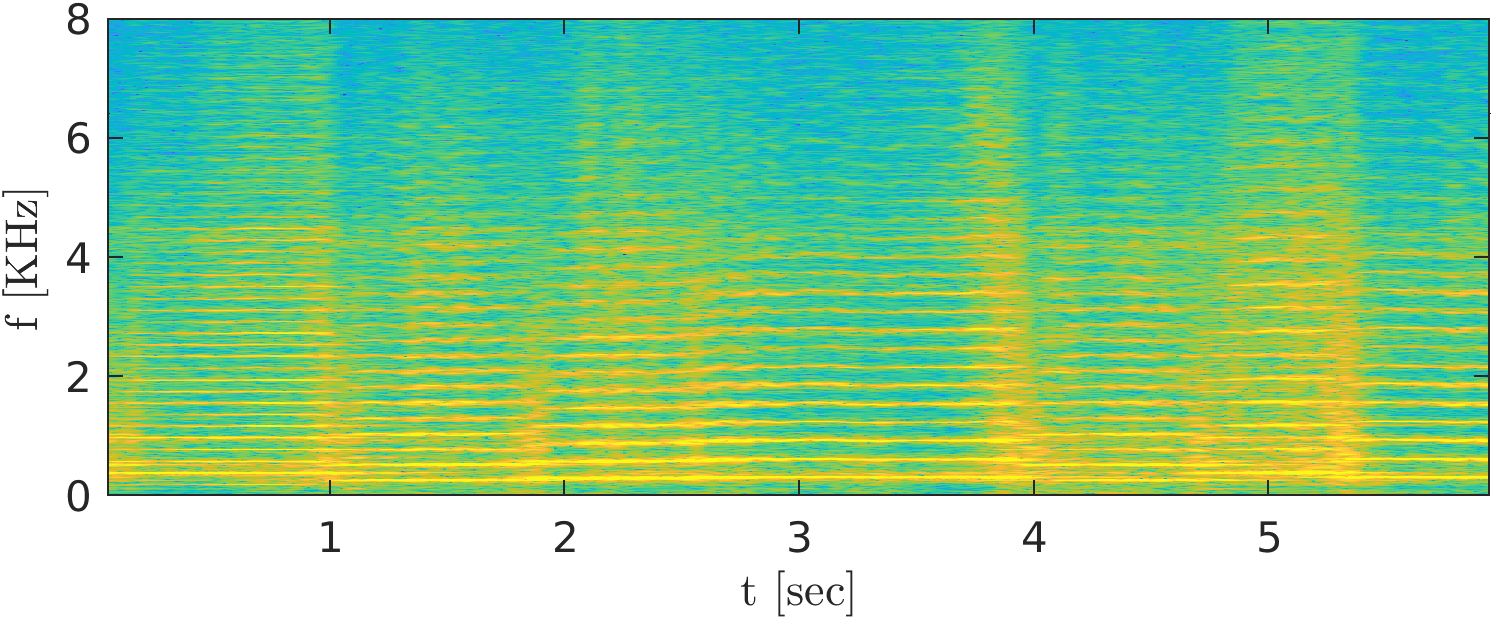}
	\end{subfigure}
	\caption{\label{fig:unconditional_example_violin} \textbf{Unconditional generation of violin signal}
	}
\end{figure*}

\begin{figure*}[h]
	\centering
	\captionsetup[subfigure]{labelformat=empty,justification=centering,aboveskip=1pt,belowskip=1pt}
	\begin{subfigure}[t]{0.98\textwidth}
	    \vspace{-.3mm}
		\centering
		\caption{Instructions to participants}
		\includegraphics[width=1\linewidth]{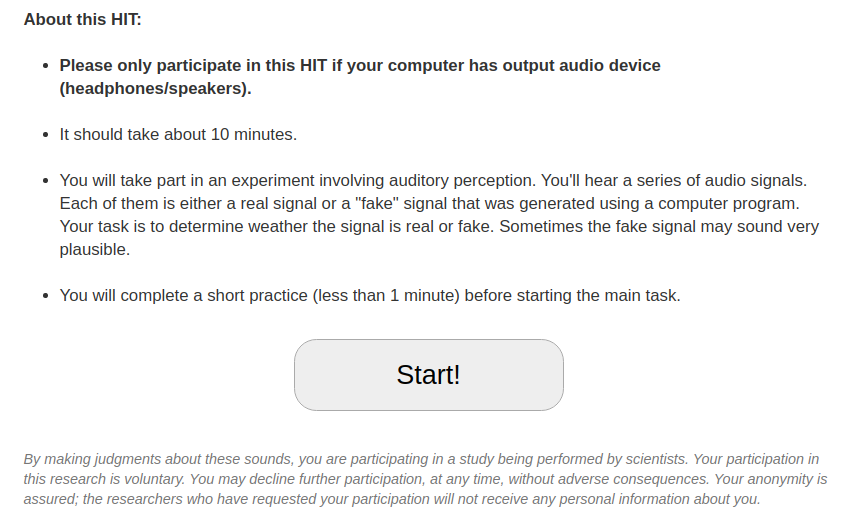}	
	\end{subfigure}
	\par\smallskip
	\begin{subfigure}[t]{0.98\textwidth}
	    \vspace{-.3mm}
		\centering
		\caption{Question presented to participants}
		\includegraphics[width=1\linewidth]{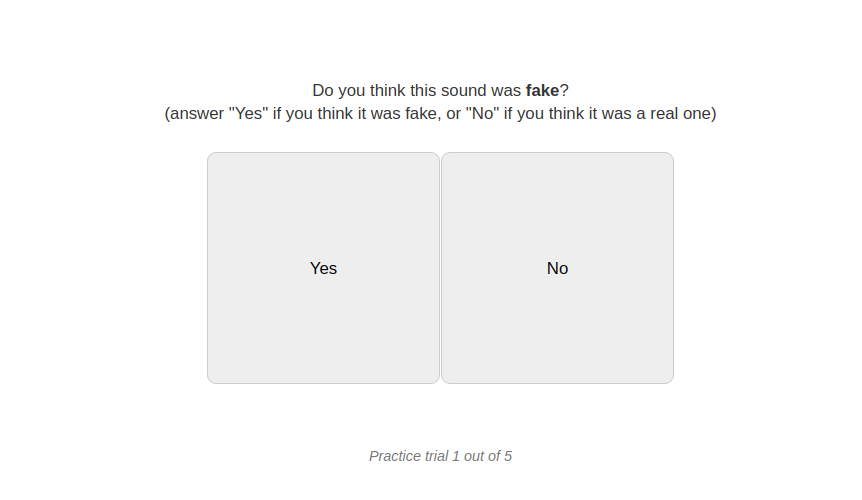}	
	\end{subfigure}
	\caption{\textbf{Unconditional generation unpaired user study.} After reading instructions (upper) and listening to a sound sample one time, the participant had to answer whether this sound was fake (bottom).\label{fig:amt_unconditional_unpaired} 
	} 
\end{figure*}

\begin{figure*}[h]
	\centering
	\captionsetup[subfigure]{labelformat=empty,justification=centering,aboveskip=1pt,belowskip=1pt}
	\begin{subfigure}[t]{0.98\textwidth}
	    \vspace{-.3mm}
		\centering
		\caption{Instructions to participants}
		\includegraphics[width=1\linewidth]{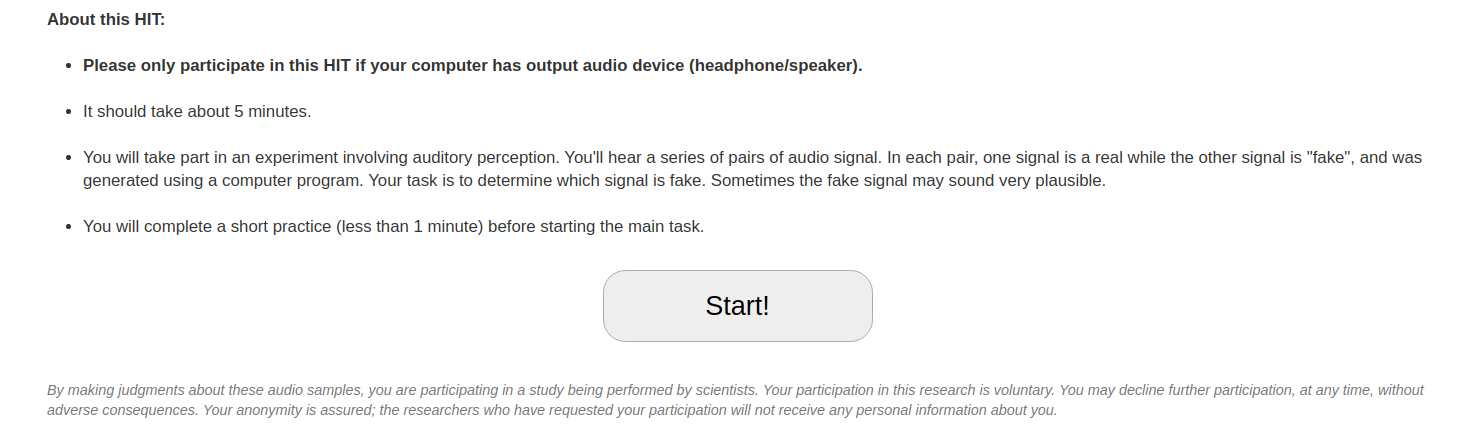}	
	\end{subfigure}
	\par\smallskip
	\begin{subfigure}[t]{0.98\textwidth}
	    \vspace{-.3mm}
		\centering
		\caption{Question presented to participants}
		\includegraphics[width=1\linewidth]{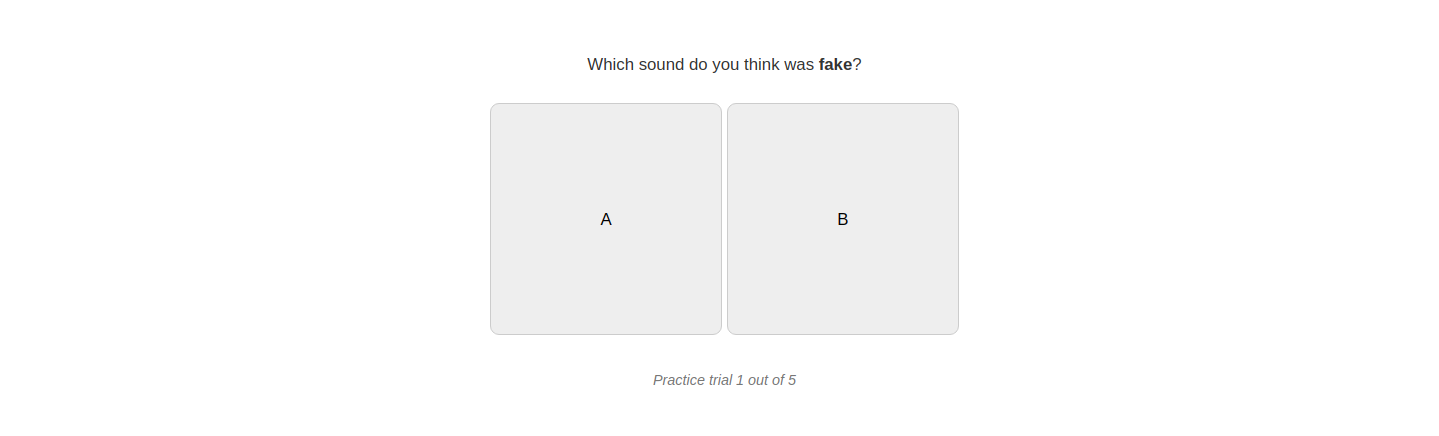}	
	\end{subfigure}
	\par\smallskip
	\begin{subfigure}[t]{0.98\textwidth}
	    \vspace{-.3mm}
		\centering
		\caption{Feedback after tutorial questions}
		\includegraphics[width=1\linewidth]{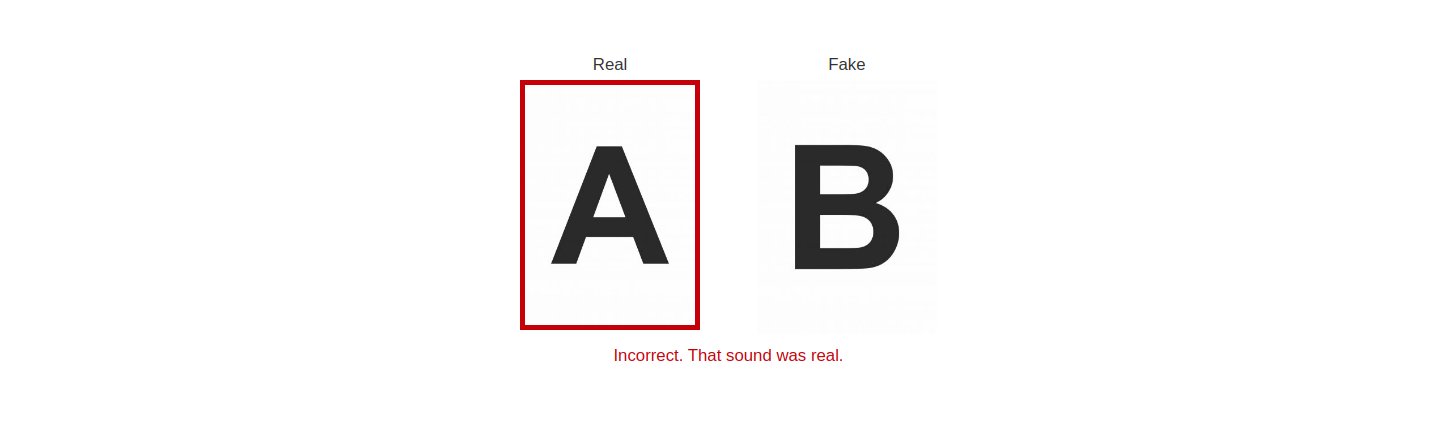}	
	\end{subfigure}
	\caption{\textbf{Unconditional generation paired user study.} After reading instructions (upper) and listening to real and fake sounds one time each, the participant had to decide which sound was fake (middle). The bottom image shows example of the paired tutorial presented to participants. \label{fig:amt_unconditional_paired} 
	} 
\end{figure*}

During training, we generate fake signals having the same length as the input. We do this by injecting to the generator noise of the same length as the input, padded with zeros of the length of the receptive field (we use no padding within the convolutional layers). To generate a signal of different length at inference time, we simply inject noise having the desired length at the coarsest scale's input.
Figures \ref{fig:unconditional_example_speech}-\ref{fig:unconditional_example_violin} show examples of real and generated signals of different types.
In order to evaluate performance and perceptual quality of our generated signals, a user study was conducted. Screenshots from the unpaired user study can be found in Fig. \ref{fig:amt_unconditional_unpaired}, and from the paired one in Fig. \ref{fig:amt_unconditional_paired}.

\begin{figure*}
	\centering
	\captionsetup[subfigure]{labelformat=empty,justification=centering,aboveskip=1pt,belowskip=1pt}
	\begin{subfigure}[t]{0.49\textwidth}
	    \vspace{.3mm}
		\centering
		\includegraphics[width=1\linewidth]{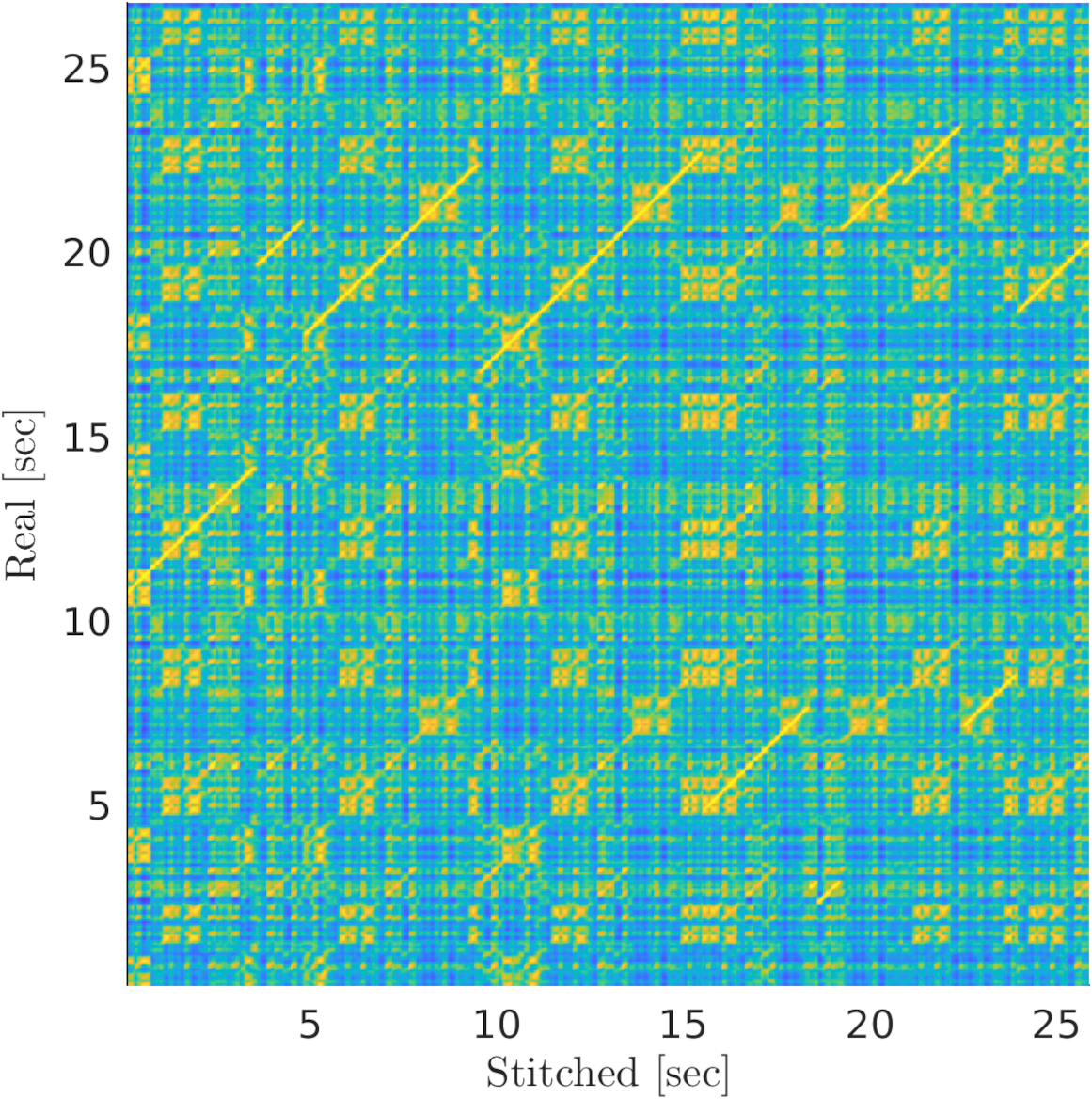}	
	\end{subfigure}
	\begin{subfigure}[t]{0.49\textwidth}
	    \vspace{.3mm}
		\centering
		\includegraphics[width=1\linewidth]{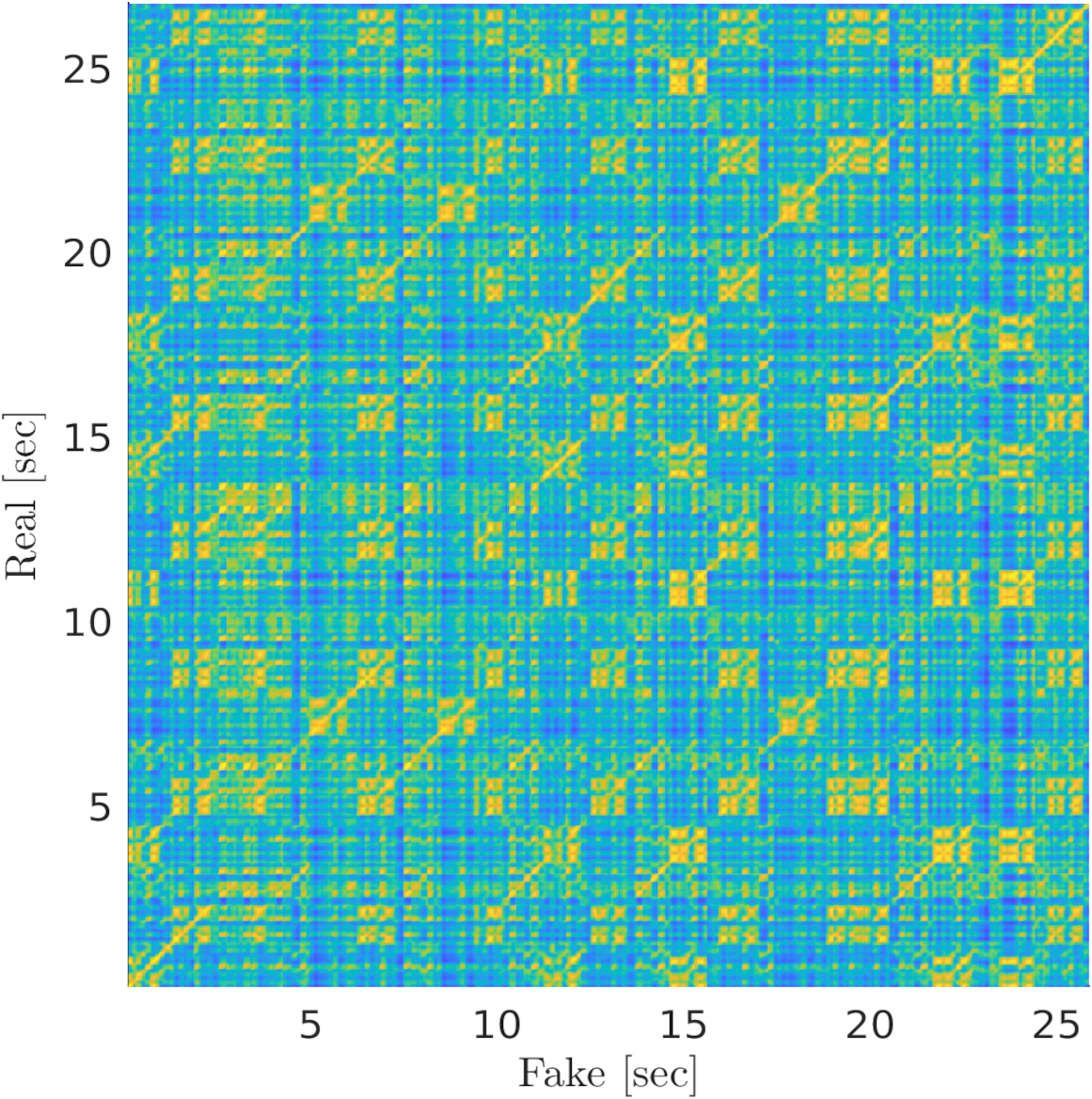}
	\end{subfigure}
	\begin{subfigure}[t]{0.49\textwidth}
	    \vspace{.3mm}
		\centering
		\includegraphics[width=1\linewidth]{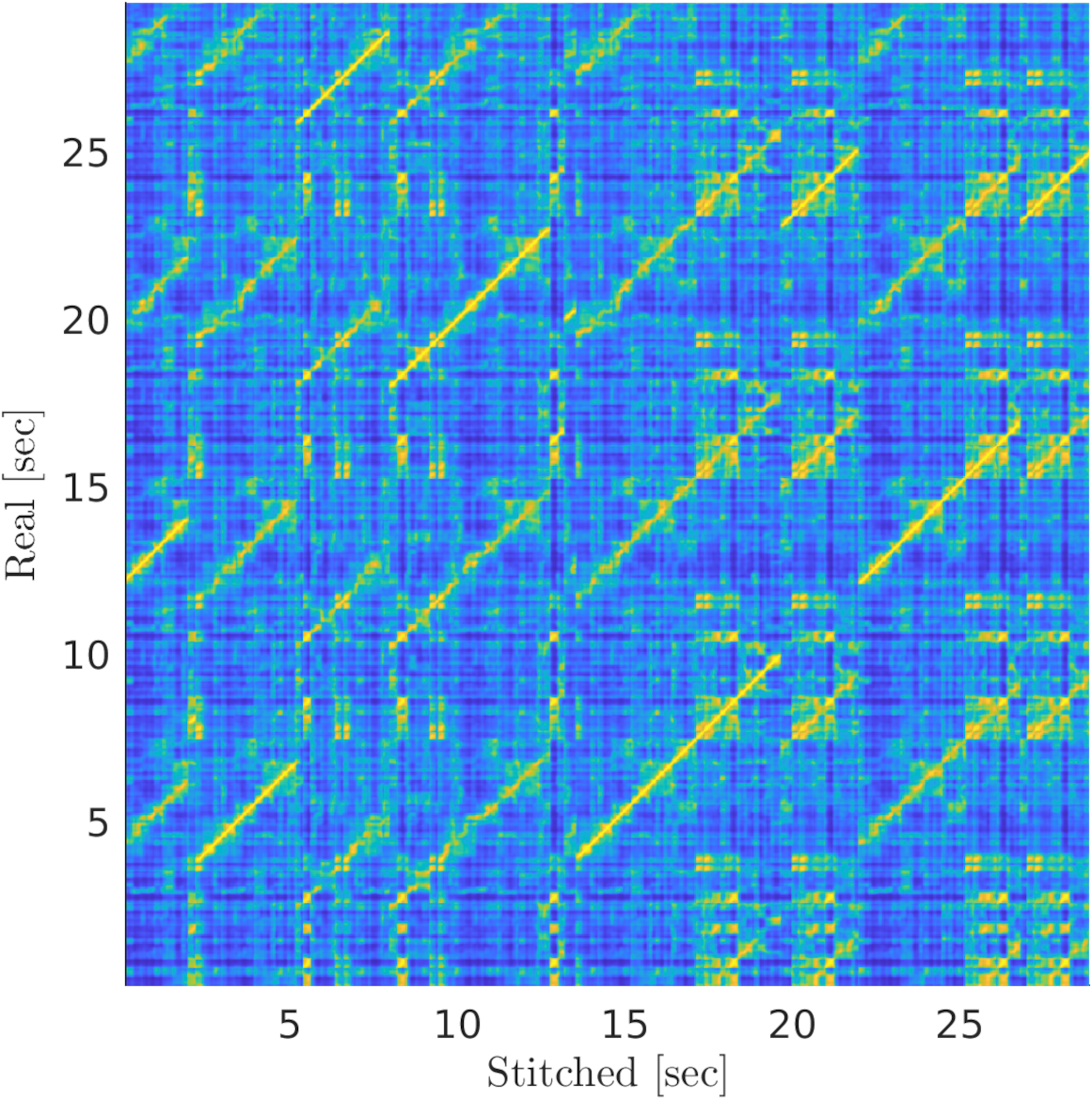}	
	\end{subfigure}
	\begin{subfigure}[t]{0.49\textwidth}
	    \vspace{.3mm}
		\centering
		\includegraphics[width=1\linewidth]{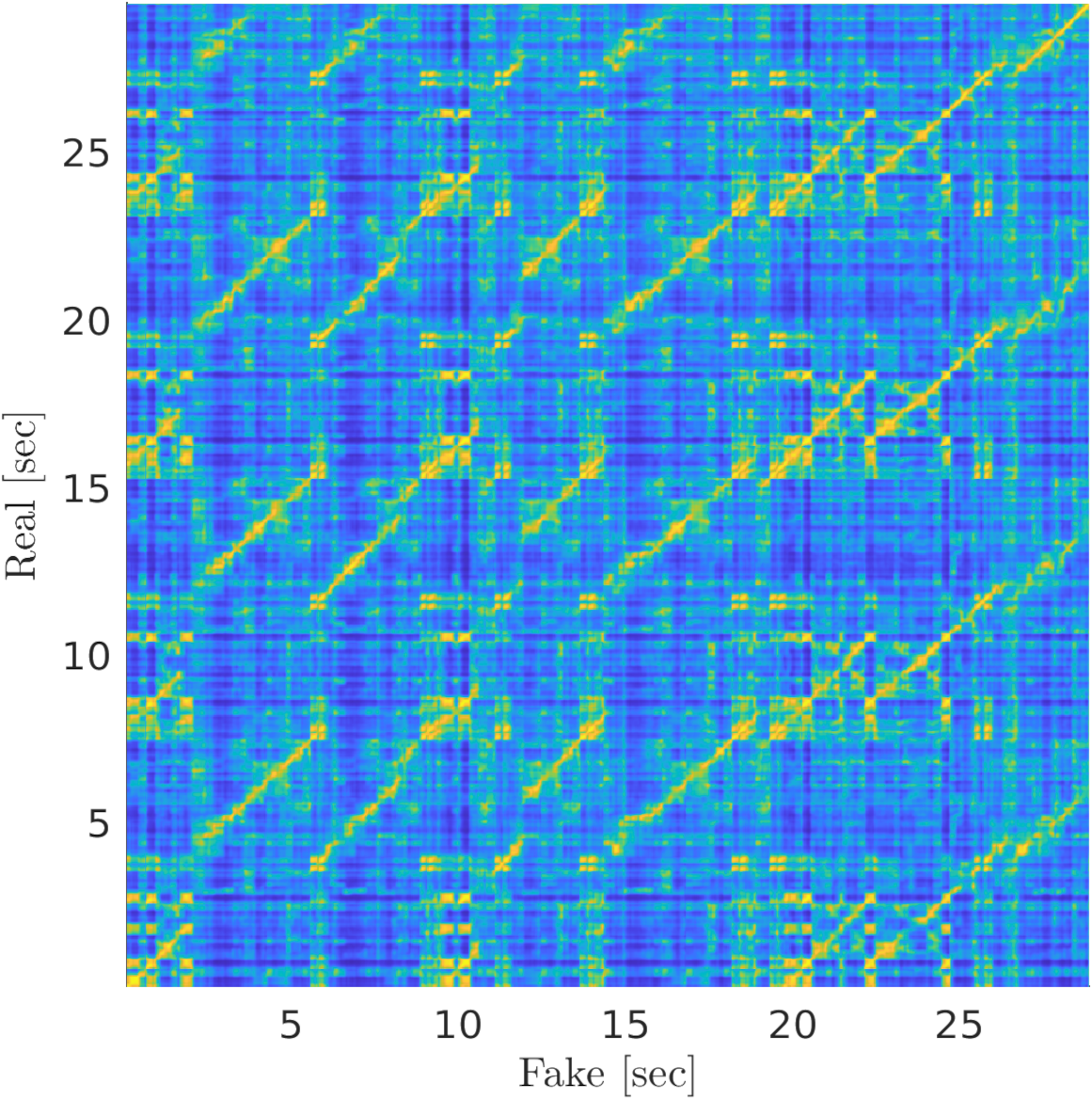}
	\end{subfigure}
	\begin{subfigure}[t]{0.49\textwidth}
	    \vspace{.3mm}
		\centering
		\includegraphics[width=1\linewidth]{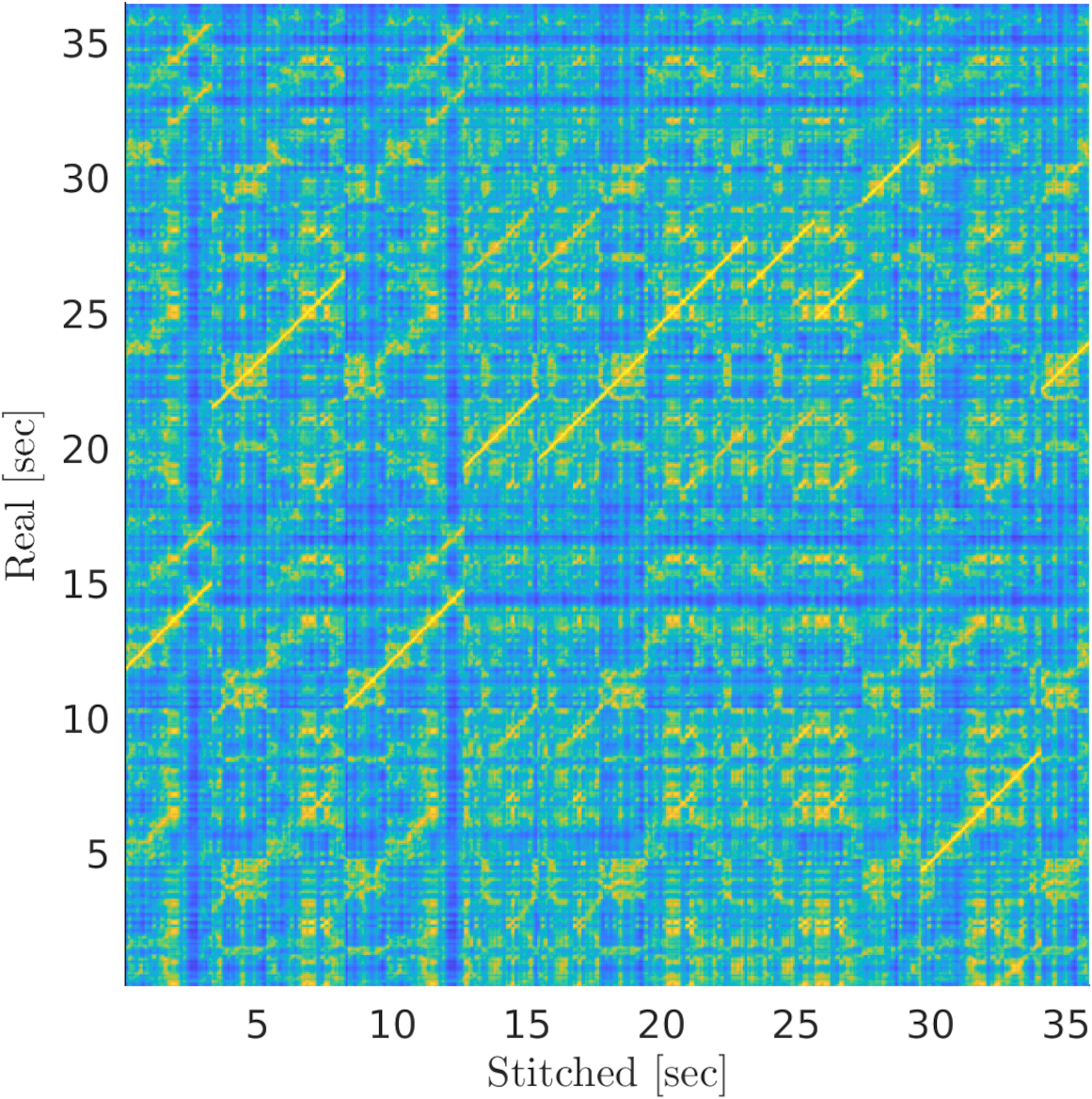}	
	\end{subfigure}
	\begin{subfigure}[t]{0.49\textwidth}
	    \vspace{.3mm}
		\centering
		\includegraphics[width=1\linewidth]{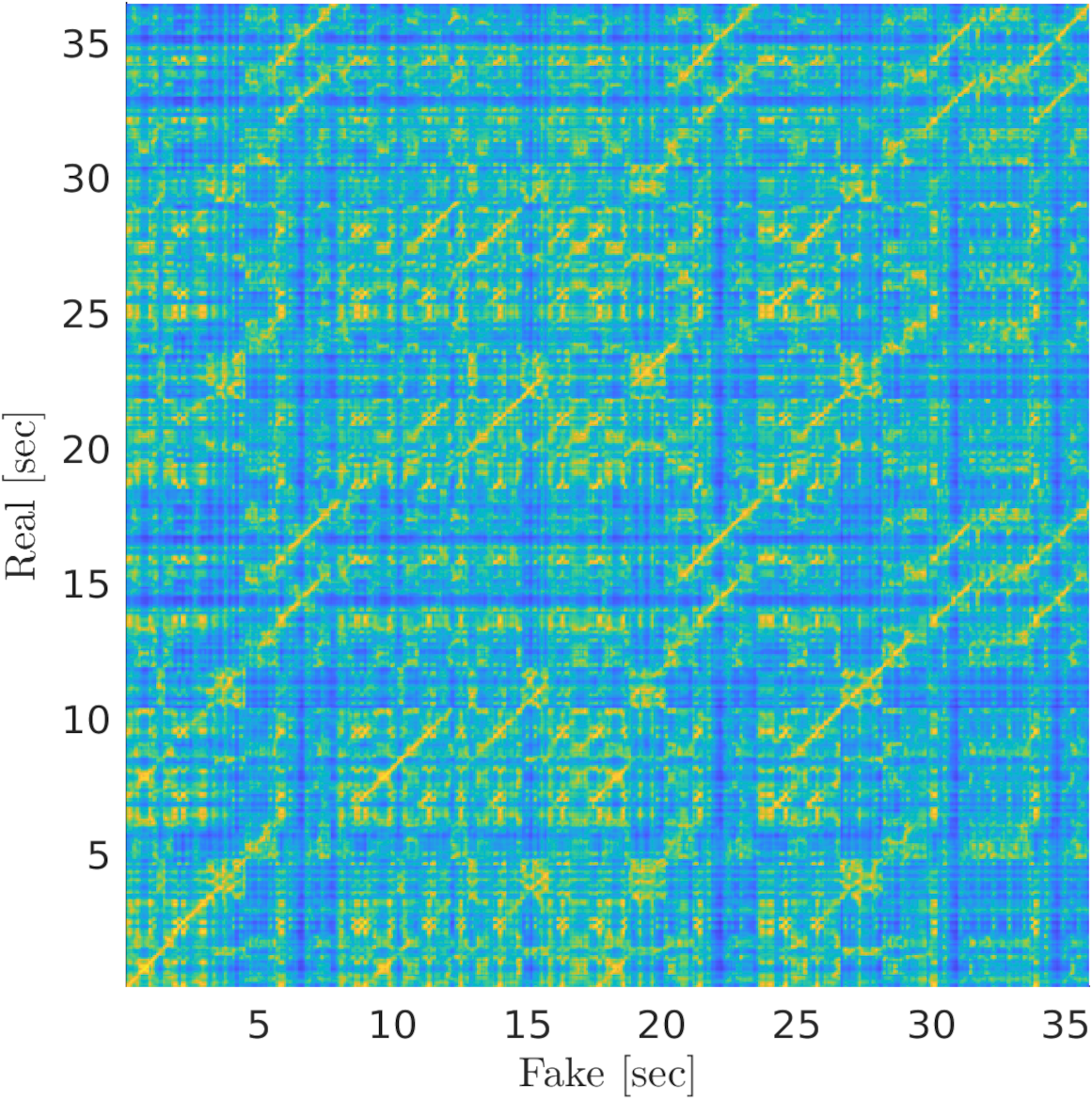}
	\end{subfigure}
	\caption{\textbf{Similarity matrices.} Matrices of signals created by naive cut and paste method (left column), and of our generated signals (right column). Our signals show more blurry lines as they can contain information from different temporal positions across frequency scales. \label{fig:sim_matrix_example_SM} 
	} 
\end{figure*}

\myparagraph{Calculation of similarity matrix.} As explained in the main text, in order to better understand the nature of our generated signals and specifically how they differ from signals generated by a naive cut-and-paste approach, we compute a similarity matrix between the fake and the real signals. The matrix is calculated as follows. First we compute STFT matrices for the real and fakes signals, and take the absolute values of their entries. We denote these by $R$ and $F$, respectively. The STFT matrix is calculated on segments of 4096 samples, multiplied by the Hann window, and with hop size of 128 samples. Next, the similarity value between frame $i$ in the real signal and frame $j$ in the fake signal is computed as the cosine similarity between the $i^{th}$ and $j^{th}$ columns in $R$ and $F$, respectively, i.e.,
\begin{equation}
    \text{sim}(i,j) = \frac{\langle R_i,F_j \rangle}{\Vert R_i\Vert\Vert F_j\Vert}
\end{equation}
In Figure \ref{fig:sim_matrix_example_SM} we show several examples of similarity matrices of naively stitched signals and of our generated signals.
{}

\subsection{Bandwidth extension}
For the bandwidth extension experiments, we trained CAW models for each speaker of the 9 test speakers in the VCTK dataset. Each speaker's sentences were divided to batches of 4-10 sentences, such that each batch contains between 20 and 25 seconds of speech. This resulted in around 50 models for each speaker. At inference time, each sentence of the speaker was extended by all of the models, except for the one the sentence was trained on. For each sentence we calculated the mean result and the standard deviation, across all models. The mean and std reported in the main text correspond to the average of all sentences of all speakers. In the \textbf{single speaker} task, we only evaluate the sentences defined as test set for the TFiLM model~\cite{birnbaum2019temporal}.

\myparagraph{Evaluation metrics.} We used two common evaluation metrics in order to evaluate the BE results: SNR and LSD. These are defined as
\begin{equation*}
    \text{SNR}(x, \hat{x}) = 20  \log_{10}\left(\frac{||x||_2}{||x-\hat{x}||_2}\right)
\end{equation*}
\begin{equation*}
    \text{LSD}(x, \hat{x}) = \frac{1}{L}\sum_{l=1}^{L}{\sqrt{\frac{1}{K}\sum_{k=1}^{K}{\left(X(l,k)-\hat{X}(l,k)\right)^2}}}
\end{equation*}
where $X$ and $\hat{X}$ are the log magnitudes of the STFTs of the ground truth signal $x$ and the output extended signal $\hat{x}$, respectively. $L$ is the number of STFT frames and $K$ is the window size, which is 2048 samples in our case, calculated without overlaps.

\subsection{Audio inpainting}
\begin{figure*}
	\centering
	\captionsetup[subfigure]{labelformat=empty,justification=centering,aboveskip=1pt,belowskip=1pt}
	\begin{subfigure}[t]{0.49\textwidth}
	    \vspace{-.3mm}
		\centering
		\caption{Spectogram of a signal with a silent gap}
		\includegraphics[width=1\linewidth]{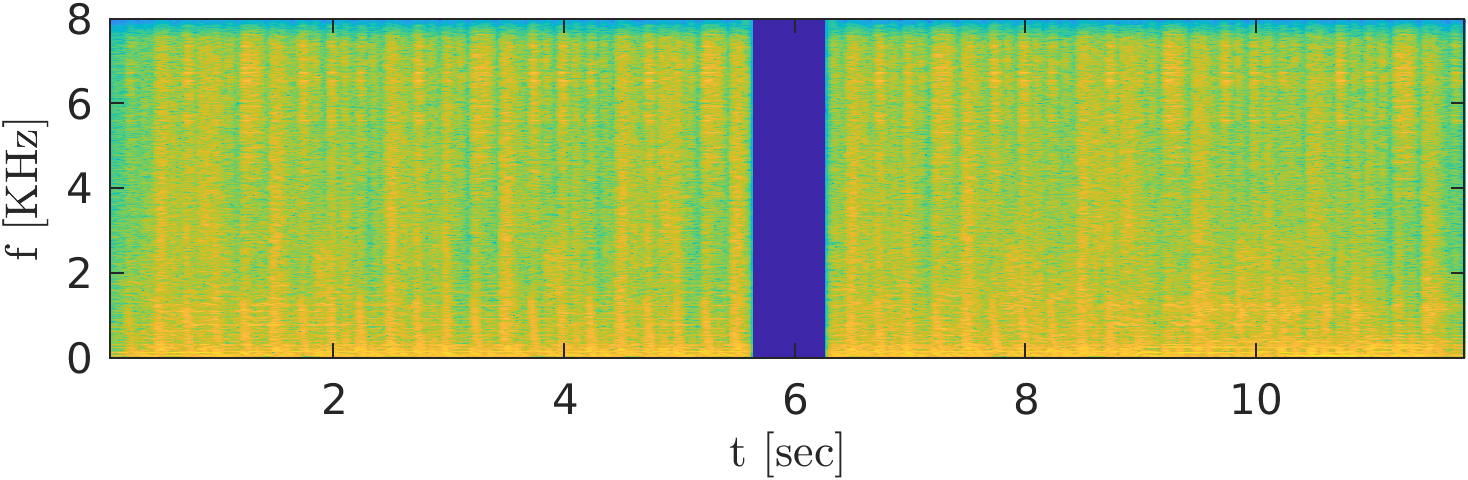}	
	\end{subfigure}
	\begin{subfigure}[t]{0.49\textwidth}
	    \vspace{.3mm}
		\centering
		\caption{Spectogram of a signal inpainted by our model}
		\includegraphics[width=0.95\linewidth]{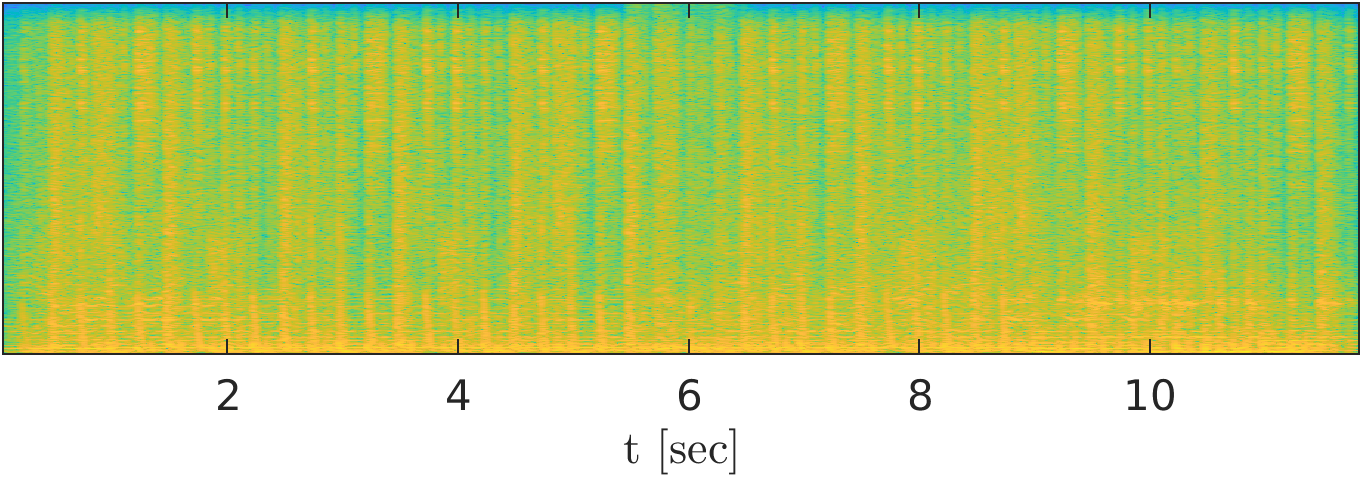}
	\end{subfigure}
	\par\smallskip
	\begin{subfigure}[t]{0.49\textwidth}
	    \vspace{-.3mm}
		\centering
		\caption{Spectogram of a signal with a silent gap}
		\includegraphics[width=1\linewidth]{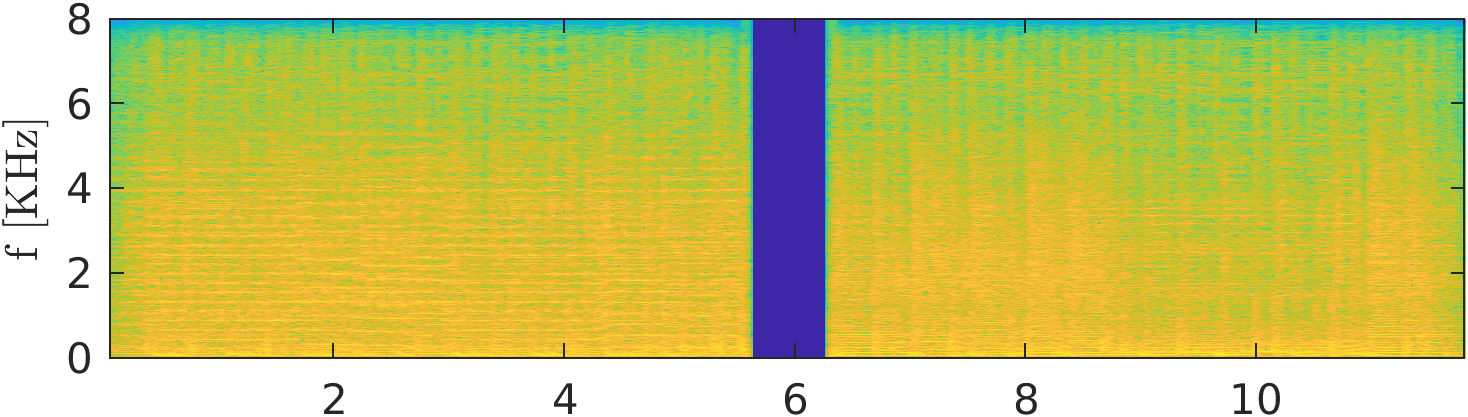}	
	\end{subfigure}
	\begin{subfigure}[t]{0.49\textwidth}
	    \vspace{.3mm}
		\centering
		\caption{Spectogram of a signal inpainted by our model}
		\includegraphics[width=0.95\linewidth]{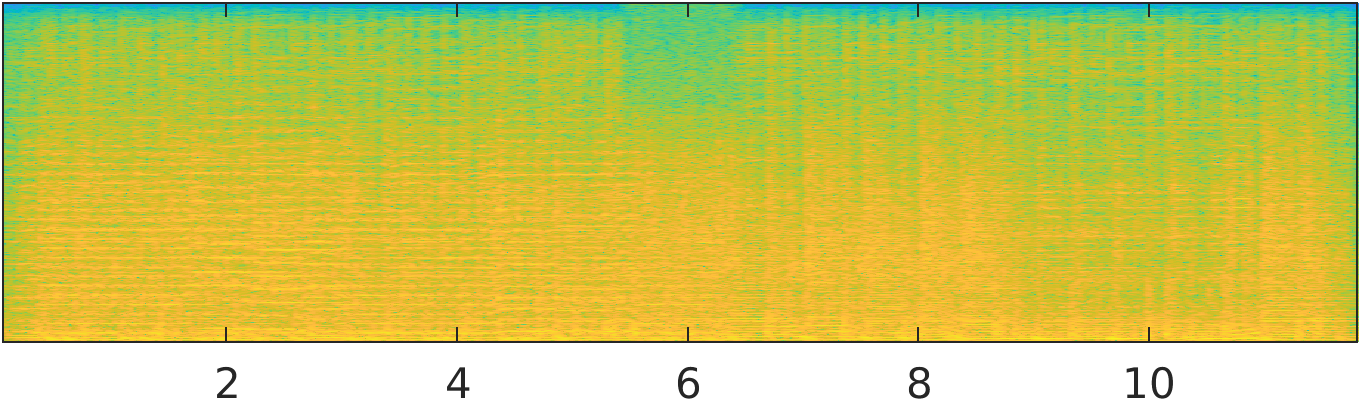}
	\end{subfigure}
	\caption{\textbf{Audio inpainting.}\label{fig:inpainting_example_SM} Examples of inpainting done by our model. The only input to the model is the signal with the missing gap, and a mask indicating the temporal location of the hole.
	} 
\end{figure*}

\begin{figure*}[h]
	\centering
	\captionsetup[subfigure]{labelformat=empty,justification=centering,aboveskip=1pt,belowskip=1pt}
	\begin{subfigure}[t]{0.98\textwidth}
	    \vspace{-.3mm}
		\centering
		\caption{Instructions to participants}
		\includegraphics[width=1\linewidth]{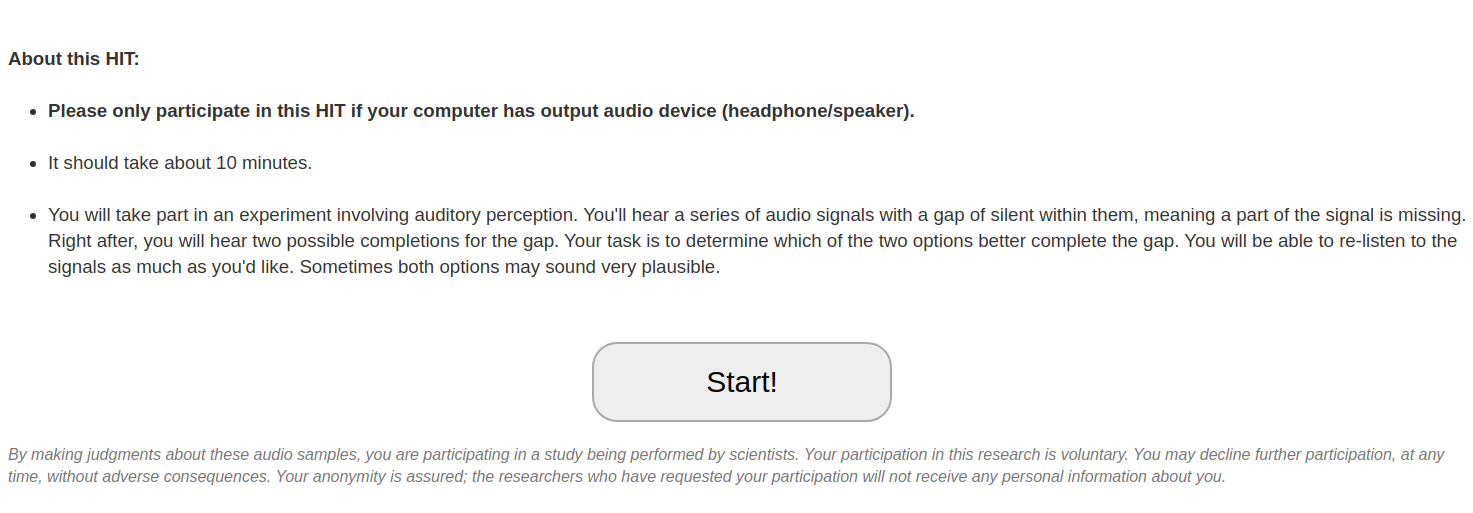}	
	\end{subfigure}
	\par\smallskip
	\begin{subfigure}[t]{0.98\textwidth}
	    \vspace{-.3mm}
		\centering
		\caption{Question presented to participants}
		\includegraphics[width=1\linewidth]{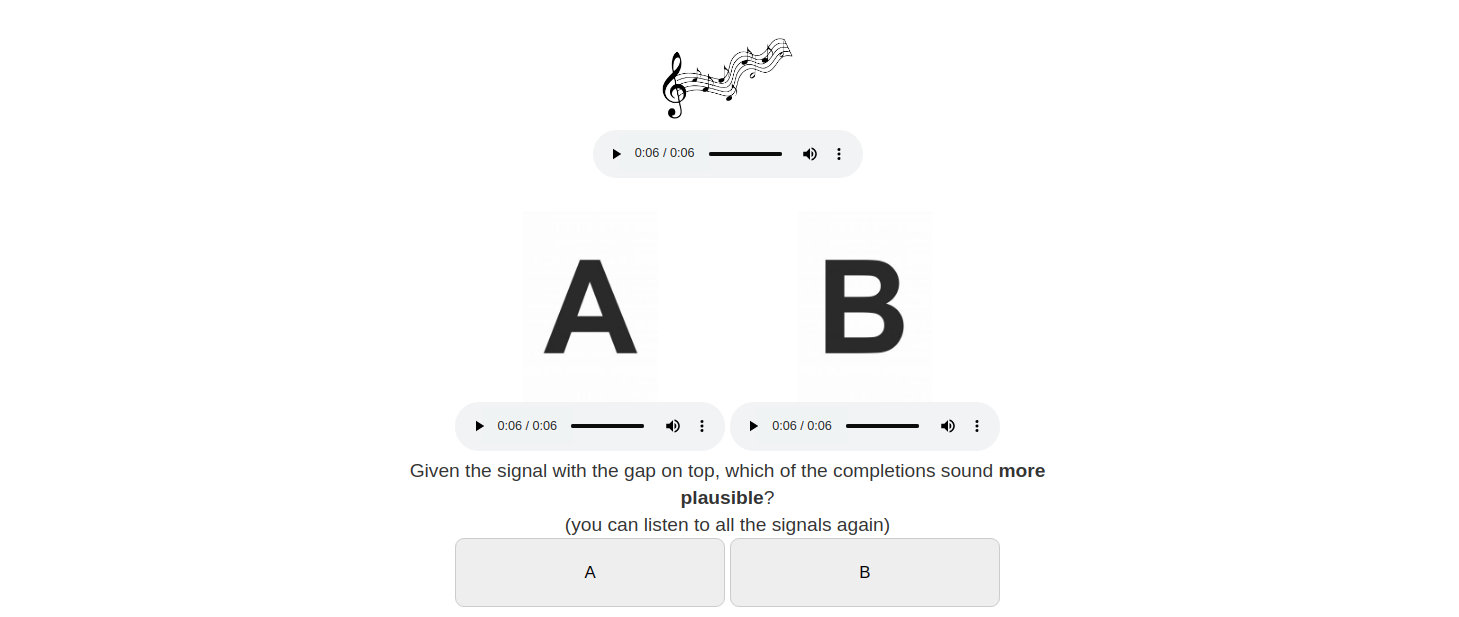}	
	\end{subfigure}
	\caption{\textbf{Inpainting user study.} After reading instructions (upper), participants were given the sound with a gap, along with two possible completions. They could listen to all three signals as many times as they wanted, and had to decide which completion sounded better. \label{fig:amt_inpainting} 
	} 
\end{figure*}

As explained in the main text, inpainting is done by training on a signal with a silent gap, where the loss terms are calculated only on the valid parts. More examples for inpainting of rock songs from the FMA dataset can be found in Fig.~\ref{fig:inpainting_example_SM}. In order to evaluate inpainting performance, a user study was conducted, comparing our results to the GACELA model~\cite{marafioti2020gacela} and to ground truth signals. Screenshots from the study can be found in Fig.
\ref{fig:amt_inpainting}
{}

\begin{figure*}
	\centering
	\captionsetup[subfigure]{labelformat=empty,justification=centering,aboveskip=1pt,belowskip=1pt}
	\begin{subfigure}[t]{0.49\textwidth}
	    \vspace{.3mm}
		\centering
		\caption{Clean signal}
		\includegraphics[width=1\linewidth]{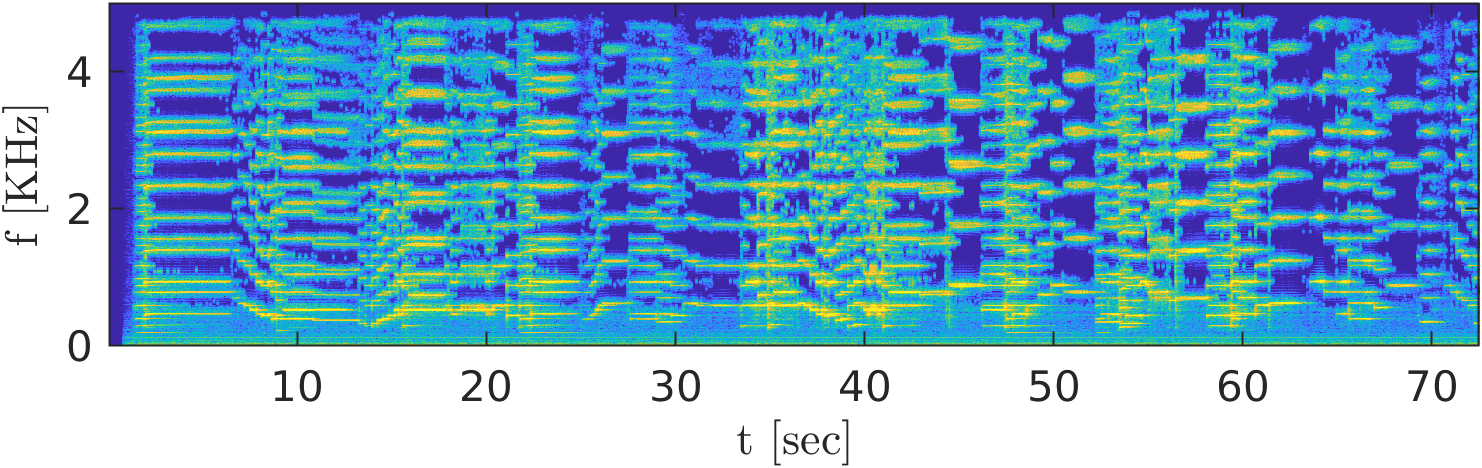}	
	\end{subfigure}
	\begin{subfigure}[t]{0.49\textwidth}
	    \vspace{.3mm}
		\centering
		\caption{Recorded gramophone noise}
		\includegraphics[width=1\linewidth]{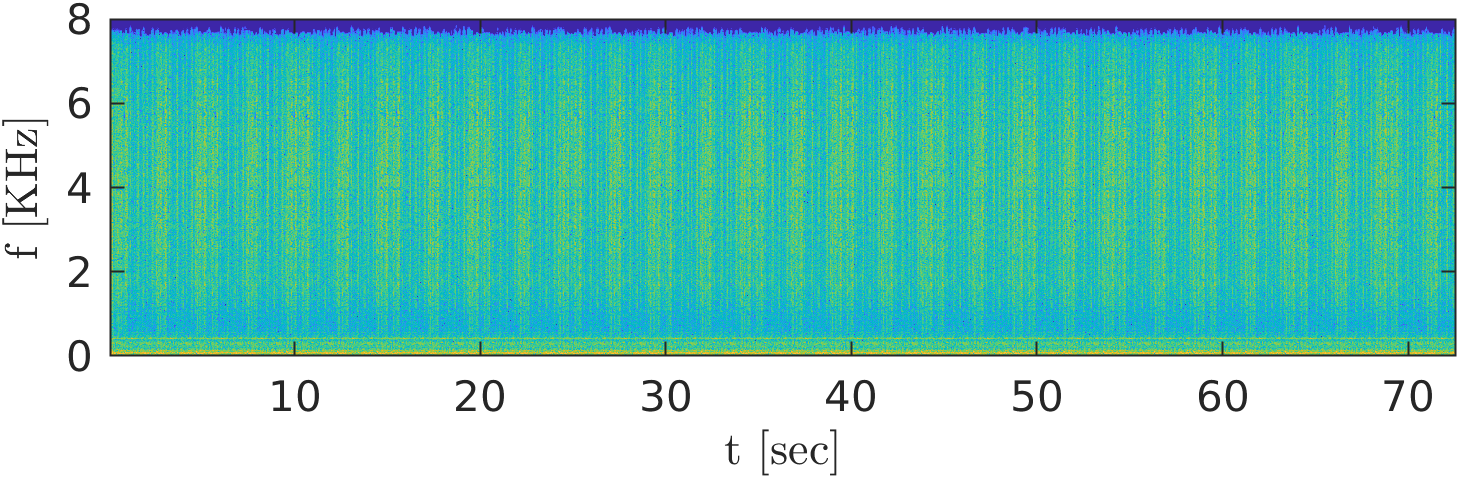}
	\end{subfigure}
	\begin{subfigure}[t]{0.49\textwidth}
	    \vspace{.3mm}
		\centering
		\caption{White noise added ($5dB$)}
		\includegraphics[width=1\linewidth]{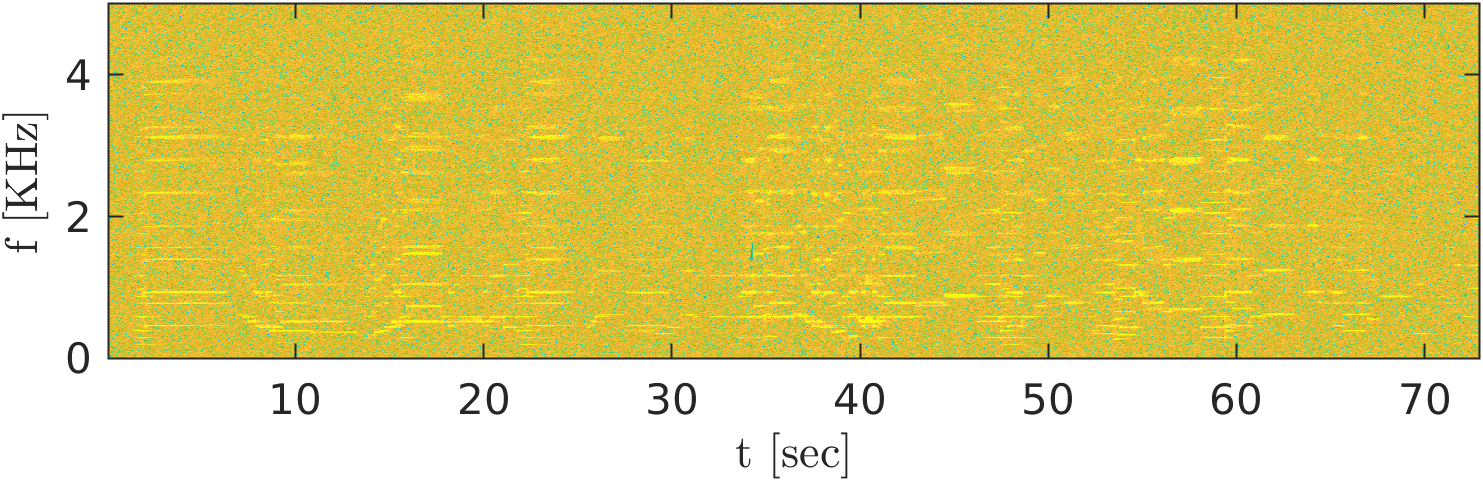}	
	\end{subfigure}
	\begin{subfigure}[t]{0.49\textwidth}
	    \vspace{.3mm}
		\centering
		\caption{Reconstructed ($9.78dB$)}
		\includegraphics[width=1\linewidth]{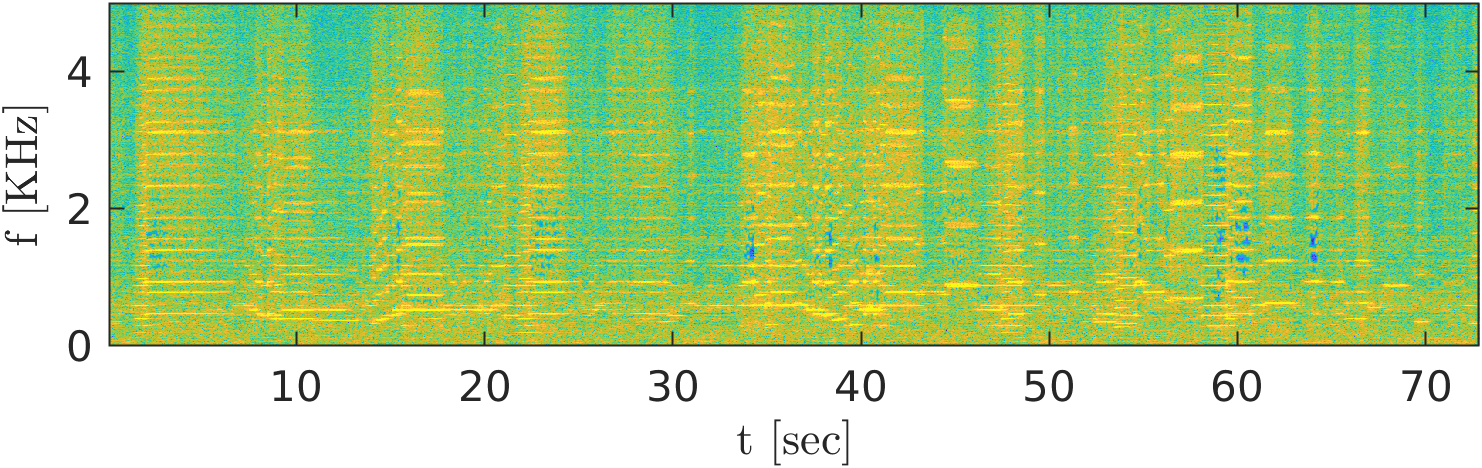}
	\end{subfigure}
	\begin{subfigure}[t]{0.49\textwidth}
	    \vspace{.3mm}
		\centering
		\caption{White noise added ($10dB$)}
		\includegraphics[width=1\linewidth]{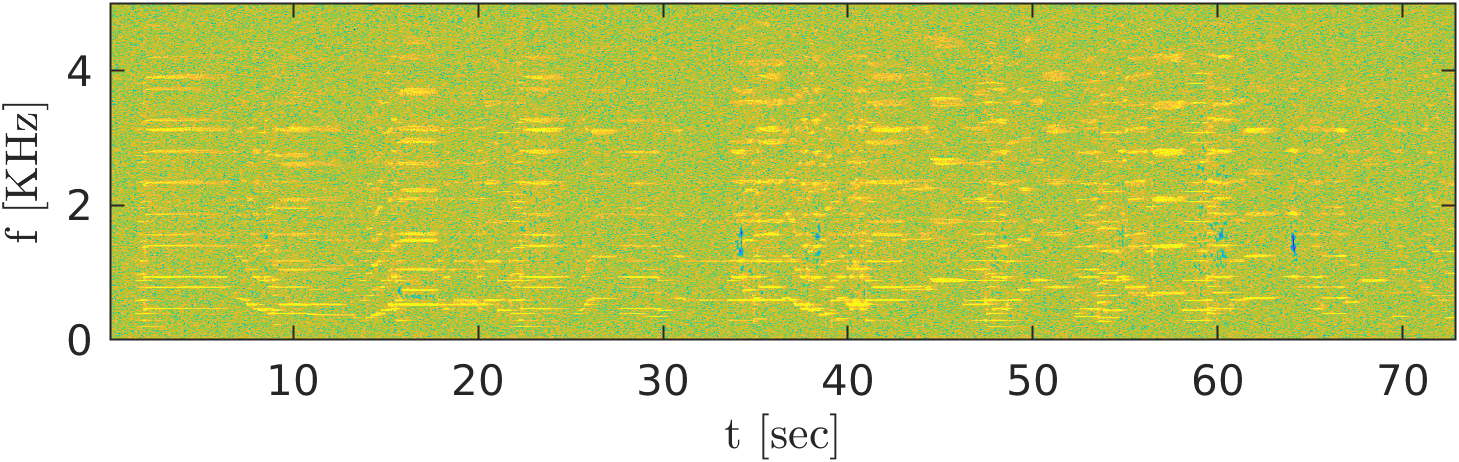}	
	\end{subfigure}
	\begin{subfigure}[t]{0.49\textwidth}
	    \vspace{.3mm}
		\centering
		\caption{Reconstructed ($11.53dB$)}
		\includegraphics[width=1\linewidth]{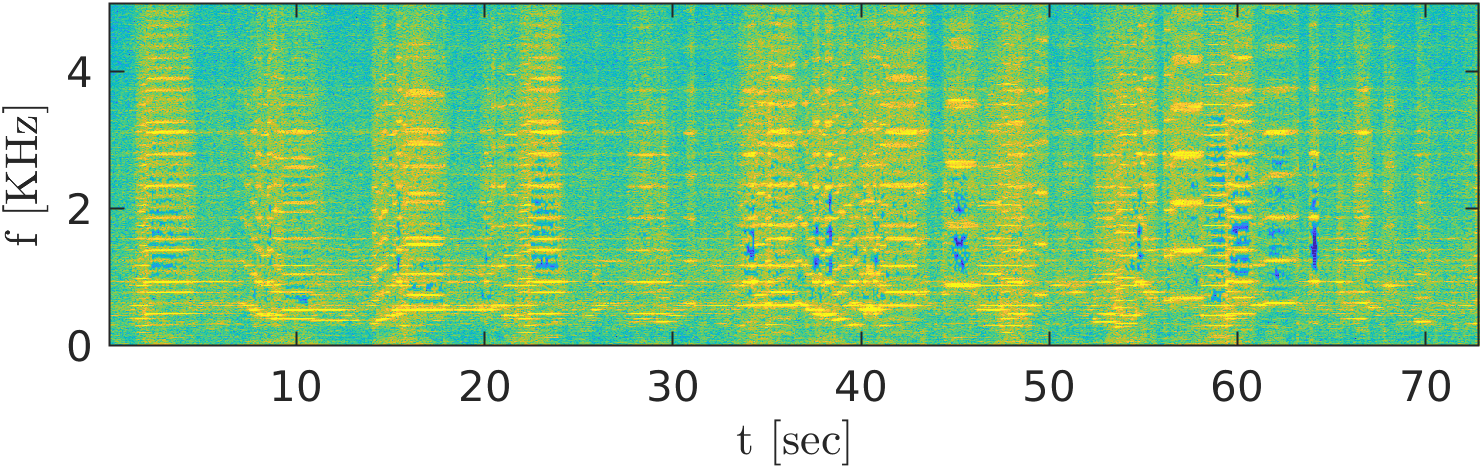}
	\end{subfigure}
	\begin{subfigure}[t]{0.49\textwidth}
	    \vspace{.3mm}
		\centering
		\caption{Gramophone noise added ($5dB$)}
		\includegraphics[width=1\linewidth]{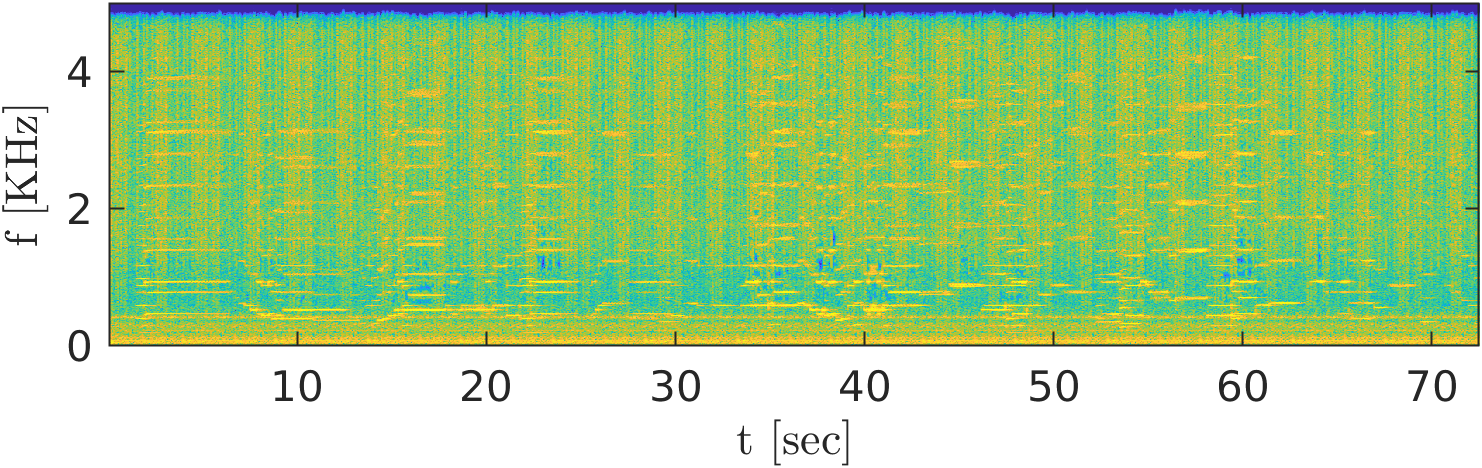}	
	\end{subfigure}
	\begin{subfigure}[t]{0.49\textwidth}
	    \vspace{.3mm}
		\centering
		\caption{Reconstructed ($6.89dB$)}
		\includegraphics[width=1\linewidth]{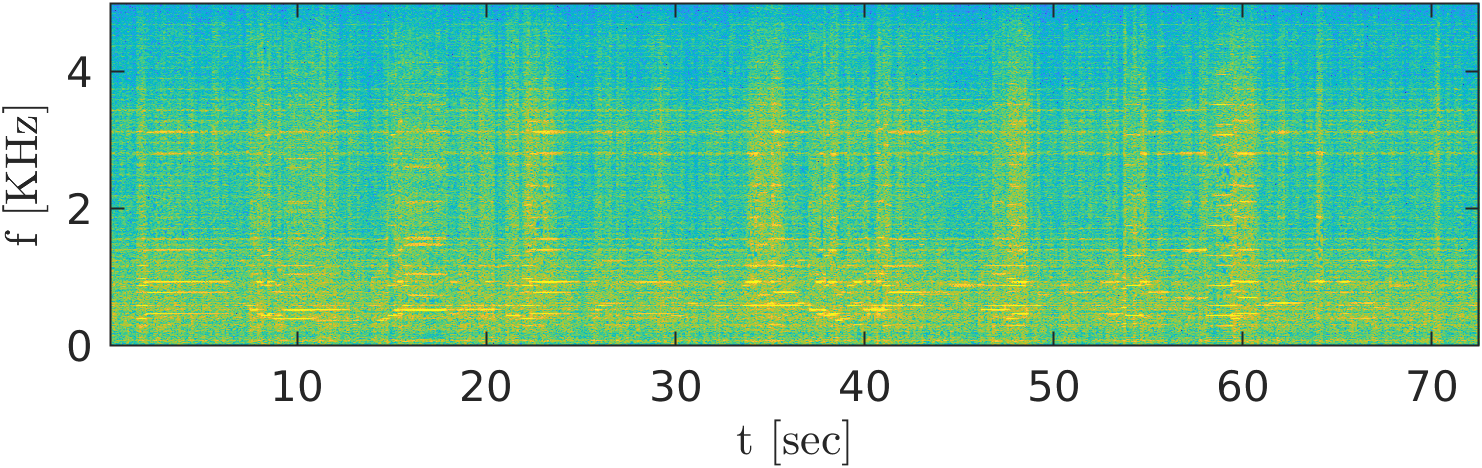}
	\end{subfigure}
	\begin{subfigure}[t]{0.49\textwidth}
	    \vspace{.3mm}
		\centering
		\caption{Gramophone noise added ($10dB$)}
		\includegraphics[width=1\linewidth]{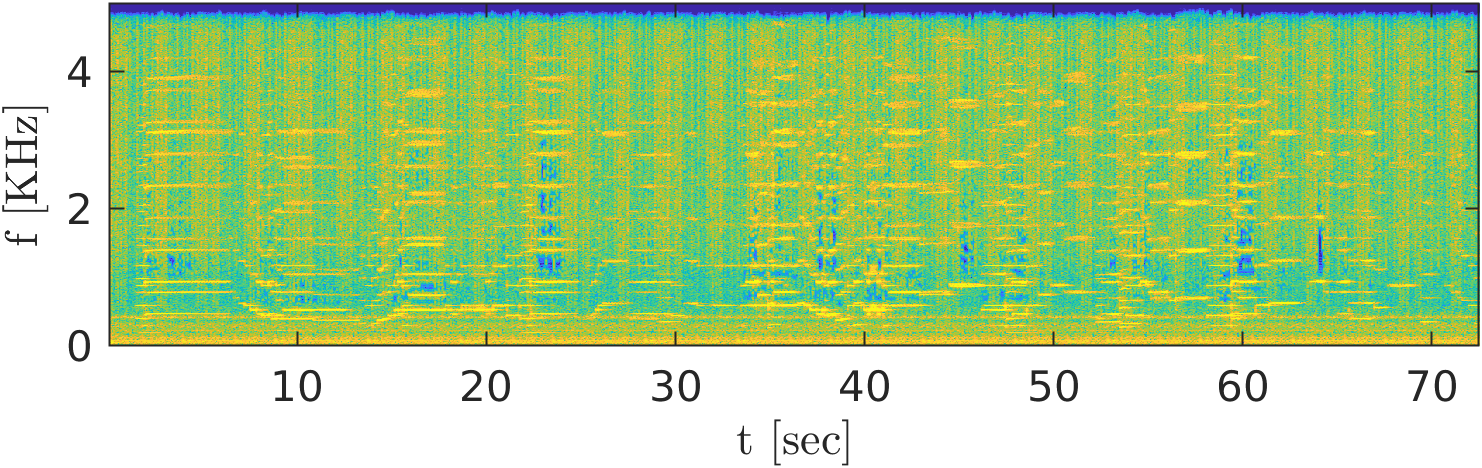}	
	\end{subfigure}
	\begin{subfigure}[t]{0.49\textwidth}
	    \vspace{.3mm}
		\centering
		\caption{Reconstructed ($11.56dB$)}
		\includegraphics[width=1\linewidth]{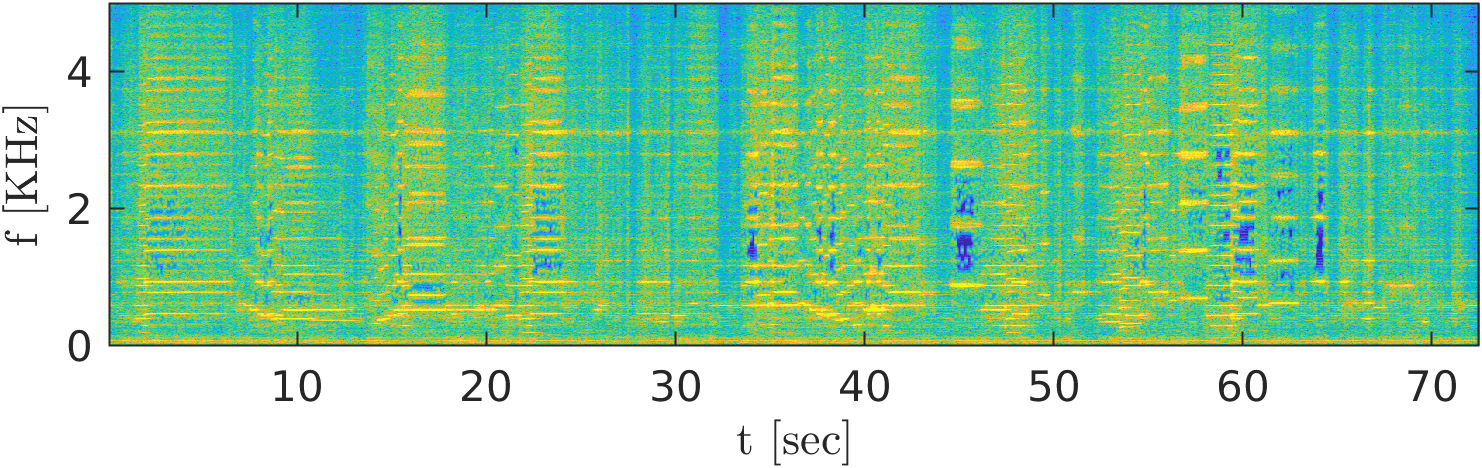}
	\end{subfigure}
	\caption{\textbf{Audio denoising.} The upper row depicts the original violin recording (left) and the recorded gramophone noise (right). The other rows show results of denoising white and recorded noise, at two levels of input SNR. \label{fig:denoising_example_SM} 
	} 
\end{figure*}{}

{}
{}
{}

\subsection{Audio denoising}
As detailed in the main text, we examined denoising of noisy signals that we created by adding white noise and recorded gramophone noise to a clean violin recording. The clean signal, recorded gramophone noise and results of the method are presented in Figure \ref{fig:denoising_example_SM}.

\end{document}